\documentclass[twocolumn,aps,prl,showpacs,superscriptaddress,tightenlines,lengthcheck]{revtex4-1}
\usepackage{amsmath}
\usepackage{amsfonts}
\usepackage{graphicx}
\usepackage{epsfig}
\usepackage{color}
\usepackage[colorlinks,citecolor=blue]{hyperref}
\begin{document}
\title{Nonreciprocal ground-state cooling of multiple mechanical resonators}

\author{Deng-Gao Lai}
\affiliation{Key Laboratory of Low-Dimensional Quantum Structures and Quantum Control of Ministry of Education, Key Laboratory for Matter Microstructure and Function of Hunan Province, Department of Physics and Synergetic Innovation Center for Quantum Effects and Applications, Hunan Normal University, Changsha 410081, China}
\affiliation{Theoretical Quantum Physics Laboratory, RIKEN, Saitama 351-0198, Japan}
\author{Jin-Feng Huang}
\affiliation{Key Laboratory of Low-Dimensional Quantum Structures and Quantum Control of Ministry of Education, Key Laboratory for Matter Microstructure and Function of Hunan Province, Department of Physics and Synergetic Innovation Center for Quantum Effects and Applications, Hunan Normal University, Changsha 410081, China}
\author{Xian-Li Yin}
\affiliation{Key Laboratory of Low-Dimensional Quantum Structures and Quantum Control of Ministry of Education, Key Laboratory for Matter Microstructure and Function of Hunan Province, Department of Physics and Synergetic Innovation Center for Quantum Effects and Applications, Hunan Normal University, Changsha 410081, China}
\author{Bang-Pin Hou}
\affiliation{College of Physics and Electronic Engineering, Institute of Solid State Physics, Sichuan Normal University, Chengdu 610068, China}
\author{Wenlin Li}
\affiliation{School of Science and Technology, Physics Division, University of Camerino, I-62032 Camerino (MC), Italy}
\author{David Vitali}
\affiliation{School of Science and Technology, Physics Division, University of Camerino, I-62032 Camerino (MC), Italy}
\affiliation{INFN, Sezione di Perugia, I-06123 Perugia, Italy}
\affiliation{CNR-INO, L.go Enrico Fermi 6, I-50125 Firenze, Italy}
\author{Franco Nori}
\affiliation{Theoretical Quantum Physics Laboratory, RIKEN, Saitama 351-0198, Japan}
\affiliation{Physics Department, The University of Michigan, Ann Arbor, Michigan 48109-1040, USA}
\author{Jie-Qiao Liao}
\email{jqliao@hunnu.edu.cn}
\affiliation{Key Laboratory of Low-Dimensional Quantum Structures and Quantum Control of Ministry of Education, Key Laboratory for Matter Microstructure and Function of Hunan Province, Department of Physics and Synergetic Innovation Center for Quantum Effects and Applications, Hunan Normal University, Changsha 410081, China}

\begin{abstract}
The simultaneous ground-state cooling of multiple degenerate or near-degenerate mechanical modes coupled to a common cavity-field mode has become an outstanding challenge in cavity optomechanics. This is because the dark modes formed by these mechanical modes decouple from the cavity mode and prevent extracting energy from the dark modes through the cooling channel of the cavity mode. Here we propose a universal and reliable dark-mode-breaking method to realize the simultaneous ground-state cooling of two degenerate or nondegenerate mechanical modes by introducing a phase-dependent phonon-exchange interaction, which is used to form a loop-coupled configuration. We find an asymmetrical cooling performance for the two mechanical modes and expound this phenomenon based on the nonreciprocal energy transfer mechanism, which leads to the directional flow of phonons between the two mechanical modes. We also generalize this method to cool multiple mechanical modes. The physical mechanism in this cooling scheme has general validity and this method can be extended to break other dark-mode and dark-state effects in physics.
\end{abstract}
%42.50.Wk     Mechanical effects of light on material media, microstructures and particles
%42.50.Pq     Cavity quantum electrodynamics; micromasers
%42.50.Dv     Quantum state engineering and measurements
\maketitle

\emph{Introduction.}---Mechanical resonators in cavity optomechanical systems~\cite{Kippenberg2008Science,Meystre2013AP,Aspelmeyer2014RMP} have the advantages of easy resonance, wide compatibility, and tunable coupling to diverse physical devices. These resonators not only provide a promising platform for investigating macroscopic mechanical coherence~\cite{Liao2016PRL,Mancini2002PRL,Tian2004PRL,Hartmann2008PRL,Massel2012Nc,Xiang2013RMP,Mari2013PRL,Matheny2014PRL,Zhang2015PRL,Riedinger2018Nature,Ockeloen-Korppi2018,Stefano2019PRL}, quantum many-body effects~\cite{Heinrich2011PRL,Xuereb2012PRL,Ludwig2013PRL,Xuereb2014PRL,Xuereb2015NJP,Mahmoodian2018PRL}, and topological energy transfer~\cite{Xu2016Nature}, but also can be used as high-performance sensors~\cite{Massel2011Nature,Huang2013PRL,Bernier2016PRL}, transducers~\cite{Rabl2010NP}, and mechanical computers~\cite{Masmanidis2007science,Yamaguchi2008NN}. To suppress thermal noise in those applications, the simultaneous ground-state cooling of these mechanical resonators becomes an obligatory and important task. Though great advances have been made in ground-state cooling of a single mechanical resonator~\cite{Wilson-Rae2007PRL,Marquardt2007PRL,Genes2008PRA,Xia2009PRL,Tian2009PRB,Xuereb2010PRL,Chan2011Nature,Teufel2011Nature,Liu2013PRL,Clarkl2017Nature,Xu2017PRL,Rossi2017PRL,Genes2019PRL}, the simultaneous ground-state cooling of multiple mechanical resonators remains an outstanding challenge in cavity optomechanics~\cite{Genes2008NJP,Sommer2019PRL,Ockeloen-Korppi2019PRA}. The physical origin behind this obstacle is the existence of the dark-mode effect~\cite{Genes2008NJP,Massel2012Nc,Shkarin2014PRL,Sommer2019PRL,Ockeloen-Korppi2019PRA} induced by the multiple mechanical resonators (modes) coupled to a common cavity field, as demonstrated theoretically~\cite{Genes2008NJP,Sommer2019PRL} and experimentally~\cite{Shkarin2014PRL,Ockeloen-Korppi2019PRA}.

In this Rapid Communication, we propose a reliable method to realize the \emph{simultaneous ground-state cooling} of multiple mechanical modes by \emph{breaking the dark-mode effect} in an optomechanical system consisting of a cavity mode coupled to two mechanical modes. This is realized by introducing a phase-dependent phonon-exchange interaction between the two mechanical modes~\cite{seeSM}. Owing to the phase-dependent phonon-exchange interaction in this loop-coupled system, there is no dark mode anymore, and asymmetrical ground state cooling of the two mechanical resonators is realized via an interference effect. We find that the asymmetrical cooling performance is caused by nonreciprocal excitation transfer between the two mechanical modes~\cite{Fang2017NP,Bernier2017Nc,Malz2018PRL,Mathew2018arXiv,Fang2012NP,Fang2012PRL,Hafezi2012OE,Metelmann2015PRX,Shen2016NP,Peterson2017PRX,Kim2017NC,Lodahl2017Nature,Shen2018NC,Seif2018NC,Xu2019Nature}. We also extend this method to the simultaneous cooling of $N$ mechanical resonators and this advance will be helpful for the miniaturization of quantum devices~\cite{Partanen2016NP,Barzanjeh2018PRL}. This dark-mode-breaking mechanism is universal and can be generalized to break the dark-state or dark-mode effects in other physical systems~\cite{seeSM}.

%%%%%%%%%%%%%%%%%%%%%%%%%%%%%%
\begin{figure}[tbp]
\center
\includegraphics[bb=0 0 228 110 width=0.45 \textwidth]{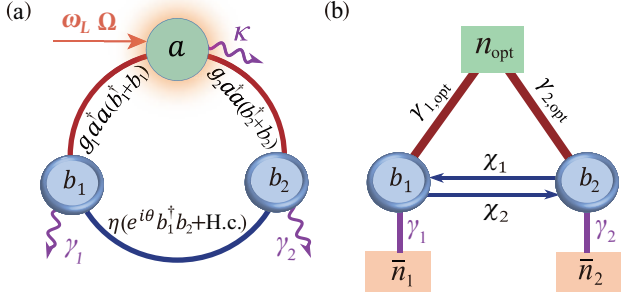}
\caption{(a) A loop-coupled optomechanical system consists of one cavity-field mode $a$ optomechanically coupled to two mechanical modes $b_{1}$ and $b_{2}$, which are coupled with each other via a phase-dependent phonon-exchange coupling (with the coupling strength $\eta$ and phase $\theta$). (b) The reduced two-mechanical-mode system with the effective phonon-exchange channel ($\chi_{l=1,2}$), the common optomechanical-cooling channel ($\gamma_{l,\text{opt}}$, $n_{\text{opt}}$), and the mechanical dissipations ($\gamma_{l=1,2}$, $\bar{n}_{l}$).}
\label{Figmodel}
\end{figure}
%%%%%%%%%%%%%%%%%%%%%%%%%%%%%%

\emph{System.}---We consider a three-mode optomechanical structure [Fig.\ref{Figmodel}(a)] consisting of a cavity field optomechanically coupled to two mechanical modes, which are coupled with each other via a phase-dependent phonon-exchange interaction~\cite{seeSM}. A monochromatic driving field with frequency $\omega_{L}$ and amplitude $\Omega$ is applied to the optical cavity. In a rotating frame defined by $\exp(-i\omega_{L}ta^{\dagger}a)$, the system Hamiltonian reads ($\hbar=1$)~\cite{seeSM}
\begin{eqnarray}
H_{I}&=&\Delta_{c}a^{\dagger}a+\sum_{l=1,2}[\omega_{l}b_{l}^{\dagger }b_{l}+g_{l}a^{\dagger}a(b_{l}+b_{l}^{\dagger})]\nonumber \\
&&+(\Omega a+\Omega^{\ast}a^{\dagger})+\eta(e^{i\theta}b_{1}^{\dagger}b_{2}+e^{-i\theta}b_{2}^{\dagger}b_{1}), \label{eq1iniH}
\end{eqnarray}
where $a$ ($a^{\dagger}$) and $b_{l=1,2}$ ($b^{\dagger}_{l}$) are, respectively, the annihilation (creation) operators of the cavity mode ($\omega_{c}$) and the $l$th mechanical mode ($\omega_{l}$). The $g_{l=1,2}$ terms describe the optomechanical couplings. The $\Omega$ term denotes the cavity-field driving with detuning $\Delta_{c}=\omega_{c}-\omega_{L}$, and the $\eta$ term describes a phase-dependent phonon-exchange interaction between the two mechanical resonators, with the real coupling strength $\eta$ and phase $\theta$. Note that this model can be implemented with either circuit electromechanical systems~\cite{Massel2011Nature,Massel2012Nc} or photonic crystal optomechanical cavity systems~\cite{Fang2017NP}. The phase-dependent phonon-hopping coupling in the electromechanical system can be indirectly induced by coupling to a charge qubit~\cite{seeSM}. In the photonic crystal optomechanical setup, the phase-dependent phonon-hopping coupling has been suggested by using two assistant cavity fields~\cite{Fang2017NP}. In addition, we mention that the two mechanical modes could be either bare mechanical modes in individual mechanical resonators or supermodes of coupled mechanical resonators~\cite{Ramos2014APL,Barzanjeh2016PRA}. For the latter case, the phase-dependent phonon-exchange coupling should be implemented between these supermodes accordingly.

By expressing the operators $o\in$\{$a$, $b_{l=1,2}$, $a^{\dagger}$, $b^{\dagger}_{l=1,2}$\} with their steady-state average values and fluctuations $o=\langle o\rangle_{\text{ss}}+\delta o$, the system can be linearized in the strong-driving regime, and the linearized Hamiltonian in the rotating-wave approximation (RWA) reads
\begin{eqnarray}
H_{\text{RWA}}&=&\Delta \delta a^{\dagger}\delta a+\sum_{l=1,2}[\omega_{l}\delta b_{l}^{\dagger}\delta b_{l}+G_{l}(\delta a\delta b_{l}^{\dagger}+\delta b_{l}\delta a^{\dagger})]\nonumber \\
&&+\eta(e^{i\theta}\delta b_{1}^{\dagger}\delta b_{2}+e^{-i\theta}\delta b_{2}^{\dagger}\delta b_{1}), \label{HRWA1}
\end{eqnarray}
where $\Delta$ is the normalized driving detuning and $G_{l=1,2}=g_{l}\alpha$ are the linearized optomechanical-coupling strengths. The displacement $\alpha\equiv\langle a\rangle_{\text{ss}}=-i\Omega^{*}/(\kappa+i\Delta)$ is assumed to be real by choosing a proper driving amplitude $\Omega$, where $\kappa$ is the decay rate of the cavity field. When $\omega_{1}=\omega_{2}$ and $\eta=0$, there exists a bright mode $B_{+}$ and a dark mode $B_{-}$ defined by~\cite{seeSM}
\begin{align}
B_{\pm}=(G_{1(2)}\delta b_{1}\pm G_{2(1)}\delta b_{2})/\sqrt{G^{2}_{1}+G^{2}_{2}}.
\end{align}
Then $H_{\text{RWA}}=\Delta\delta a^{\dagger}\delta a+\omega_{+}B_{+}^{\dagger}B_{+}+\omega_{-}B_{-}^{\dagger }B_{-}+G_{+}(\delta aB_{+}^{\dagger}+B_{+}\delta a^{\dagger })$ with $G_{+}=\sqrt{G^{2}_{1}+G^{2}_{2}}$. Here the dark mode $B_{-}$ decouples from the cavity mode and the ground-state cooling of the two resonators is unaccessible.
%%%%%%%%%%%%%%%%%%%%%%%%%%%%%%
\begin{figure}[tbp]
\centering
\includegraphics[bb=0 0 371 313, width=0.45 \textwidth]{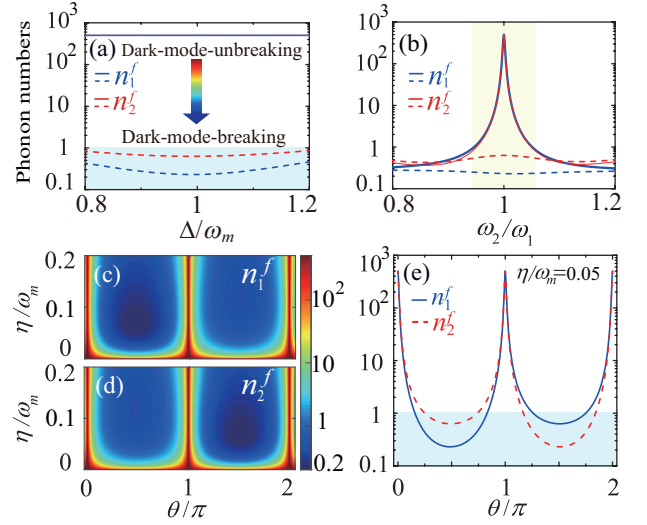}
\caption{(Color online) (a) The final average phonon numbers $n^{f}_{1}$ (blue curves) and $n^{f}_{2}$ (red curves) in the two mechanical resonators versus the effective driving detuning $\Delta$ in the dark-mode-unbreaking ($\eta=0$, solid curves) and -breaking ($\eta=0.05\omega_{m}$ and $\theta=\pi/2$, dashed curves) cases when $\omega_{1}=\omega_{2}=\omega_{m}$. (b) $n^{f}_{1}$ and $n^{f}_{2}$ as functions of $\omega_{2}/\omega_{1}$ in both the dark-mode-unbreaking (solid curves) and -breaking (dashed curves) cases when $\Delta=\omega_{1}$. (c) $n^{f}_{1}$ and (d) $n^{f}_{2}$ vs $\eta$ and $\theta$ under the optimal driving $\Delta=\omega_{m}$ and $\omega_{1}=\omega_{2}=\omega_{m}$. (e) $n^{f}_{1}$ and $n^{f}_{2}$ vs $\theta$ at $\eta=0.05\omega_{m}$. Other used parameters are given by $G_{1}/\omega_{m}=G_{2}/\omega_{m}=0.1$, $\gamma_{1}/\omega_{m}=\gamma_{2}/\omega_{m}=10^{-5}$, $\kappa/\omega_{m}=0.2$, and $\bar{n}_{1}=\bar{n}_{2}=10^{3}$.}
\label{Fig2}
\end{figure}
%%%%%%%%%%%%%%%%%%%%%%%%%%%%%%

\emph{Ground-state cooling by breaking the dark mode.}---To analyze the action of the phonon-exchange interaction, we introduce two bosonic modes $\tilde{B}_{+}=f\delta b_{1}-e^{i\theta}h\delta b_{2}$ and $\tilde{B}_{-}=e^{-i\theta}h\delta b_{1}+f\delta b_{2}$, where the coefficients are given by $f =\vert\tilde{\omega}_{-}-\omega_{1}\vert/\sqrt{(\tilde{\omega}_{-}-\omega_{1})^{2}+\eta^{2}}$ and $h =\eta f/(\tilde{\omega}_{-}-\omega_{1})$, with the resonance frequencies $\tilde{\omega}_{\pm} =\frac{1}{2}(\omega _{1}+\omega _{2}\pm \sqrt{(\omega_{1}-\omega_{2})^{2}+4\eta ^{2}})$ and the coupling strengths $\tilde{G}_{+}=fG_{1}-e^{-i\theta }hG_{2}$ and $\tilde{G}_{-}=e^{i\theta }hG_{1}+fG_{2}$.
The linearized optomechanical Hamiltonian becomes
\begin{eqnarray}
H_{\text{RWA}}&=&\Delta \delta a^{\dagger }\delta a+\tilde{\omega}_{+}\tilde{B}_{+}^{\dagger}\tilde{B}_{+}+\tilde{\omega}_{-}\tilde{B}_{-}^{\dagger }\tilde{B}_{-}+(\tilde{G}_{+}^{\ast}\delta a\tilde{B}_{+}^{\dagger}\nonumber \\
&&+\tilde{G}_{+}\tilde{B}_{+}\delta a^{\dagger})+(\tilde{G}_{-}^{\ast}\delta a\tilde{B}_{-}^{\dagger}+\tilde{G}_{-}\tilde{B}_{-}\delta a^{\dagger}). \label{HRWA3}
\end{eqnarray}
In the degenerate-resonator ($\omega_{1}=\omega_{2}=\omega_{m}$) and symmetric-coupling ($G_{1}=G_{2}=G$) cases, the coupling strengths become
$\tilde{G}_{+} =\sqrt{2}G(1+e^{-i\theta })/2$ and $\tilde{G}_{-}=\sqrt{2}G(1-e^{i\theta })/2$. When $\theta=n\pi$ for an integer $n$, the cavity field is decoupled from one of the two hybrid mechanical modes $\tilde{B}_{-}$ (for even $n$) and $\tilde{B}_{+}$ (for odd $n$). However, in the general case $\theta\neq n\pi$, the dark-mode effect is broken~\cite{seeSM}, and then the simultaneous ground-state cooling becomes accessible under proper parameter conditions. We emphasize that the dark-mode-breaking mechanism is universal and it can be proved by analyzing the eigenstates of a $3\times3$ matrix, which is used to describe either a three-mode system or a three-level system~\cite{seeSM}.

To study the cooling performance of the two mechanical resonators, we calculate the final average phonon numbers $n_{1}^{f}$ and $n_{2}^{f}$ by solving the steady-state covariance matrix governed by the Lyapunov equation~\cite{seeSM}. Figure~\ref{Fig2}(a) shows the phonon numbers $n_{1}^{f}$ and $n_{2}^{f}$ as functions of the driving detuning $\Delta$ when the system works in both the dark-mode-unbreaking ($\eta=0$) and -breaking ($\eta/\omega_{m}=0.05$ and $\theta=\pi/2$) regimes. The results indicate that ground-state cooling of the two mechanical resonators is unfeasible when the system possesses the dark mode [the upper solid curves in Fig.~\ref{Fig2}(a)]. When the dark mode is broken by adding the phonon-exchange coupling [the dashed curves in Fig.~\ref{Fig2}(a)], the emergence of the valley corresponds to ground-state cooling ($n^{f}_{1,2}\ll1$). The phonon-exchange coupling provides the physical origin for breaking the dark mode and builds the channel to transfer the excitation energy between the two mechanical resonators. The optimal driving detuning is located at $\Delta=\omega_{m}$, which is consistent with a typical resolved-sideband cooling~\cite{Wilson-Rae2007PRL,Marquardt2007PRL,Genes2008PRA,Chan2011Nature,Teufel2011Nature}, because the phonons exactly compensate the energy mismatch between the scattered photons and the driving light.

When the phonon-exchange coupling is absent, though the dark mode exists theoretically only in the degenerate-resonator case  (i.e., $\omega_{1}=\omega_{2}$), the dark-mode effect actually works for a wider detuning range in the near-degenerate-resonator case [as marked by the shadow area in Fig.~\ref{Fig2}(b)]~\cite{seeSM}. The width of the shadow area can be characterized by the effective mechanical linewidth $\Gamma_{l}$ ($\Delta\omega=|\omega_{2}-\omega_{1}|\leq\Gamma_{l}$). The cooling of the individual mechanical resonators is suppressed in this region, i.e., the individual mechanical resonators have significant spectral overlap and become effectively degenerate. When the phonon-exchange coupling is applied, the dark-mode effect is broken and the ground-state cooling for the degenerate and near-degenerate resonators becomes feasible [the dashed curves in Fig.~\ref{Fig2}(b)].

The dependence of the final average phonon numbers $n_{1}^{f}$ and $n_{2}^{f}$ on the phonon-exchange parameters $\eta$ and $\theta$ is displayed in Figs.~\ref{Fig2}(c) and~\ref{Fig2}(d). The ground-state cooling of the two mechanical resonators is achievable in the region $0<\theta<\pi$ ($\pi<\theta<2\pi$) for a wide range of $\eta$, and the cooling performance of the first (second) resonator is better than the other one $n^{f}_{1}<n^{f}_{2}$ ($n^{f}_{1}>n^{f}_{2}$). In particular, at $\theta=n\pi$, the two mechanical resonators cannot be cooled to their ground states, which corresponds to the dark-mode-unbreaking case, as shown in Figs.~\ref{Fig2}(c-e).

%%%%%%%%%%%%%%%%%%%%%%%%%%%%%%
\begin{figure}[tbp]
\centering
\includegraphics[bb=0 0 353 141, width=0.45 \textwidth]{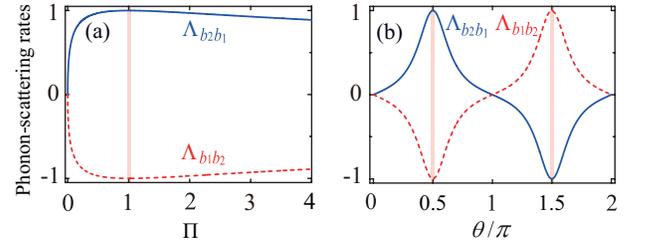}
\caption{(Color online) The relative resonant-phonon-scattering rate $\Lambda_{b_{2}b_{1}}$ (blue solid curves) and $\Lambda_{b_{1}b_{2}}$ (red dashed curves) versus (a) the ratio $\Pi$ of the optomechanical cooperativities  when $\theta=\pi/2$ and (b) the phase $\theta$ when $\Pi=1$. Here $\Delta=\omega_{m}$ and $\omega_{1}=\omega_{2}=\omega_{m}$. Other parameters used are the same as those in Fig.~\ref{Fig2}.}\label{Fig3}
\end{figure}
%%%%%%%%%%%%%%%%%%%%%%%%%%%%%%
\emph{Nonreciprocal phonon transfer.}---To explain the asymmetrical cooling phenomenon in Fig.~\ref{Fig2}(e), we introduce a relative resonant-phonon-scattering rate $\Lambda_{vw}=(T_{vw}-T_{wv})/(T_{vw})_{\text{max}}$ corresponding to the transfer of a phonon with frequency $\omega_{m}$ from modes $w$ to $v$, where $T_{vw}$ denotes the transmittance from modes $w$ to $v$ [$v,w\in\{b_{1}, b_{2}\}$]. The relative resonant-phonon-scattering rates can be expressed as~\cite{seeSM}
\begin{eqnarray}
\Lambda_{b_{2}b_{1}} &=&\frac{4\sqrt{\Pi }\sin \theta }{(1+\sqrt{\Pi })^{2}}\left[1+\frac{4\Pi \cos ^{2}\theta }{\left(\frac{{\mathcal{C}}_{1}+{\mathcal{C}}_{2}+1}{{\mathcal{C}}_{1}{\mathcal{C}}_{2}}+\Pi \right) ^{2}}\right]^{-1},\label{b2b11}
\end{eqnarray}
and $\Lambda_{b_{1}b_{2}}=-\Lambda_{b_{2}b_{1}}$, where $\Pi=\mathcal{C}_{3}/(\mathcal{C}_{1}\mathcal{C}_{2})$ with ${\mathcal{C}}_{l=1,2}=G_{l}^{2}/\gamma_{l}\kappa$ and ${\mathcal{C}}_{3}=\eta ^{2}/\gamma _{1}\gamma_{2}$ being the cooperativities associated with the optomechanical couplings and the phonon-exchange coupling ($\gamma_{l=1,2}$ denoting the decay rate of the $l$th resonator), respectively. The dependence of the relative resonant-phonon-scattering rates $\Lambda_{vw}$ on the ratio $\Pi$ of the optomechanical cooperativities and the phase $\theta$ is shown in Fig.~\ref{Fig3}. In panel (a), we find that in the region $0<\Pi<1$ ($\Pi>1$), $\Lambda_{b_{2}b_{1}}$ increases (decreases) with increasing $\Pi$, and the optimal nonreciprocity ($\Lambda_{b_{2}b_{1}}=1$) emerges at $\Pi=1$, which indicates directional flow of phonons between the two mechanical resonators. As shown in Fig.~\ref{Fig3}(b), when $0<\theta<\pi$, $\Lambda_{b_{2}b_{1}}>0$, i.e., $T_{b_{2}b_{1}}>T_{b_{1}b_{2}}$, the phonon transmission from mechanical mode $b_{1}$ to $b_{2}$ is enhanced, while the transmission in the backward direction is suppressed (see blue solid curves); In the range $\pi<\theta<2\pi$, it exhibits $\Lambda_{b_{1}b_{2}}>0$, i.e., $T_{b_{1}b_{2}}>T_{b_{2}b_{1}}$ (see red dashed curves). Meanwhile, the phonon transmission satisfies the Lorentz reciprocal theorem [$\Lambda_{b_{2}b_{1}}=\Lambda_{b_{1}b_{2}}=0$, i.e., $T_{b_{1}b_{2}} =T_{b_{2}b_{1}}$] at $\theta=n\pi$. Moreover, the transmittance is optimal for the process from $b_{1}$ ($b_{2}$) to $b_{2}$ ($b_{1}$) and is zero for the opposite process when $\theta=\pi/2$ ($\theta=3\pi/2$). We see from Eq.~(\ref{b2b11}) that, when $\Pi=1$ and $\theta=\pi/2$, an excellent nonreciprocal phonon transfer ($\Lambda_{b_{2}b_{1}}=1$) is realized.

%%%%%%%%%%%%%%%%%%%%%%%%%%%%%%
\begin{figure}[tbp]
\centering
\includegraphics[bb=0 0 358 285, width=0.45 \textwidth]{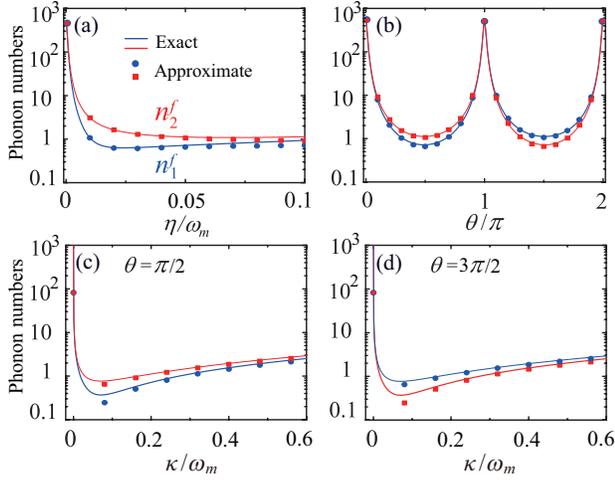}
\caption{(Color online) The exact and approximate final average phonon numbers $n^{f}_{1}$ (blue) and $n^{f}_{2}$ (red) versus (a) the phonon-exchange coupling strength $\eta$ when $\theta=\pi/2$ and $\kappa/\omega_{m}=0.2$, (b) the phase $\theta$ when $\eta/\omega_{m}=0.05$ and $\kappa/\omega_{m}=0.2$, and (c,d) the cavity-field decay rate $\kappa$ when $\eta/\omega_{m}=0.05$ for (c) $\theta=\pi/2$ and (d) $\theta=3\pi/2$. The solid curves and the symbols correspond to the exact (numerical) and approximate (analytical) results, respectively. Here $\Delta/\omega_{m}=1$ and $G_{1}/\omega_{m}=G_{2}/\omega_{m}=0.05$. Other parameters used are the same as those in Fig.~\ref{Fig2}.}\label{Fig4}
\end{figure}
%%%%%%%%%%%%%%%%%%%%%%%%%%%%%%
\emph{Cooling limits.}---The cooling limits can be analytically obtained in the large cavity-field-decay regime, in which the cavity field is eliminated adiabatically such that the three-mode optomechanical system is reduced to a two-mode system described by the Hamiltonian $\tilde{H}_{\text{eff}}=\sum_{l=1}^{2}(\Omega _{l}-i\Gamma_{l})b_{l}^{\dagger}b_{l}+i\xi_{1}b_{1}^{\dagger}b_{2}+i\xi_{2}b_{2}^{\dagger}b_{1}$ [Fig.~\ref{Figmodel}(b)], where $\Gamma_{l}=\gamma_{l}+\gamma_{l,\text{opt}}$ and $\Omega_{l}=\omega_{l}-\omega_{l,\text{opt}}$ are, respectively, the effective decay rate and resonance frequency for the $l$th mechanical resonator, with the optical induced decay rates $\gamma_{l,\text{opt}}=G_{l}^{2}/\kappa$ and mechanical frequency shifts $\omega_{l,\text{opt}}=G_{l}^{2}/2\omega_{l}$. In addition, $i\xi_{l}$ is the effective phonon-exchange coupling strength between the two mechanical modes with $\xi_{1(2)}=-[G_{1}G_{2}/\kappa +i(\eta e^{\pm i\theta}-G_{1}G_{2}/2\omega_{2(1)})]$. The mechanical mode $b_{l}$ is contacted to an effective optomechanical cooling bath ($\gamma_{l,\text{opt}}$ and $n_{\text{opt}}$) and a heat bath ($\gamma_{l}$ and $\bar{n}_{l}$). Considering the parameter relations $\omega_{1,2}\gg\kappa\gg G_{1,2}\gg\{\gamma_{1,\text{opt}}\approx\gamma_{2,\text{opt}}\}\gg\gamma_{1,2}$, the final average phonon occupations can be obtained as~\cite{seeSM}
\begin{eqnarray}
n_{l=1,2}^{f}&\approx&\frac{\gamma_{l}\bar{n}_{l}+\gamma_{l,\text{opt}}n_{\text{opt}}}{\Gamma_{l}+\chi_{+}}+\frac{(-1)^{l-1}\sqrt{\chi_{l}}}{\Gamma_{l}+\chi_{-}}\nonumber\\
&&\times(\sqrt{\chi_{1}}n_{\chi_{1}}-\sqrt{\chi_{2}}n_{\chi_{2}}),\label{coolimit1}
\end{eqnarray}
where $n_{\text{opt}}=4\kappa^{2}/(\omega_{1}+\omega_{2}+2\Delta)^{2}$, $n_{\chi_{1(2)}}=2(\gamma _{2(1)}\bar{n}_{2(1)}+\gamma_{2(1),\text{opt}}n_{\text{opt}})/(\Gamma_{1}+\Gamma_{2}+2\chi_{+})$, and $\chi_{\pm}=\mp\sqrt{\chi_{1}\chi_{2}}-\mathrm{Re}[\xi_{1}\xi_{2}/(\Gamma_{1}+\Gamma_{2})]$, with $\chi_{l=1,2}=\vert \xi_{l}\vert ^{2}/(\Gamma_{1}+\Gamma_{2})$ being the effective phonon-transfer rate from $b_{2}$ ($b_{1}$) to $b_{1}$ ($b_{2}$). The cooling limits ($n_{l}^{\lim}$) are obtained at $\Delta=\omega_{l}$. In Fig.~\ref{Fig4}, we plot the exact final average phonon numbers (solid lines) and the cooling limits (symbols) given by Eq.~(\ref{coolimit1}) as functions of the phonon-exchange parameters $\eta$ and $\theta$. Figure~\ref{Fig4} shows asymmetrical ground-state cooling and excellent agreement between numerical and analytical results.

The first term in Eq.~(\ref{coolimit1}) is caused by the thermal bath and the effective optical bath connected by the $l$th mechanical mode, while the phonon extraction by the phonon-exchange channel is described by the last term. Physically, the nonreciprocity of the phonon transfer is determined by the phonon-exchange rate $\chi_{l}$ which depends on the phase $\theta$. For the case: $\bar{n}_{1}\approx\bar{n}_{2}$ and $\gamma_{1}\approx \gamma_{2}$, we have $n_{\chi_{1}}\approx n_{\chi_{2}}=n_{\chi}$ and thus $(\sqrt{\chi_{1}}n_{\chi_{1}}-\sqrt{\chi_{2}}n_{\chi_{2}})\approx (\sqrt{\chi_{1}}-\sqrt{\chi_{2}})n_{\chi}$ [see Eq.~(\ref{coolimit1})]. In the range $0<\theta<\pi$ ($\pi<\theta<2\pi$), we obtain $\sqrt{\chi_{1}}<\sqrt{\chi_{2}}$ ($\sqrt{\chi_{1}}>\sqrt{\chi_{2}}$). This means that the phonon-transfer efficiency from $b_{1}$ ($b_{2}$) to $b_{2}$ ($b_{1}$) is larger than that for the opposite case, i.e., $n^{f}_{1}<n^{f}_{2}$ ($n^{f}_{1}>n^{f}_{2}$) [see Fig.~\ref{Fig4}(b)]. When $\theta=\pi/2$ ($3\pi/2$) and $\sqrt{{\mathcal{C}}_{1}{\mathcal{C}}_{2}}=\sqrt{{\mathcal{C}}_{3}}$, the unidirectional flow of the phonons between the two mechanical resonators is obtained [$\chi_{1}\approx0$ ($\chi_{2}\approx0$)]. For $\theta=n\pi$, the phonon transfer between the two mechanical resonators is reciprocal ($\sqrt{\chi_{1}}=\sqrt{\chi_{2}}$), due to the emergence of the dark mode. In the absence of the phonon-transfer interaction ($\eta=0$), the ground-state cooling is unfeasible due to the invalid effective cooling channel ($\Gamma_{l}+\chi_{+}\rightarrow\gamma_{l}$) [see Fig.~\ref{Fig4}(a)]. In the absence of the optomechanical cooling channels ($G_{1,2}=0$), Eq.~(\ref{coolimit1}) becomes $n_{l=1,2}^{f}\approx \bar{n}_{l}+(-1)^{l-1}(n_{\chi_{1}}-n_{\chi_{2}})/2$, which indicates quantum thermalization in the coupled mechanical system.

\emph{Cooling $N$ mechanical resonators.}---Our proposal can be extended to the cooling of a net-coupled system: a cavity mode coupled to $N\geq 3$ mechanical modes via the optomechanical couplings $H_{\text{opc}}=\sum_{j=1}^{N}g_{j}a^{\dagger}a(b_{j}+b_{j}^{\dagger})$, and the nearest-neighboring mechanical modes are coupled through the phase-dependent phonon-exchange couplings $H_{\text{pec}}=\sum_{j=1}^{N-1}\eta_{j}(e^{i\theta_{j}}b_{j}^{\dagger}b_{j+1}+\mathrm{H.c.})$. We find that the function of these phases in the optomechanical interactions is determined by the term $\sum_{\nu =1}^{j-1}\theta_{\nu}$~\cite{seeSM} and hence, for convenience, we assume $\theta_{1}=\pi/2$ and $\theta_{j}=0$ for $j=2$ - $(N-1)$ in our simulations. In the dark-mode-unbreaking case ($\eta_{j}=0$), the ground-state cooling of the mechanical resonators is unfeasible, with the final average phonon numbers $\bar{n}(N-1)/N$ in the case of $\bar{n}_{j}=\bar{n}$~\cite{seeSM}. When the dark modes are broken, simultaneous ground-state cooling can be realized in this system ($n^{f}_{j}<1$).

\emph{Conclusions.}---We proposed a dark-mode-breaking method to realize simultaneous ground-state cooling of multiple mechanical modes coupled to a common cavity mode by constructing a loop-coupled optomechanical system with a phase drop. We found an asymmetric cooling phenomenon and expounded it using the nonreciprocal phonon exchange mechanism. The present physical mechanism is universal and hence it will motivate the manipulation of various dark-state related physical effects.

\emph{Acknowledgments.}---D.-G.L. thanks Yue-Hui Zhou and Dr. Wei Qin for valuable discussions. J.-Q.L. is supported in part by National Natural Science Foundation of China (Grants No.~11822501, No.~11774087, and No.~11935006), Natural Science Foundation of Hunan Province, China (Grant No.~2017JJ1021), and Hunan Science and Technology Plan Project (Grant No.~2017XK2018). D.-G.L. is supported in part by Hunan Provincial Postgraduate Research and Innovation project (Grant No.~CX2018B290). J.-F.H. is supported in part by the National Natural Science Foundation of China (Grant No. 11505055) and Scientific Research Fund of Hunan Provincial Education Department (Grant No. 18A007). B.-P.H. is supported in part by NNSFC (Grant No.~11974009). W.L. and D.V. are supported by the European Union Horizon 2020 Programme for Research and Innovation through the Project No. 732894 (FET Proactive HOT) and the Project QuaSeRT funded by the QuantERA ERA-NET Cofund in Quantum Technologies. F.N. is supported in part by: NTT Research, Army Research Office (ARO) (Grant No. W911NF-18-1-0358), Japan Science and Technology Agency (JST) (via the CREST Grant No. JPMJCR1676), Japan Society for the Promotion of Science (JSPS) (via the KAKENHI Grant No. JP20H00134, and the JSPS-RFBR Grant No. JPJSBP120194828), and the Foundational Questions Institute Fund (FQXi) (Grant No. FQXi-IAF19-06), a donor advised fund of the Silicon Valley Community Foundation.

%%%%%%%%%% Merge with supplemental materials %%%%%%%%%%

\onecolumngrid
\newpage
%%%%%%%%%% Prefix a "S" to all equations, figures, tables and reset the counter %%%%%%%%%%
\setcounter{equation}{0} \setcounter{figure}{0}
\setcounter{table}{0}
\setcounter{page}{1}\setcounter{secnumdepth}{3} \makeatletter
\renewcommand{\theequation}{S\arabic{equation}}
\renewcommand{\thefigure}{S\arabic{figure}}
\renewcommand{\bibnumfmt}[1]{[S#1]}
\renewcommand{\citenumfont}[1]{S#1}
\renewcommand\thesection{S\arabic{section}}
%%%%%%%%%% Prefix a "S" to all equations, figures, tables and reset the counter %%%%%%%%%%

\begin{center}
{\large \bf Supplementary Material for ``Nonreciprocal Ground-State Cooling of Multiple Mechanical Resonators"}
\end{center}

\begin{center}
Deng-Gao Lai$^{1,2}$, Jin-Feng Huang$^{1}$, Xian-Li Yin$^{1}$, Bang-Pin Hou$^{3}$, Wenlin Li$^{4}$, David Vitali$^{4,5,6}$,
Franco Nori$^{2,7}$, and Jie-Qiao Liao$^{1,^*}$
\end{center}

\begin{minipage}[]{18cm}
\small{\it
\centering $^{1}$Key Laboratory of Low-Dimensional Quantum Structures and Quantum Control of Ministry of Education,  \\
\centering Key Laboratory for Matter Microstructure and Function of Hunan Province, \\
\centering Department of Physics and Synergetic Innovation Center for Quantum  Effects and Applications, \\
\centering Hunan Normal University, Changsha 410081, China \\
\centering $^{2}$Theoretical Quantum Physics Laboratory, RIKEN, Saitama 351-0198, Japan \\
\centering $^{3}$College of Physics and Electronic Engineering, Institute of Solid State Physics,
\centering Sichuan Normal University, Chengdu 610068, China \\
\centering $^{4}$School of Science and Technology, Physics Division, University of Camerino, I-62032 Camerino (MC), Italy \\
\centering $^{5}$INFN, Sezione di Perugia, I-06123 Perugia, Italy \\
\centering $^{6}$CNR-INO, L.-argo Enrico Fermi 6, I-50125 Firenze, Italy\\
\centering $^{7}$Physics Department, The University of Michigan, Ann Arbor, Michigan 48109-1040, USA\\}

\end{minipage}

\vspace{8mm}

This document consists of ten parts: (I) The dark-mode effect and its breaking in a two-mechanical-resonator optomechanical system;
(II) Ground-state cooling of the two mechanical resonators; (III) Phonon scattering probability and nonreciprocal phonon transfer; (IV) Derivation of the cooling limits of the two mechanical resonators; (V) Analyzing the dark-mode effect and breaking the dark-mode effect in a multi-mechanical-resonator optomechanical system; (VI) Ground-state cooling of the multiple mechanical resonators; (VII) Discussions on the justification of performing the rotating-wave approximation (RWA); (VIII) Simultaneous cooling of the mechanical supermodes; (IX) Physical mechanism for breaking the dark-state effect in a Lambda-type three-level system; (X) A possible experimental realization and derivation of a phase-dependent phonon-hopping interaction between two mechanical resonators.

\section{The dark-mode effect and its breaking in a two-mechanical-resonator optomechanical system \label{darkmode}}

In this section, we analyze the dark-mode effect in a two-mechanical-resonator optomechanical system, which is composed of one cavity-field mode and two mechanical resonators. Note that here we only consider one mechanical mode in each mechanical resonator. We also show that the dark-mode effect can be broken by introducing a phase-dependent phonon-exchange interaction between the two mechanical resonators. In a rotating frame defined by the transform operator $\exp(-i\omega_{L}ta^{\dagger}a)$, the total Hamiltonian of the system reads ($\hbar =1$)
\begin{eqnarray}
H_{I}&=&\Delta_{c}a^{\dagger}a+\omega_{1}b_{1}^{\dagger}b_{1}+\omega_{2}b_{2}^{\dagger}b_{2}+g_{1}a^{\dagger}a(b_{1}+b_{1}^{\dagger})+g_{2}a^{\dagger}a(b_{2}+b_{2}^{\dagger})+(\Omega a+\Omega^{\ast}a^{\dagger})+\eta(e^{i\theta}b_{1}^{\dagger}b_{2}+e^{-i\theta}b_{2}^{\dagger}b_{1}), \label{eq1iniH}
\end{eqnarray}
where $\Delta_{c}=\omega_{c}-\omega_{L}$ is the detuning of the cavity-field resonance frequency $\omega_{c}$ with respect to the cavity-field driving frequency $\omega_{L}$. The operators $a$ ($a^{\dagger}$) and $b_{l=1,2}$ ($b^{\dagger}_{l}$) are, respectively, the annihilation (creation) operators of the cavity-field mode and the $l$th mechanical resonator, with the corresponding resonance frequencies $\omega_{c}$ and $\omega_{l}$. The $g_{1}$ and $g_{2}$ terms in Hamiltonian~(\ref{eq1iniH}) describe the optomechanical coupling between the cavity mode and the $l$th mechanical resonator, with $g_{l=1,2}$ being the single-photon optomechanical-coupling strength. The $\Omega$ term denotes the cavity-field driving with the driving amplitude $\Omega$. To control the energy exchange between the two mechanical resonators, we introduce a phase-dependent phonon-exchange interaction between the two mechanical resonators, with the coupling strength $\eta$ and the phase $\theta$.

According to Hamiltonian~(\ref{eq1iniH}), the Langevin equations for the annihilation operators of the optical and mechanical modes can be obtained by phenologically adding the dissipation and noise terms into the Heisenberg equations of motion as
\begin{subequations}
\label{Langevineqorig}
\begin{align}
\dot{a}=&-\{\kappa+i[ \Delta_{c}+g_{1}(b_{1}+b_{1}^{\dagger})+g_{2}(b_{2}+b_{2}^{\dagger})]\}a-i\Omega^{\ast}+\sqrt{2\kappa}a_{\text{in}},\\
\dot{b}_{1}=&-(\gamma_{1}+i\omega_{1})b_{1}-ig_{1}a^{\dagger}a-i\eta e^{i\theta}b_{2}+\sqrt{2\gamma_{1}}b_{1,\text{in}},\\
\dot{b}_{2}=&-(\gamma_{2}+i\omega_{2})b_{2}-ig_{2}a^{\dagger}a-i\eta e^{-i\theta}b_{1}+\sqrt{2\gamma_{2}}b_{2,\text{in}},
\end{align}
\end{subequations}
where $\kappa$ and $\gamma_{l=1,2}$ are the decay rates of the cavity-field mode and the $l$th mechanical resonator, respectively. The operators $a_{\text{in}}$ and $b_{l=1,2,\text{in}}$ ($a^{\dagger}_{\text{in}}$ and $b^{\dagger}_{l,\text{in}}$) are the noise operators associated with the cavity-field mode and the $l$th mechanical resonator, respectively. These noise operators have zero mean values and the following correlation functions,
\begin{subequations}
\label{correlationfun}
\begin{align}
\langle a_{\text{in}}(t) a_{\text{in}}^{\dagger}(t^{\prime})\rangle=&\delta(t-t^{\prime}),\\
\langle a_{\text{in}}^{\dagger}(t) a_{\text{in}}(t^{\prime})\rangle=&0,\\
\langle b_{l,\text{in}}(t) b_{l,\text{in}}^{\dagger}(t^{\prime})\rangle=&(\bar{n}_{l}+1)\delta(t-t^{\prime}),\\
\langle b_{l,\text{in}}^{\dagger}(t) b_{l,\text{in}}(t^{\prime})\rangle=&\bar{n}_{l}\delta(t-t^{\prime}),
\end{align}
\end{subequations}
where $\bar{n}_{l=1,2}$ is the average thermal-phonon occupation number associated with the heat bath of the $l$th mechanical resonator. In this paper we consider a vacuum bath for the cavity field and a heat bath (with $\bar{n}_{l=1,2}$) for each mechanical resonator. The vacuum bath of the cavity field provides the cooling reservoir to absorb the thermal excitations extracted from the two mechanical resonators.

To cool the mechanical resonators, we consider the strong-driving regime of the cavity such that the average photon number in the cavity is sufficiently large and then the linearization procedure can be used to simplify the physical model. To this end, we expand the quantum fluctuations of the system around their steady-state values and express the operators in Eq.~(\ref{Langevineqorig}) as a summation of their steady-state mean values and quantum fluctuations, namely $o=\langle o\rangle_{\text{ss}}+\delta o$
for operators $o=a$, $a^{\dagger}$, $b_{l=1,2}$, and $b^{\dagger}_{l=1,2}$. By separating the classical motion and quantum fluctuations, the linearized equations of motion for quantum fluctuations can be written as
\begin{subequations}
\label{lineLangevineq}
\begin{align}
\delta\dot{a}=&-(\kappa +i\Delta) \delta a-iG_{1}(\delta b_{1}+ \delta b_{1}^{\dagger})-iG_{2}(\delta b_{2}+\delta b_{2}^{\dagger})+\sqrt{2\kappa}a_{\text{in}},\\
\delta\dot{b}_{1}=&-iG_{1}^{\ast}\delta a-(\gamma_{1}+i\omega_{1}) \delta b_{1}-i\eta e^{i\theta}\delta b_{2}-iG_{1}\delta a^{\dagger}+\sqrt{2\gamma_{1}}b_{1,\text{in}},\\
\delta\dot{b}_{2}=&-iG_{2}^{\ast}\delta a-i\eta e^{-i\theta}\delta b_{1}-(\gamma_{2}+i\omega_{2})\delta b_{2}-iG_{2}\delta a^{\dagger}+\sqrt{2\gamma_{2}}b_{2,\text{in}},
\end{align}
\end{subequations}
where $\Delta=\Delta_{c}+2(g_{1}\text{Re}[\beta_{1}]+g_{2}\text{Re}[\beta_{2}])$ is the normalized driving detuning of the cavity field with $\text{Re}[\beta_{l}]$ extracting the real part of $\beta_{l}$, and $G_{l=1,2}=g_{l}\alpha$ is the strength of the linearized optomechanical coupling between the cavity field and the $l$th mechanical resonator. Here, the steady-state solutions of the classical motion (namely the steady-state average values of the operators of the system) can be obtained as
\begin{subequations}
\begin{align}
\alpha \equiv& \langle a\rangle_{\text{ss}}=\frac{-i\Omega^{\ast}}{\kappa +i\Delta},\\
\beta_{1}\equiv&\langle b_{1}\rangle_{\text{ss}}=\frac{-i\left(g_{1}\vert\alpha\vert^{2}+\eta e^{i\theta}\beta_{2}\right)}{\gamma_{1}+i\omega_{1}},\\
\beta_{2}\equiv&\langle b_{2}\rangle_{\text{ss}}=\frac{-i\left(g_{2}\vert\alpha\vert^{2}+\eta e^{-i\theta}\beta_{1}\right)}{\gamma_{2}+i\omega_{2}}.
\end{align}
\end{subequations}
For simplicity, in the following discussions we consider the case where $\alpha$ is real, which is accessible by choosing a proper driving amplitude $\Omega$. Then the linearized optomechanical coupling strengths $G_{1}$ and $G_{2}$ are real.

A linearized optomechanical Hamiltonian can be inferred according to Eqs.~(\ref{lineLangevineq}). For studying quantum cooling of the two mechanical resonators, the beam-splitting-type interactions (i.e., the rotating-wave interaction term) between these bosonic modes are expected to dominate the linearized couplings in this system, and hence we can simplify the Hamiltonian of the system by making the rotating-wave approximation (RWA). The linearized optomechanical Hamiltonian in the RWA takes the following form (discarding the noise terms)
\begin{eqnarray}
H_{\text{RWA}}&=&\Delta \delta a^{\dagger}\delta a+\omega_{1}\delta b_{1}^{\dagger}\delta b_{1}+\omega_{2}\delta b_{2}^{\dagger}\delta b_{2}+G_{1}(\delta a\delta b_{1}^{\dagger}+\delta b_{1}\delta a^{\dagger}) +G_{2}(\delta a\delta b_{2}^{\dagger}+\delta b_{2}\delta a^{\dagger})+\eta(e^{i\theta}\delta b_{1}^{\dagger}\delta b_{2}+e^{-i\theta}\delta b_{2}^{\dagger}\delta b_{1}),\label{RWAH}\nonumber\\
\end{eqnarray}
where $\delta a$ ($\delta a^{\dagger}$) and $\delta b_{l=1,2}$ ($\delta b_{l}^{\dagger}$) are the fluctuation operators of the cavity-field mode and the $l$th mechanical resonator, respectively.

To see the dark-mode effect in this two-mechanical-resonator optomechanical system, we first consider the case where the phase-dependent phonon-exchange interaction between the two mechanical resonators is absent, i.e., $\eta=0$. In this case, the coupled two-mechanical-mode system forms two hybrid mechanical modes: a bright mode and a dark mode, which are expressed by the new annihilation operators as
\begin{subequations}
\begin{align}
B_{+}=&\frac{1}{\sqrt{G^{2}_{1}+G^{2}_{2}}}(G_{1}\delta b_{1}+G_{2}\delta b_{2}),\\
B_{-}=&\frac{1}{\sqrt{G^{2}_{1}+G^{2}_{2}}}(G_{2}\delta b_{1}-G_{1}\delta b_{2}).\label{BDmodedef}
\end{align}
\end{subequations}
These new operators satisfy the bosonic commutation relations $[B_{+},B^{\dagger}_{+}]=1$ and $[B_{-},B^{\dagger}_{-}]=1$. In the absence of the phonon-exchange interaction ($\eta=0$), the Hamiltonian in Eq.~(\ref{RWAH}) can be rewritten with the two hybrid modes as
\begin{eqnarray}
H_{\text{hyb}} &=&\Delta\delta a^{\dagger }\delta a+\omega_{+}B_{+}^{\dagger}B_{+}+\omega_{-}B_{-}^{\dagger }B_{-}+\zeta(B_{+}^{\dagger}B_{-}+B_{-}^{\dagger}B_{+})+G_{+}
(\delta aB_{+}^{\dagger}+B_{+}\delta a^{\dagger }),\label{Parity}
\end{eqnarray}
where we introduce the resonance frequencies $\omega_{\pm}$ and the coupling strengths $\zeta$ and $G_{+}$
\begin{subequations}
\begin{align}
\omega_{+} =&\frac{G^{2}_{1}\omega_{1}+G^{2}_{2}\omega_{2}}{G^{2}_{1}+G^{2}_{2}},\\
\omega_{-} =&\frac{G^{2}_{2}\omega_{1}+G^{2}_{1}\omega_{2}}{G^{2}_{1}+G^{2}_{2}},\\
\zeta =&\frac{G_{1}G_{2}(\omega_{1}-\omega_{2})}{G^{2}_{1}+G^{2}_{2}},\\
G_{+}=&\sqrt{G^{2}_{1}+G^{2}_{2}}.\label{ParityG}
\end{align}
\end{subequations}
When $\omega_{1}=\omega_{2}$, the two hybrid modes are decoupled from each other due to $\zeta=0$, and the mode $B_{-}$ becomes a dark mode in the sense that it is decoupled from both the cavity mode $a$ and the other hybrid mode $B_{+}$.

In order to break the dark-mode effect, we introduce a phase-dependent phonon-exchange interaction (i.e., the $\eta$ term) between the two mechanical resonators. By introducing two new bosonic modes $\tilde{B}_{+}$ and $\tilde{B}_{-}$ defined by
\begin{subequations}
\begin{align}
\delta b_{1}=& f\tilde{B}_{+}+e^{i\theta}h\tilde{B}_{-},\\
\delta b_{2}=& -e^{-i\theta}h\tilde{B}_{+}+f\tilde{B}_{-},
\end{align}
\end{subequations}
Hamiltonian~(\ref{RWAH}) becomes
\begin{eqnarray}
H_{\text{RWA}}&=&\Delta \delta a^{\dagger }\delta a+\tilde{\omega}_{+}\tilde{B}_{+}^{\dagger}\tilde{B}_{+}+\tilde{\omega}_{-}\tilde{B}_{-}^{\dagger}\tilde{B}_{-} +(\tilde{G}_{+}^{\ast}\delta a\tilde{B}_{+}^{\dagger}+\tilde{G}_{+}\tilde{B}_{+}\delta a^{\dagger})+(\tilde{G}_{-}^{\ast}\delta a\tilde{B}_{-}^{\dagger}+\tilde{G}_{-}\tilde{B}_{-}\delta a^{\dagger}),\label{DigH}
\end{eqnarray}
where we introduce the resonance frequencies $\tilde{\omega}_{\pm}$ and the coupling strengths $\tilde{G}_{\pm}$ as
\begin{subequations}
\label{DigG}
\begin{align}
\tilde{\omega}_{\pm}=&\frac{1}{2}(\omega _{1}+\omega _{2}\pm \sqrt{(\omega_{1}-\omega_{2})^{2}+4\eta ^{2}}),\\
\tilde{G}_{+}=&fG_{1}-e^{-i\theta }hG_{2},\\
\tilde{G}_{-}=&e^{i\theta }hG_{1}+fG_{2},
\end{align}
\end{subequations}
with
\begin{subequations}
\begin{align}
f =&\frac{\vert\tilde{\omega}_{-}-\omega_{1}\vert}{\sqrt{(\tilde{\omega}_{-}-\omega_{1})^{2}+\eta^{2}}},\\
h =&\frac{\eta f}{\tilde{\omega}_{-}-\omega_{1}}.
\end{align}
\end{subequations}

In the degenerate-resonator case, namely when the two mechanical resonators have the same resonance frequencies $\omega_{1}=\omega_{2}=\omega_{m}$, the coupling strengths in Eq.~(\ref{DigG}) can be simplified as
\begin{subequations}
\label{DDigGequwG12}
\begin{align}
\tilde{G}_{+}=&(G_{1}+e^{-i\theta}G_{2})/\sqrt{2},\\
\tilde{G}_{-}=&(G_{2}-e^{i\theta}G_{1})/\sqrt{2}.
\end{align}
\end{subequations}
We proceed to analyze the dependence of the dark-mode effect on the coupling strengths $G_{1}$ and $G_{2}$. Concretely, we will consider three special cases.

(i) In the symmetric-coupling case: $G_{1}=G_{2}=G$, we obtain the relations
\begin{subequations}
\label{DigGequwG12}
\begin{align}
\tilde{G}_{+}=&G(1+e^{-i\theta})/\sqrt{2},\\
\tilde{G}_{-}=&G(1-e^{i\theta})/\sqrt{2}.
\end{align}
\end{subequations}
It can be seen from Eq.~(\ref{DigGequwG12}) that, when $\theta=n\pi$ for an integer $n$, one of the two hybrid mechanical modes (the dark mode) will be decoupled from the cavity-field mode. In this case, the excitation energy stored in the dark mode cannot be extracted through the optomechanical-cooling channel. In general cases of $\theta\neq n\pi$, the dark-mode effect is broken and then ground-state cooling of the two mechanical resonators  becomes accessible under proper parameter conditions.

(ii) In the case $\theta=n\pi$ for an even number $n$, Eq.~(\ref{DDigGequwG12}) becomes
\begin{subequations}
\label{thetaeq0}
\begin{align}
\tilde{G}_{+}=&(G_{1}+G_{2})/\sqrt{2},\\
\tilde{G}_{-}=&(G_{2}-G_{1})/\sqrt{2}.
\end{align}
\end{subequations}
We can see that the dark mode (i.e., the mode $\tilde{B}_{-}$ in this case) can be broken when the two optomechanical coupling strengths are different $G_{1}\neq G_{2}$. In this case, our numerical simulation indicates that simultaneous ground-state cooing of the two mechanical resonators can be realized when $G_{2}/G_{1}\ll1$.

(iii) In the case $\theta=n\pi$ for an odd number $n$, we have
\begin{subequations}
\label{thetaeqpi}
\begin{align}
\tilde{G}_{+}=&(G_{1}-G_{2})/\sqrt{2},\\
\tilde{G}_{-}=&(G_{2}+G_{1})/\sqrt{2}.
\end{align}
\end{subequations}
In this case, the mode $\tilde{B}_{+}$ becomes the dark mode when $G_{1}=G_{2}$. The simultaneous ground-state cooing of the two mechanical resonators can be realized when $G_{2}/G_{1}\ll1$, as shown by Fig.~\ref{FigS2}(d).

\section{Ground-state cooling of the two mechanical resonators \label{phononexact}}

In this section, we study the cooling performance in this system by evaluating the final average phonon numbers in the two mechanical resonators. To this end, we proceed to rewrite the linearized Langevin equations ~(\ref{lineLangevineq}) as the following compact form
\begin{eqnarray}
\mathbf{\dot{u}}(t)=\mathbf{Au}(t)+\mathbf{N}(t),\label{MatrixLeq}
\end{eqnarray}
where the fluctuation operator vector $\mathbf{u}(t)$, the noise operator vector $\mathbf{N}(t)$, and the coefficient matrix $\mathbf{A}$ are defined as
\begin{eqnarray}
\mathbf{u}(t)=[\delta a(t),\delta b_{1}(t), \delta b_{2}(t),\delta a^{\dagger}(t),\delta b^{\dagger}_{1}(t), \delta b^{\dagger}_{2}(t)]^{T},\label{Matrixu}
\end{eqnarray}
\begin{eqnarray}
\mathbf{N}(t)=&[\sqrt{2\kappa}a_{\text{in}}(t),\sqrt{2\gamma_{1}}b_{1,\text{in}}(t), \sqrt{2\gamma_{2}}b_{2,\text{in}}(t) ,\sqrt{2\kappa}a^{\dagger}_{\text{in}}(t),\sqrt{2\gamma_{1}}b^{\dagger}_{1,\text{in}}(t), \sqrt{2\gamma_{2}}b^{\dagger}_{2,\text{in}}(t)]^{T},\label{MatrixN}
\end{eqnarray}
and
\begin{equation}
\mathbf{A}=\left(
\begin{array}{cccccc}
-(\kappa +i\Delta)  &  -iG_{1} & -iG_{2} & 0 &  -iG_{1} & -iG_{2}  \\
-iG_{1}^{\ast}  &  -(\gamma_{1}+i\omega_{1}) & -i\eta e^{i\theta} &-iG_{1}  & 0 &0   \\
-iG_{2}^{\ast}  & -i\eta e^{-i\theta} & -(\gamma_{2}+i\omega_{2}) & -iG_{2} &0  &0   \\
0 & iG_{1}^{\ast} & iG_{2}^{\ast} &-(\kappa-i\Delta)  &  iG_{1}^{\ast} &iG_{2}^{\ast}   \\
iG_{1}^{\ast} & 0 &0  &iG_{1} &  -(\gamma_{1}-i\omega_{1}) & i\eta e^{-i\theta}  \\
iG_{2}^{\ast} & 0 &0  &iG_{2} & i\eta e^{i\theta} & -(\gamma_{2}-i\omega_{2})  \\
\end{array}
\right).
\end{equation}

The formal solution of the linearized Langevin equation~(\ref{MatrixLeq}) can be written as
\begin{equation}
\mathbf{u}(t) =\mathbf{M}(t) \mathbf{u}(0)+\int_{0}^{t}\mathbf{M}(t-s) \mathbf{N}(s)ds,
\end{equation}
where the matrix $\mathbf{M}(t)$ is defined by $\mathbf{M}(t)=\exp(\mathbf{A}t)$. Based on the solution, we can calculate the steady-state average phonon numbers in the two mechanical resonators by solving the Lyapunov equation. Note that the parameters used in the following calculations satisfy the stability conditions derived from the Routh-Hurwitz criterion. Namely, the real parts of all the eigenvalues of the coefficient matrix $\mathbf{A}$ are negative.

For studying quantum cooling of the two mechanical resonators, we focus on the final average phonon numbers in the two mechanical resonators by calculating the steady-state value of the covariance matrix $\mathbf{V}$, which is defined by the matrix elements
\begin{equation}
\mathbf{V}_{ij}=\frac{1}{2}[\langle \mathbf{u}_{i}(\infty) \mathbf{u}_{j}(\infty ) \rangle +\langle \mathbf{u}_{j}( \infty) \mathbf{u}_{i}(\infty )\rangle], \hspace{1 cm}i,j=1-6.
\end{equation}
In the linearized optomechanical system, the covariance matrix $\mathbf{V}$ satisfies the Lyapunov equation
\begin{equation}
\mathbf{A}\mathbf{V}+\mathbf{V}\mathbf{A}^{T}=-\mathbf{Q}, \label{Lyapunov}
\end{equation}
where ``$T$" denotes the matrix transpose operation and the matrix $\mathbf{Q}$ is defined by
\begin{equation}
\mathbf{Q}=\frac{1}{2}(\mathbf{C}+\mathbf{C}^{T}),
\end{equation}
with $\mathbf{C}$ being the noise correlation matrix defined by the matrix elements
\begin{eqnarray}
\langle \mathbf{N}_{k}(s) \mathbf{N}_{l}(s^{\prime})\rangle =\mathbf{C}_{k,l}\delta (s-s^{\prime }).
\end{eqnarray}
%%%%%%%%%%%%%%%%%%%%%%%%%%%%%%
\begin{figure}[tbp]
\centering
\includegraphics[bb=0 0 554 652, width=0.7 \textwidth]{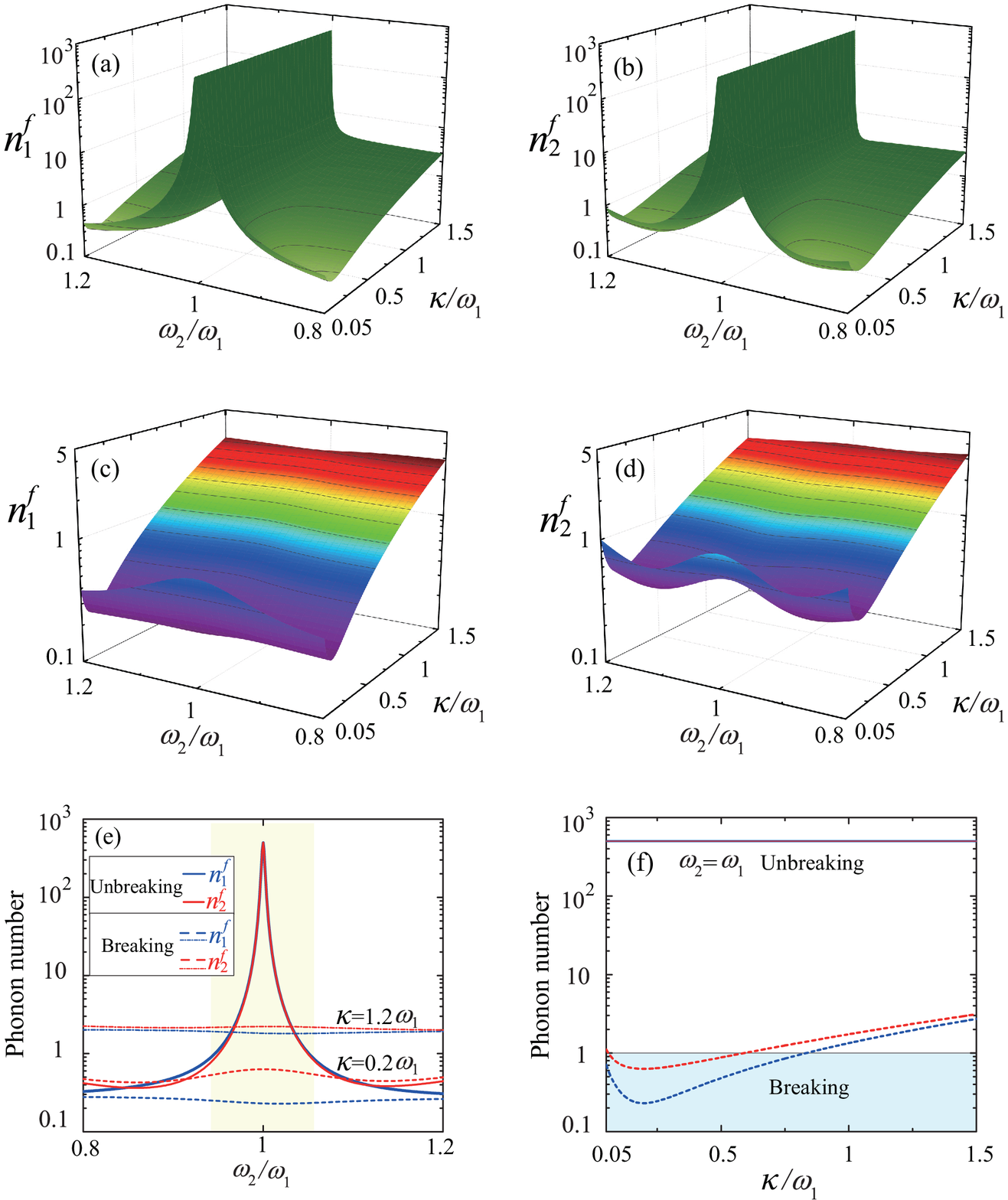}
\caption{(Color online) The final average phonon numbers $n^{f}_{1}$ and $n^{f}_{2}$ versus the resonance-frequency ratio $\omega_{2}/\omega_{1}$ and the cavity-field decay rate $\kappa$ scaled by $\omega_{1}$ in both (a,b) the dark-mode-unbreaking case ($\eta/\omega_{1}=0$) and (c,d) the dark-mode-breaking case ($\eta/\omega_{1}=0.05$ and $\theta=\pi/2$). (e) The final average phonon numbers $n^{f}_{1}$ (blue curves) and $n^{f}_{2}$ (red curves) as functions of $\omega_{2}/\omega_{1}$ in both the dark-mode-unbreaking case ($\eta/\omega_{1}=0$, solid curves) and the dark-mode-breaking case ($\eta/\omega_{1}=0.05$ and $\theta=\pi/2$, dashed curves) under either $\kappa/\omega_{1}=0.2$ or $\kappa/\omega_{1}=1.2$. (f) The final average phonon numbers $n^{f}_{1}$ (blue curves) and $n^{f}_{2}$ (red curves) versus $\kappa/\omega_{1}$ in both the dark-mode-unbreaking case ($\eta/\omega_{1}=0$, solid curves) and the dark-mode-breaking case ($\eta/\omega_{1}=0.05$ and $\theta=\pi/2$, dashed curves) when $\omega_{1}=\omega_{2}$. Here, we consider red-sideband resonance driving $\Delta=\omega_{1}$. Other used parameters are given by $G_{1}/\omega_{1}=G_{2}/\omega_{1}=0.1$, $\gamma_{1}/\omega_{1}=\gamma_{2}/\omega_{1}=10^{-5}$, and $\bar{n}_{1}=\bar{n}_{2}=10^{3}$.}\label{FigS1}
\end{figure}
%%%%%%%%%%%%%%%%%%%%%%%%%%%%%%
For the Markovian baths considered in this work, the constant matrix $\mathbf{C}$ is given by
\begin{equation}
\mathbf{C}=\left(
\begin{array}{cccccc}
0 & 0 & 0 & 2\kappa & 0 & 0\\
0 & 0 & 0 & 0 & 2\gamma_{1}(\bar{n}_{1}+1) & 0\\
0 & 0 & 0 & 0 & 0 & 2\gamma_{2}(\bar{n}_{2}+1)\\
0 & 0 & 0 & 0 & 0 & 0\\
0 & 2\gamma _{1}\bar{n}_{1} & 0 & 0 & 0 & 0\\
0 & 0 & 2\gamma _{2}\bar{n}_{2} & 0 & 0 & 0
\end{array}
\right).
\end{equation}
Based on the covariance matrix $\mathbf{V}$, the final average phonon numbers in the two mechanical resonators are obtained by
\begin{subequations}
\label{finalexact}
\begin{align}
n^{f}_{1}=&\langle \delta b_{1}^{\dagger}\delta b_{1}\rangle=\mathbf{V}_{52}-\frac{1}{2},\\
 n^{f}_{2}=&\langle \delta b_{2}^{\dagger}\delta b_{2}\rangle=\mathbf{V}_{63}-\frac{1}{2},
\end{align}
\end{subequations}
where $\mathbf{V}_{52}$ and $\mathbf{V}_{63}$ can be obtained by solving the Lyapunov equation ~(\ref{Lyapunov}).

In Figs.~\ref{FigS1}(a) and~\ref{FigS1}(b), we plot the final average phonon numbers $n_{1}^{f}$ and $n_{2}^{f}$ as functions of the ratio $\omega_{2}/\omega_{1}$ (the resonance frequency of the second mechanical resonator over that of the first mechanical resonator) and the scaled cavity-field decay rate $\kappa/\omega_{1}$ when the phase-dependent phonon-exchange coupling is absent ($\eta=0$), i.e., in the dark-mode-unbreaking case. Here, we can see that there exists a peak around $\omega_{2}=\omega_{1}$, which means that the two mechanical resonators cannot be cooled in the degenerate and near-degenerate two-resonator cases. This phenomenon can be clearly explained based on the dark-mode effect. When $\omega_{1}=\omega_{2}$, the two mechanical resonators form two hybrid mechanical modes: a bright mode and a dark mode. The dark mode is decoupled from both the cavity-field mode and the bright mechanical mode and hence the excitation energy stored in the dark mode cannot be extracted through the optomechanical-cooling channel. When the two mechanical resonators are far-off-resonant with each other, there is no dark mode, then the ground-state cooling can be realized when this system works in the resolved-sideband regime and under proper driving condition (red-sideband resonance).

The dark-mode effect can be broken by introducing a phase-dependent phonon-exchange interaction between the two mechanical resonators, and then the ground-state cooling can be realized in the degenerate and near-degenerate two-mechanical-resonator cases. In Figs.~\ref{FigS1}(c) and~\ref{FigS1}(d), we plot the final average phonon numbers $n_{1}^{f}$ and $n_{2}^{f}$ in the two mechanical resonators as functions of the ratio $\omega_{2}/\omega_{1}$ and the scaled cavity-field decay rate $\kappa/\omega_{1}$ in the dark-mode-breaking case ($\eta/\omega_{1}=0.05$ and $\theta=\pi/2$). Different from the results in Figs.~\ref{FigS1}(a) and~\ref{FigS1}(b), here we can see that the simutaneous ground-state cooling can be realized ($n^{f}_{1,2}\ll1$) in the resolved-sideband regime ($\kappa\ll\omega_{1}$), which is consistent with the sideband-cooling results in a typical optomechanical system. In addition, simultaneous ground-state cooling of the two mechanical resonators can be reached in a wide parameter range of $\omega_{2}/\omega_{1}$. We also see that the cooling performance of the first resonator is better than that of the second resonator ($n^{f}_{1}<n^{f}_{2}$). This is because the phase $\theta=\pi/2$ is chosen in this case. As we will see in the following section, the nonreciprocal phonon transfer is more helpful to cool the first (second) resonator when $0<\theta<\pi$ ($\pi<\theta<2\pi$).
%%%%%%%%%%%%%%%%%%%%%%%%%%%%%%
\begin{figure}[tbp]
\centering
\includegraphics[bb=0 0 589 351, width=0.95  \textwidth]{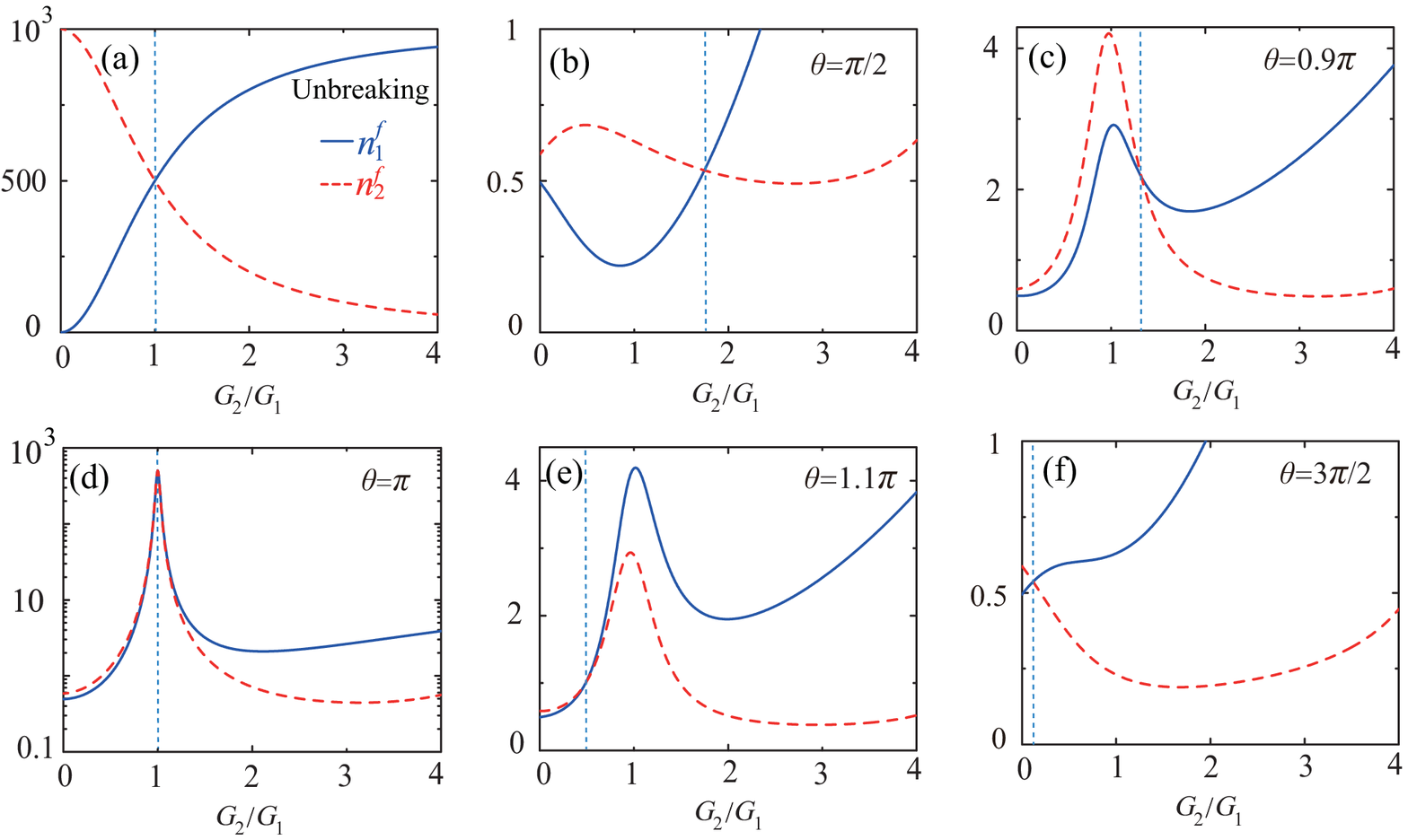}
\caption{(Color online) The final average phonon numbers $n^{f}_{1}$ (blue solid curves) and $n^{f}_{2}$ (red dashed curves) as functions of the ratio $G_{2}/G_{1}$ when the phonon-exchange coupling parameters $\theta$ and $\eta$ take various values: (a) $\eta/\omega_{1}=0$, (b-f) $\eta/\omega_{1}=0.05$ and $\theta=\pi/2$, $0.9\pi$, $\pi$, $1.1\pi$, and $3\pi/2$. Here we choose the optimal driving $\Delta=\omega_{1}=\omega_{2}=\omega_{m}$, $\kappa/\omega_{m}=0.2$, $G_{1}/\omega_{m}=0.1$, $\gamma_{1}/\omega_{m}=\gamma_{2}/\omega_{m}=10^{-5}$, and $\bar{n}_{1}=\bar{n}_{2}=10^{3}$.}\label{FigS2}
\end{figure}
%%%%%%%%%%%%%%%%%%%%%%%%%%%%%%

We note that though the dark mode exists theoretically only in the degenerate-resonator case of this optomechanical system, i.e., $\omega_{1}=\omega_{2}$, the dark-mode effect works within a finite parameter range of the near-degenerate-resonator case. To know the width of the frequency-detuning window associated with the dark-mode effect, in Fig.~\ref{FigS1}(e) we show the final average phonon numbers $n_{1}^{f}$ and $n_{2}^{f}$ as functions of the ratio $\omega_{2}/\omega_{1}$ in both the dark-mode-unbreaking ($\eta/\omega_{1}=0$) and -breaking ($\eta/\omega_{1}=0.05$ and $\theta=\pi/2$) cases. For the dark-mode-unbreaking case, the ground-state cooling cannot be reached in the degenerate and near-degenerate-resonator cases, as marked by the shadow area. The width of the shadow area can be characterized by the effective mechanical linewidth ($\Delta\omega=|\omega_{2}-\omega_{1}|\leq\Gamma_{l}=\gamma_{l}+\gamma_{l,\text{opt}}$). This is because the cooling of the two mechanical resonators is suppressed in this region, i.e., the two mechanical resonators have significant spectral overlap and become effectively degenerate. In the dark-mode-breaking case, we can see that the ground-state cooling can be realized irrespective of the value of the ratio $\omega_{2}/\omega_{1}$ in the resolved-sideband regime ($\kappa/\omega_{1}=0.2$). When the phonon sidebands cannot be resolved, the ground-state cooling is unaccessible in this system (see the curves corresponding to $\kappa/\omega_{1}=1.2$). Especially, in this shadow area shown in Fig.~\ref{FigS1}(e), the emergences of a small valley (the blue dashed curve) and a small hill (the red dashed curve) can be explained based on the nonreciprocical phonon transfer. At an optimal nonreciprocical phonon-transfer point ($\omega_{1}=\omega_{2}$, $\theta=\pi/2$), the phonons in the first mechanical resonator are extracted through both the optomechanical-cooling channel and the phonon-exchange channel, while the phonons in the second mechanical resonator are extracted only through the optomechanical-cooling channel. This is because the phonon transmission rate from modes $b_{2}$ ($b_{1}$) to $b_{1}$ ($b_{2}$) is zero (a finite value) in this case.

We also investigate the influence of the cavity-field decay rate $\kappa$ on the cooling efficiency in both the dark-mode-breaking and -unbreaking cases. In Fig.~\ref{FigS1}(f), we plot the final average phonon numbers $n_{1}^{f}$ and $n_{2}^{f}$ as functions of the scaled cavity-field decay rate $\kappa/\omega_{1}$ in both the dark-mode-unbreaking and -breaking cases when the two mechanical resonators have the same resonance frequencies $\omega_{1}=\omega_{2}$. Here, we can see that, in the dark-mode-unbreaking case, the final phonon numbers $n_{1}^{f}$ and $n_{2}^{f}$ are approximately $500$. This is because the energy (half of the thermal phonons) stored in the dark mode cannot be extracted and hence the mechanical resonators cannot be cooled. In the dark-mode-breaking case, the ground-state cooling can be reached when the system works in the resolved-sideband regime. The optimal working parameter of the cavity-field decay rate (corresponding to the minimal value of the final mean phonon numbers) is around $\kappa/\omega_{1}\approx0.2$. This optimal value is reached under the combined competition between the optomechanical-cooling rate (i.e., the excitation-energy extraction efficiency through the cavity-field decay channel) and the phonon-sideband resolution condition.
%%%%%%%%%%%%%%%%%%%%%%%%%%%%%%
\begin{figure}[tbp]
\centering
\includegraphics[bb=0 0 585 483, width=0.7 \textwidth]{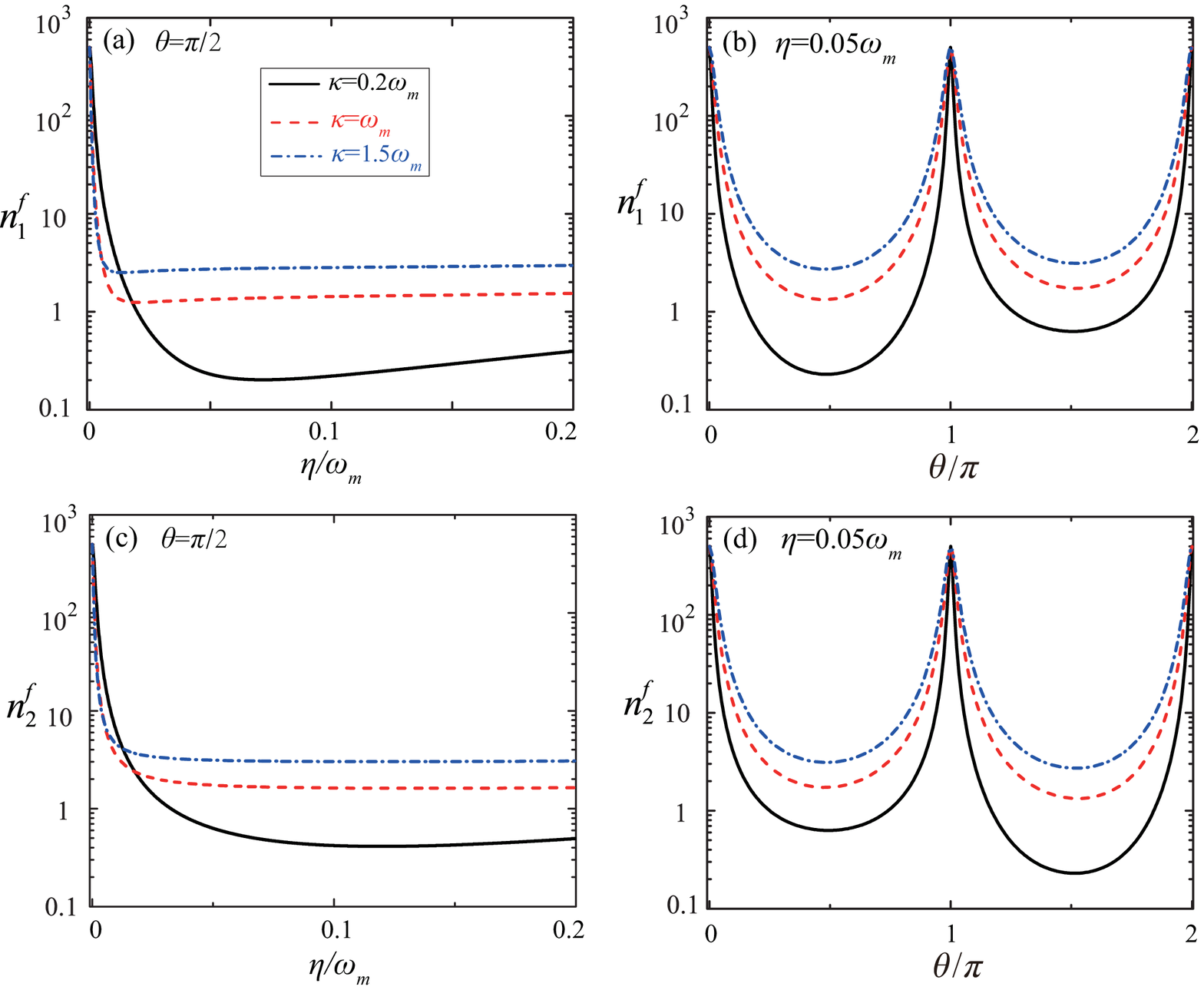}
\caption{(Color online) The final average phonon numbers (a,c) $n^{f}_{1}$ and (b,d) $n^{f}_{2}$ versus either (a,b) the coupling strength $\eta$ at $\theta=\pi/2$ or (c,d) the phase $\theta$ at $\eta=0.05\omega_{m}$ when the cavity-field decay rate takes various values: $\kappa/\omega_{m}=0.2$, $1$, and $1.5$. Here we choose $\Delta=\omega_{1}=\omega_{2}=\omega_{m}$, $G_{1}/\omega_{m}=G_{2}/\omega_{m}=0.1$, $\gamma_{1}/\omega_{m}=\gamma_{2}/\omega_{m}=10^{-5}$, and $\bar{n}_{1}=\bar{n}_{2}=10^{3}$.}\label{FigS3}
\end{figure}
%%%%%%%%%%%%%%%%%%%%%%%%%%%%%%

In the above discussions concerning Fig.~\ref{FigS1}, we only consider the symmetric-coupling case, i.e., $G_{1}=G_{2}$. To better understand quantum cooling in this system, we also investigate the dependence of the final average phonon numbers $n^{f}_{1}$ and $n^{f}_{2}$ on the linearized optomechanical-coupling strengths $G_{1}$ and $G_{2}$. In Fig.~\ref{FigS2}, we plot the final average phonon numbers $n^{f}_{1}$ and $n^{f}_{2}$ as functions of the ratio $G_{2}/G_{1}$ when the phonon-exchange coupling parameters $\eta$ and $\theta$ take various values: (a) $\eta/\omega_{m}=0$, (b-f) $\eta/\omega_{m}=0.05$ and $\theta=\pi/2$, $0.9\pi$, $\pi$, $1.1\pi$, and $3\pi/2$. When the phonon-exchange coupling is absent, i.e., $\eta=0$ [Fig.~\ref{FigS2}(a)], the final average phonon number in the first (second) mechanical resonator increases (decreases) with the increase of $G_{2}/G_{1}$. However, we point out that, due to the dark-mode effect, the ground-state cooling of the two mechanical resonators are unfeasible for finite values of the ratio $G_{2}/G_{1}$. When $G_{2}/G_{1}<1$, the bright mechanical mode is dominated by mode $b_{1}$. When $G_{2}/G_{1}>1$, the bright mechanical mode is dominated by mode $b_{2}$. As a result, the cooling efficiency of the first mechanical resonator is better (worse) than that of the second one in the parameter range $G_{2}/G_{1}<1$ ($G_{2}/G_{1}>1$). The cooling performance of the two resonators is exchanged when the value of the ratio $G_{2}/G_{1}$ changes across the point $G_{2}/G_{1}=1$. In the symmetric-coupling case $G_{2}/G_{1}=1$, the same cooling performance is achieved for the two mechanical resonators ($n^{f}_{1}=n^{f}_{2}\approx500$). The physical reason is that the optomechanical-cooling channels for the two mechanical resonators take the same role when $G_{1}=G_{2}$. At this point, the superposition amplitudes of the two mechanical modes $b_{1}$ and $b_{2}$ in the bright and dark modes are the same, as shown in Eq.~(\ref{BDmodedef}). In the presence of the phonon-exchange coupling, the ground-state cooling can be realized in a wide parameter range of the ansymmetric couplings $G_{2}\neq G_{1}$ when $\theta\neq n\pi$ for integer $n$. In addition, we can see a similar intersection phenomenon for the cooling performance of the two resonators with the increase of the ratio $G_{2}/G_{1}$. However, the location of the intersection point moves to the right (left) from the point $G_{2}/G_{1}=1$ when the phase $\theta$ takes the value in the range $0<\theta<\pi$ ($\pi<\theta<2\pi$). This shift is caused by the phase-dependent phonon-exchange coupling between the two mechanical resonators. When $0<\theta<\pi$, the phonon-exchange coupling assists the cooling of the first mechanical resonator (i.e., decreasing $n^{f}_{1}$ and increasing $n^{f}_{2}$). Hence the phonon-exchange coupling pushes the intersection point moving right. When $\pi<\theta<2\pi$, the phonon-exchange coupling assists the cooling of the second mechanical resonator (i.e., decreasing $n^{f}_{2}$ and increasing $n^{f}_{1}$). As a result, the phonon-exchange coupling pushes the intersection point moving left. At $\theta=\pi$ [panel (d)], the dark mode appears in this system when $G_{1}=G_{2}$, then the two mechanical modes cannot be cooled. In this case, the dark-mode effect can be broken by choosing different values of the coupling strengths $G_{1}\neq G_{2}$, i.e., simultaneous ground-state cooling of the two mechanical resonators can only be realized when $G_{2}/G_{1}\leq0.5$.

%%%%%%%%%%%%%%%%%%%%%%%%%%%%%%
\begin{figure}[tbp]
\centering
\includegraphics[bb=0 0 568 488, width=0.7 \textwidth]{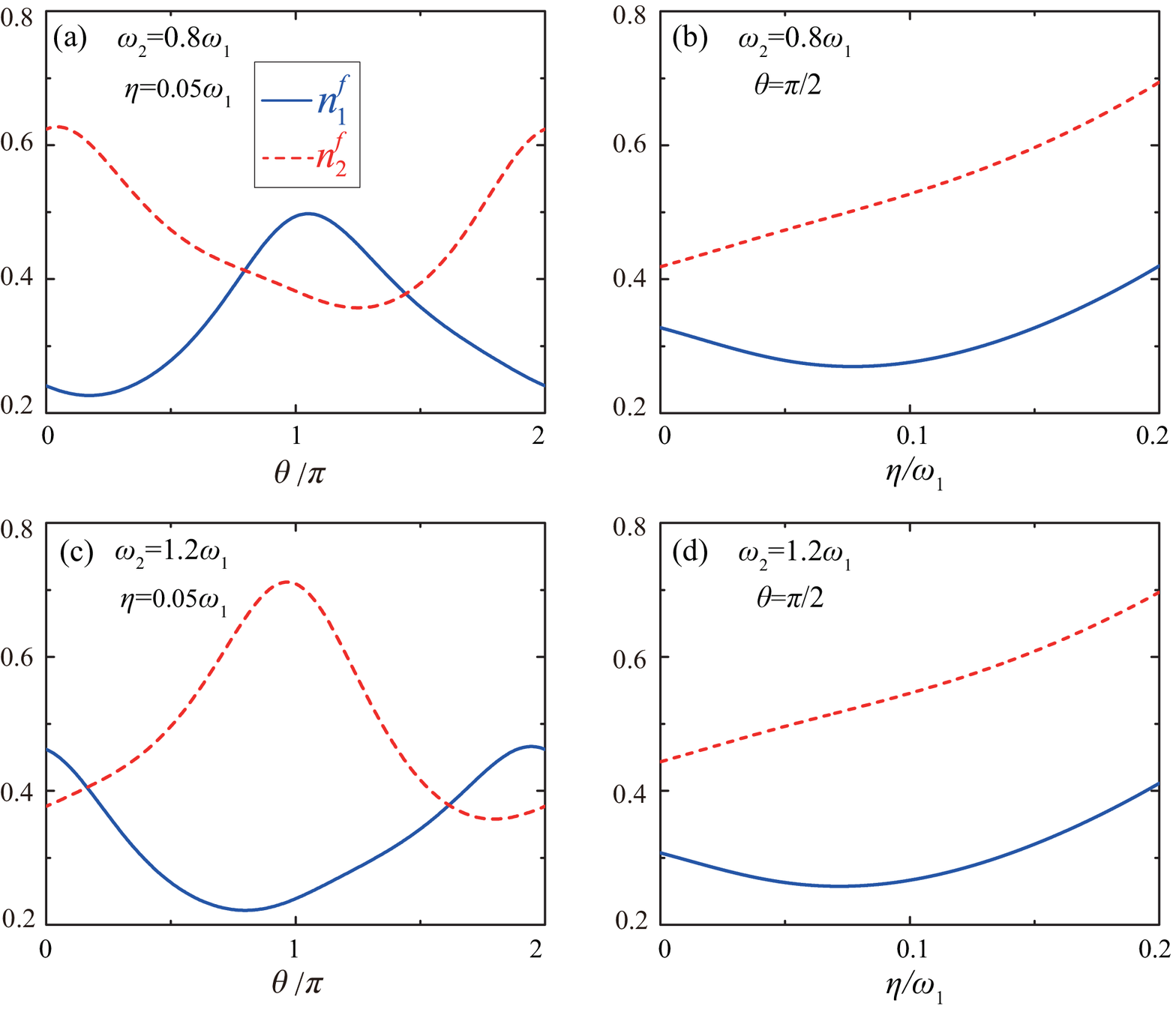}
\caption{(Color online) The final average phonon numbers $n^{f}_{1}$ (blue solid curves) and $n^{f}_{2}$ (red dashed curves) versus the phase $\theta$ and the phonon-exchange coupling strength $\eta$ in the nondegenerate two-resonator cases (a,b) $\omega_{2}=0.8\omega_{1}$ and (c,d) $\omega_{2}=1.2\omega_{1}$. Here we choose $\Delta=\omega_{1}$, $G_{1}/\omega_{1}=G_{2}/\omega_{1}=0.1$, $\kappa/\omega_{1}=0.2$, $\gamma_{1}/\omega_{1}=\gamma_{2}/\omega_{1}=10^{-5}$, and $\bar{n}_{1}=\bar{n}_{2}=10^{3}$.}\label{FigS4}
\end{figure}
%%%%%%%%%%%%%%%%%%%%%%%%%%%%%%

The phase-dependent phonon-exchange interaction plays a critical role in the ground-state cooling of the multiple mechanical resonators. Below we investigate the dependence of the cooling performance on the coupling parameters $\eta$ and $\theta$ of the phase-dependent phonon-exchange interaction between the two mechanical resonators. In Figs.~\ref{FigS3}(a) and~\ref{FigS3}(b), we plot the final average phonon numbers $n^{f}_{1}$ and $n^{f}_{2}$ as functions of the coupling strength $\eta$ and phase $\theta$ when the cavity-field decay rate takes various values: $\kappa/\omega_{m}=0.2$, $1$, and $1.5$. Here, we can see that the two mechanical resonators can be cooled efficiently (from the initial phonon number $1000$ to the final phonon number below $10$) when $\eta/\omega_{m}>0.02$. In addition, the cooling performance becomes worse for a larger value of the cavity-field decay rate $\kappa$. The ground-state cooling can only be realized in the resolved-sideband regime $\kappa/\omega_{m}<1$. We also show the dependence of the final average phonon numbers $n^{f}_{1}$ and $n^{f}_{2}$ on the phase $\theta$ for several values of $\kappa/\omega_{m}$, as shown in Figs.~\ref{FigS3}(c) and~\ref{FigS3}(d). The plots show that the cooling performance depends on the phase $\theta$. The final average phonon numbers $n^{f}_{1}$ and $n^{f}_{2}$ can be largely decreased when $0<\theta<\pi$ and $\pi<\theta<2\pi$. When $\theta=n\pi$ for an integer $n$, the two mechanical resonators cannot be cooled due to the dark-mode effect. The cooling performance becomes worse with the increase of the cavity-field decay rate. In addition, the results show that $n^{f}_{1}<n^{f}_{2}$ ($n^{f}_{1}>n^{f}_{2}$) in the parameter range $0<\theta<\pi$ ($\pi<\theta<2\pi$), which can be explained based on the nonreciprocal phonon transfer induced by quantum interference in the loop-coupled system.

In Fig.~\ref{FigS3}, we have investigated the dependence of the final average phonon numbers $n^{f}_{1}$ and $n^{f}_{2}$ on the phonon-exchange coupling parameters $\eta$ and $\theta$ in the degenerate two-mechanical-resonator case, i.e., $\omega_{1}=\omega_{2}$. In the following we also consider a nondegenerate mechanical-resonator case. In Fig.~\ref{FigS4} we plot the final average phonon numbers $n^{f}_{1}$ and $n^{f}_{2}$ versus the parameters $\eta$ and $\theta$ in the nondegenerate two-resonator cases, i.e., $\omega_{2}=0.8\omega_{1}$ or $\omega_{2}=1.2\omega_{1}$. The plots show that the simultaneous ground-state cooling of the two mechanical resonators can be realized in the nondegenerate mechanical-resonator case. In both the cases $\omega_{2}=0.8\omega_{1}$ and $\omega_{2}=1.2\omega_{1}$, the dependence of $n^{f}_{1}$ and $n^{f}_{2}$ on the phase $\theta$ has an inverse tendency, as shown in Figs.~\ref{FigS4}(a) and~\ref{FigS4}(c). In addition, the dependence of $n^{f}_{l=1,2}$ on the phase $\theta$ in the case $\omega_{2}=0.8\omega_{1}$ is inverse to that in the case of $\omega_{2}=1.2\omega_{1}$. In Figs.~\ref{FigS4}(b) and~\ref{FigS4}(d), we can see $n^{f}_{1}<n^{f}_{2}$ and the dependence of $n^{f}_{l=1,2}$ on the coupling strength $\eta$ has a similar tendency for the cases $\omega_{2}=0.8\omega_{1}$ and $\omega_{2}=1.2\omega_{1}$. In the nondegenerate-resonator case, the cooling performance can be controlled by choosing proper phonon-exchange coupling parameters $\eta$ and $\theta$. The same value of the final phonon numbers $n^{f}_{1}$ and $n^{f}_{2}$ can be obtained by choosing the intersection points in Figs.~\ref{FigS4}(a) and~\ref{FigS4}(c).

%%%%%%%%%%%%%%%%%%%%%%%%%%%%%%
\begin{figure}[tbp]
\centering
\includegraphics[bb=0 0 548 437, width=0.7 \textwidth]{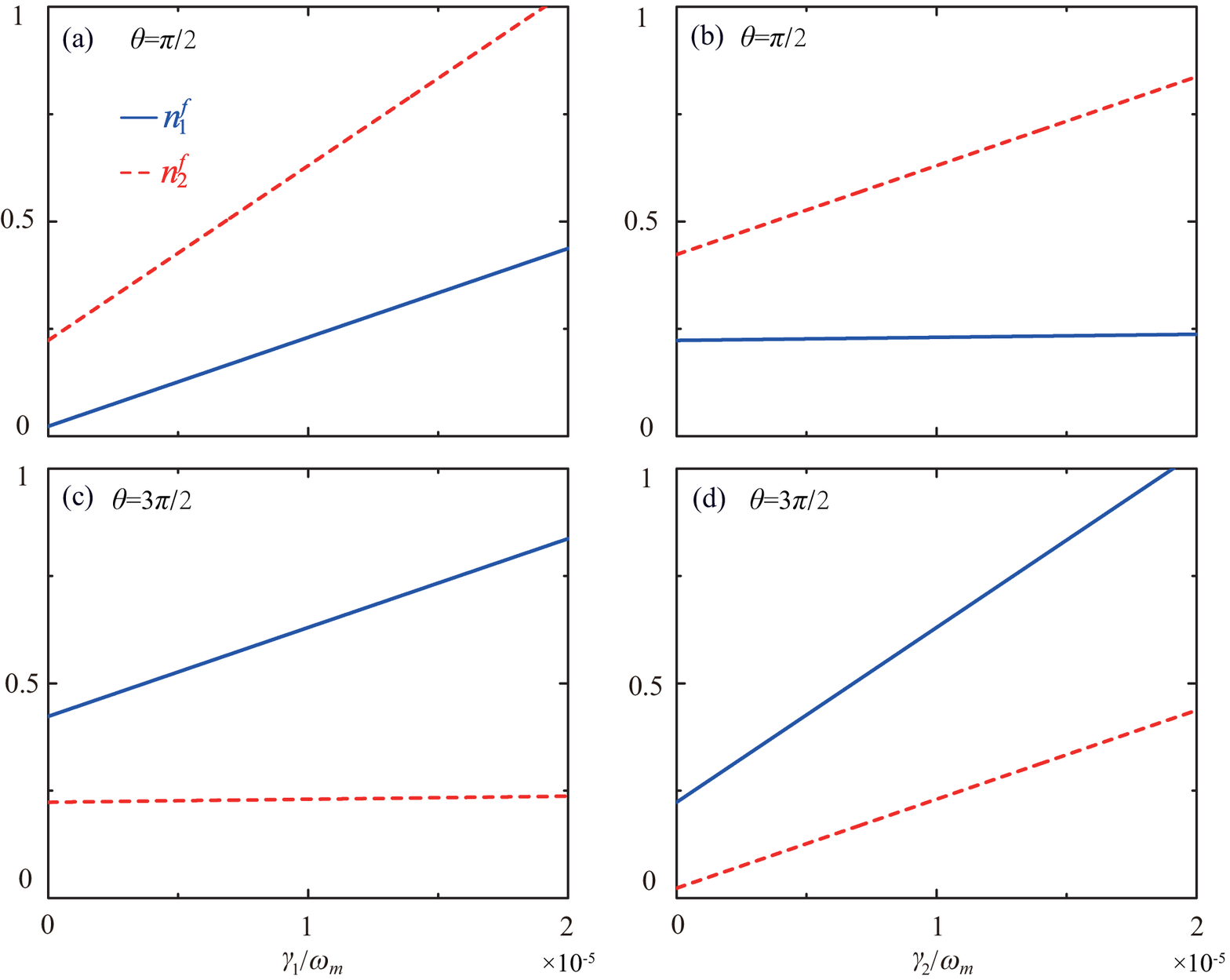}
\caption{(Color online) The final average phonon numbers $n^{f}_{1}$ and $n^{f}_{2}$ as functions of (a,c) $\gamma_{1}$ and (b,d) $\gamma_{2}$ when the phase $\theta$ takes different values: (a,b) $\theta=\pi/2$ and (c,d) $\theta=3\pi/2$. In panels (a,c) and (b,d), we choose $\gamma_{2}/\omega_{1}=10^{-5}$ and $\gamma_{1}/\omega_{1}=10^{-5}$, respectively. Other used parameters are $\Delta=\omega_{1}=\omega_{2}=\omega_{m}$, $G_{1}/\omega_{m}=G_{2}/\omega_{m}=0.1$, $\eta/\omega_{m}=0.05$, $\kappa/\omega_{m}=0.2$, and $\bar{n}_{1}=\bar{n}_{2}=10^{3}$.}\label{r12}
\end{figure}
%%%%%%%%%%%%%%%%%%%%%%%%%%%%%%

In quantum cooling of the mechanical resonators, the optomechanical cavity and its vacuum bath provide the cooling channel to extract the excitation energy in the mechanical resonators. Here, the mechanical resonators are thermalized by their thermal baths through the mechanical dissipation channels. As a result, the final average phonon numbers $n^{f}_{1}$ and $n^{f}_{2}$ in the two mechanical resonators depend on the mechanical decay rates $\gamma_{1}$ and $\gamma_{2}$. In Fig.~\ref{r12}, we show the final average phonon numbers $n^{f}_{1}$ and $n^{f}_{2}$ as functions of the decay rates $\gamma_{1}$ and $\gamma_{2}$. We can see that $n^{f}_{1}$ and $n^{f}_{2}$ increase with the increase of the mechanical decay rates. This is because the energy exchange rates between the mechanical resonators and their heat baths are faster for larger values of the decay rates, and then the thermal excitation in the heat baths will raise the total phonon numbers in the mechanical resonators. In Figs.~\ref{r12}(a) and ~\ref{r12}(b), we have $n^{f}_{1}<n^{f}_{2}$ because the phase angle $\theta=\pi/2$ is taken, then the cooling performance of the first resonator is better than that of the second resonator. However, an opposite cooling effect compared with the case of $\theta=\pi/2$ emerges when $\theta=3\pi/2$, as shown in Figs.~\ref{r12}(c) and~\ref{r12}(d). These interesting cooling phenomena can be explained according to the phonon scattering process between the two mechanical resonators, which will be studied in the next section.

\section{Phonon scattering probability and nonreciprocal phonon transfer \label{appendixa}}

In this section, we study the scattering probabilities of the phonon transport between the two mechanical resonators coupled by a phase-dependent phonon-exchange interaction. We calculate the transmission spectrum of the phonon transport based on the Langevin equation~(\ref{MatrixLeq}). To this end, we rewrite the matrix $\mathbf{N}(t)$ defined in Eq.~(\ref{MatrixN}) as
\begin{equation}
\mathbf{N}(t)=\mathbf{\Gamma}\mathbf{u_{\text{in}}}(t),\label{newdefn}
\end{equation}
where the damping matrix $\mathbf{\Gamma}$ is defined as
\begin{equation}
\mathbf{\Gamma}=\mathbf{diag}[\sqrt{2\kappa },\sqrt{2\gamma _{1}},\sqrt{2\gamma _{2}},\sqrt{2\kappa },\sqrt{2\gamma_{1}},\sqrt{2\gamma_{2}}],
 \end{equation}
with $\mathbf{diag}[x]$ giving a matrix with the elements of the list $x$ on the leading diagonal, and $0$ elsewhere.
The input noise vector $\mathbf{u_{\text{in}}}(t)$ in Eq.~(\ref{newdefn}) is given by
\begin{equation}
\mathbf{u_{\text{in}}}(t)=[a_{\text{in}}(t),b_{1,\text{in}}(t),b_{2,\text{in}}(t),a_{\text{in}}^{\dagger}(t),b_{1,\text{in}}^{\dagger}(t),b_{2,\text{in}}^{\dagger}(t)]^{T}.\label{Matrixuin}
\end{equation}

Making use of the Fourier transformation for operator $r\in\{\delta a, \delta b_{1}, \delta b_{2}, \delta a_{\text{in}}, \delta b_{1,\text{in}}, \delta b_{2,\text{in}}\}$ and its conjugate $r^{\dagger}$,
\begin{subequations}
\begin{align}
\tilde{r}(\omega)=&\frac{1}{2\pi }\int_{-\infty}^{\infty }e^{i\omega t}r(t)dt,\\
\tilde{r}^{\dagger}(\omega)=&\frac{1}{2\pi }\int_{-\infty }^{\infty}e^{i\omega t}r^{\dagger }(t)dt,
\end{align}
\end{subequations}
the solutions to the linearized quantum Langevin equation~(\ref{MatrixLeq}) in the frequency domain can be obtained as
\begin{equation}
\mathbf{\tilde{u}}(\omega)=(-i\omega\mathbf{I}-\mathbf{A)}^{-1}\mathbf{\Gamma}\mathbf{\tilde{u}}_{\text{in}}(\omega),\label{tildeuomega}
\end{equation}
where $\mathbf{\tilde{u}}(\omega)$ and $\mathbf{\tilde{u}}_{\text{in}}(\omega)$ are, respectively, the Fourier transformation of the operator vectors $\mathbf{u}(t)$ defined in Eq.~(\ref{Matrixu}) and $\mathbf{u}_{\text{in}}(t)$ defined in Eq.~(\ref{Matrixuin}). The matrix $\mathbf{I}$ in Eq.~(\ref{tildeuomega}) is an identity matrix. Using the input-output relation
\begin{equation}
o_{\text{in}}+o_{\text{out}}=\sqrt{2\gamma_{o}}\delta o
\end{equation}
for $o\in \{a, b_{1}, b_{2}\}$ and $\gamma_{o}\in\{\kappa, \gamma_{1}, \gamma_{2}\}$, we obtain the output field in the frequency domain as
\begin{equation}
\mathbf{\tilde{u}}_{\text{out}}(\omega) =\mathbf{U}(\omega) \mathbf{\tilde{u}}_{\text{in}}(\omega),\label{inout}
\end{equation}
where the transformation matrix is given by
\begin{eqnarray}
\mathbf{U}(\omega)=\mathbf{\Gamma}(-i\omega\mathbf{I}-\mathbf{A})^{-1}\mathbf{\Gamma}-\mathbf{I},
\end{eqnarray}
and
\begin{eqnarray}
\mathbf{\tilde{u}_{\text{out}}}(\omega)=[\tilde{a}_{\text{out}}(\omega), \tilde{b}_{1, \text{out}}(\omega), \tilde{b}_{2,\text{out}}(\omega), \tilde{a}_{\text{out}}^{\dagger}(\omega), \tilde{b}_{1,\text{out}}^{\dagger}(\omega), \tilde{b}_{2,\text{out}}^{\dagger}(\omega)]^{T}
\end{eqnarray}
denotes the Fourier transformation of $\mathbf{u}_{\text{out}}(t)$.

To analyze the excitation energy transfer in this system, we introduce the spectra for the input and output signals as
\begin{subequations}
\begin{align}
\mathbf{S}_{\text{in}}(\omega)=&[s_{a,\text{in}}(\omega),s_{b_{1},\text{in}}(\omega),s_{b_{2},\text{in}}(\omega)]^{T},\\
\mathbf{S}_{\text{out}}(\omega)=&[s_{a,\text{out}}(\omega), s_{b_{1},\text{out}}(\omega), s_{b_{2},\text{out}}(\omega)]^{T},
\end{align}
\end{subequations}
where the elements are defined by
\begin{subequations}
\begin{align}
\langle\tilde{o}^{\dagger}_{\text{out}}(\omega^{'})\tilde{o}_{\text{out}}(\omega)\rangle=&s_{o,\text{out}}\delta(\omega+\omega^{'}),\\
\langle\tilde{o}^{\dagger}_{\text{in}}(\omega^{'})\tilde{o}_{\text{in}}(\omega)\rangle=&s_{o,\text{in}}\delta(\omega+\omega^{'}),\\ \langle\tilde{o}_{\text{in}}(\omega^{'})\tilde{o}^{\dagger}_{\text{in}}(\omega)\rangle=&(1+s_{o,\text{in}})\delta(\omega+\omega^{'}).
\end{align}
\end{subequations}
We also define the spectrum for the input vacuum noise as
\begin{eqnarray}
\mathbf{S}_{\text{vac}}(\omega)=[s_{a,\text{vac}}(\omega),s_{b_{1},\text{vac}}(\omega),s_{b_{2},\text{vac}}(\omega)]^{T},
\end{eqnarray}
with
\begin{subequations}
\begin{align}
s_{a,\text{vac}}(\omega)=&|U_{14}(\omega)|^{2}+|U_{15}(\omega)|^{2}+|U_{16}(\omega)|^{2},\\
s_{b_{1},\text{vac}}(\omega)=&|U_{24}(\omega)|^{2}+|U_{25}(\omega)|^{2}+|U_{26}(\omega)|^{2},\\
s_{b_{2},\text{vac}}(\omega)=&|U_{34}(\omega)|^{2}+|U_{35}(\omega)|^{2}+|U_{36}(\omega)|^{2}.
\end{align}
\end{subequations}

Then the relation between these spectra can be obtained as
\begin{equation}
\mathbf{S}_{\text{out}}(\omega)=\mathbf{T}(\omega)\mathbf{S}_{\text{in}}(\omega)+\mathbf{S}_{\text{vac}}(\omega),
\end{equation}
where the transmission matrix $\mathbf{T}(\omega)$ is defined by
\begin{equation}
\mathbf{T}(\omega)=\left(
\begin{array}{ccc}
T_{aa}(\omega) & T_{ab_{1}}(\omega) & T_{ab_{2}}(\omega) \\
T_{b_{1}a}(\omega) & T_{b_{1}b_{1}}(\omega) & T_{b_{1}b_{2}}(\omega) \\
T_{b_{2}a}(\omega) & T_{b_{2}b_{1}}(\omega) & T_{b_{2}b_{2}}(\omega)
\end{array}
\right),
\end{equation}
with these matrix elements
\begin{eqnarray}
T_{aa}(\omega) &=&|U_{11}(\omega)|^{2}+|U_{14}(\omega)|^{2},\nonumber \\
T_{ab_{1}}(\omega) &=&|U_{12}(\omega)|^{2}+|U_{15}(\omega)|^{2},\nonumber \\
T_{ab_{2}}(\omega) &=&|U_{13}(\omega)|^{2}+|U_{16}(\omega)|^{2},\nonumber \\
T_{b_{1}a}(\omega) &=&|U_{21}(\omega)|^{2}+|U_{24}(\omega)|^{2},\nonumber \\
T_{b_{1}b_{1}}(\omega) &=&|U_{22}(\omega)|^{2}+|U_{25}(\omega)|^{2},  \nonumber \\
T_{b_{1}b_{2}}(\omega) &=&|U_{23}(\omega)|^{2}+|U_{26}(\omega)|^{2},  \nonumber \\
T_{b_{2}a}(\omega) &=&|U_{31}(\omega)|^{2}+|U_{34}(\omega)|^{2},\nonumber \\
T_{b_{2}b_{1}}(\omega) &=&|U_{32}(\omega)|^{2}+|U_{35}(\omega)|^{2},  \nonumber \\
T_{b_{2}b_{2}}(\omega) &=&|U_{33}(\omega)|^{2}+|U_{36}(\omega)|^{2}.
\end{eqnarray}
The element $T_{vw}(\omega)$ ($v,w\in \{a, b_{1}, b_{2}\}$) denotes the transmittance from the input mode $w$ to the output mode $v$. To explore the phonon-transfer nonreciprocity between the two mechanical modes, we only focus on the transmittance $T_{b_{1}b_{2}}(\omega)$ and $T_{b_{2}b_{1}}(\omega)$ between the two mechanical modes. Then, we numerically evaluate the transmittance between the two mechanical modes to show the nonreciprocal phonon transfer. Physically, the transmittance $T_{b_{1}b_{2}}(\omega )$ and $T_{b_{2}b_{1}}(\omega)$ can be used to analyze the thermal excitations extracted from one mechanical mode to the other one.

%%%%%%%%%%%%%%%%%%%%%%%%%%%%%%
\begin{figure}[tbp]
\centering
\includegraphics[bb=0 0 566 490, width=0.7\textwidth]{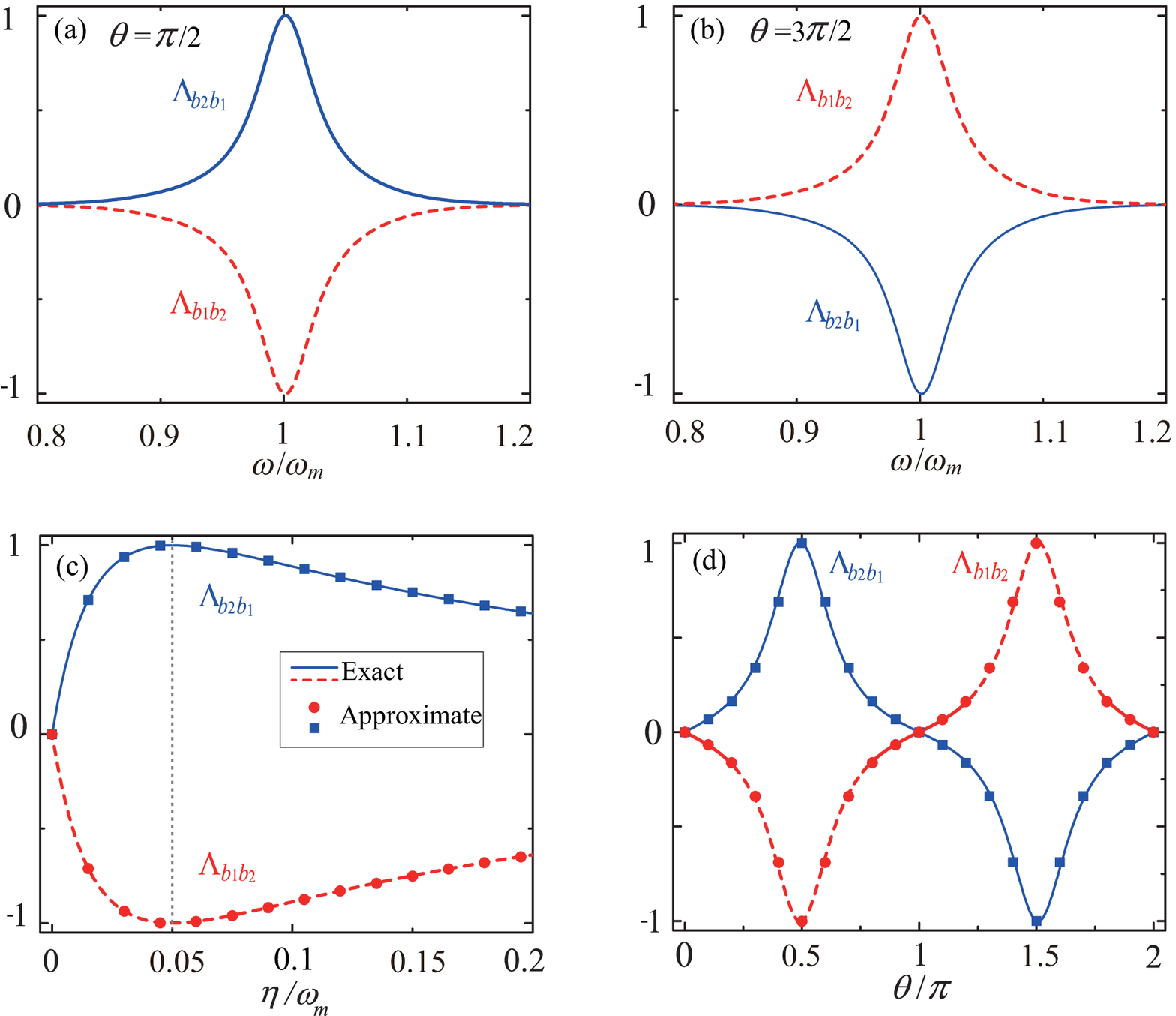}
\caption{(Color online) (a,b) The relative phonon-scattering rate $\Lambda_{b_{2}b_{1}}$ (blue curves) and $\Lambda_{b_{1}b_{2}}$ (red curves) as functions of $\omega$ when the phase $\theta$ takes different values: (a) $\theta=\pi/2$ and (b) $\theta=3\pi/2$. In panels (a,b), we choose the phonon-exchange coupling $\eta/\omega_{m}=0.05$. (c,d) The exact (solid/dashed lines) and approximate (symbols) relative resonant-phonon-scattering rates $\Lambda_{b_{2}b_{1}}$ and $\Lambda_{b_{1}b_{2}}$ vs (c) the phonon-exchange coupling $\eta$ when $\theta=\pi/2$ and (d) the phase $\theta$ when $\eta/\omega_{m}=0.05$ under the parameter $\omega=\omega_{m}$. Here we take $\Delta=\omega_{1}=\omega_{2}=\omega_{m}$, $G_{1}/\omega_{m}=G_{2}/\omega_{m}=0.1$, $\kappa/\omega_{m}=0.2$, $\gamma_{1}/\omega_{m}=\gamma_{2}/\omega_{m}=10^{-5}$, and $\bar{n}_{1}=\bar{n}_{2}=10^{3}$.}\label{scattering}
\end{figure}
%%%%%%%%%%%%%%%%%%%%%%%%%%%%%%

%%%%%%%%%%%%%%%%%%%%%%%%%%%%%%
\begin{figure}[tbp]
\centering
\includegraphics[bb=0 0 566 490, width=0.7\textwidth]{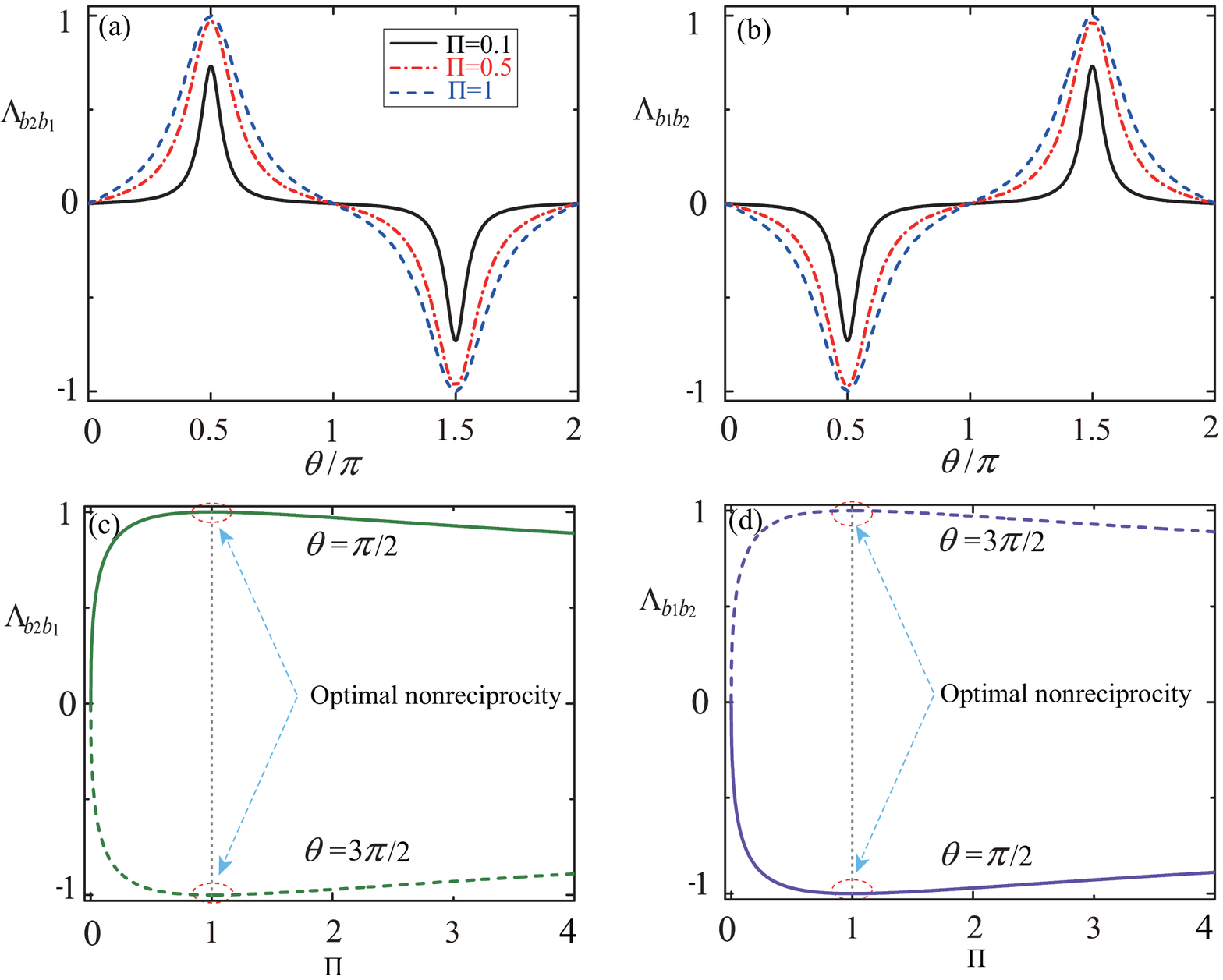}
\caption{(Color online) Dependence of the relative phonon-scattering rates (a) $\Lambda_{b_{2}b_{1}}$ and (b) $\Lambda_{b_{1}b_{2}}$ on the phase $\theta$ when $\omega=\omega_{m}$ and the optomechanical cooperativity takes various values: $\Pi=0.1, 0.5$, and $1$. The relative phonon-scattering rates (c) $\Lambda_{b_{2}b_{1}}$ and (d) $\Lambda_{b_{1}b_{2}}$ versus the ratio of the optomechanical cooperativities $\Pi$ when $\theta=\pi/2$ and $\theta=3\pi/2$. Here we take $\Delta=\omega_{1}=\omega_{2}=\omega_{m}$, $G_{1}/\omega_{m}=G_{2}/\omega_{m}=0.1$, $\kappa/\omega_{m}=0.2$, $\gamma_{1}/\omega_{m}=\gamma_{2}/\omega_{m}=10^{-5}$, and $\bar{n}_{1}=\bar{n}_{2}=10^{3}$.}\label{scattering2}
\end{figure}
%%%%%%%%%%%%%%%%%%%%%%%%%%%%%%

The above results concerning the phonon transmission are exact. Below we derive some approximate analytical results under the RWA and the resonance condition $\Delta=\omega_{1}=\omega_{2}=\omega_{m}$. Note that under the RWA, we have the approximate relations $T_{b_{1}b_{2}}(\omega)\approx|U_{23}(\omega)|^{2}$ and $T_{b_{2}b_{1}}(\omega) \approx |U_{32}(\omega)|^{2}$.
In particular, we focus on the resonant phonon transmission at the mechanical frequency $\omega_{m}$, then an analytical transmittance between the two mechanical modes can be obtained as
\begin{subequations}
\begin{align}
T_{b_{1}b_{2}}\approx &\vert U_{23}\vert^{2}=\frac{4\gamma _{1}\gamma _{2}[(G_{1}G_{2})^{2}+(\kappa \eta)^{2}-2G_{1}G_{2}\kappa \eta \sin \theta ]}{(G_{2}^{2}\gamma_{1}+G_{1}^{2}\gamma _{2}+\kappa \gamma _{1}\gamma _{2}+\kappa \eta^{2})^{2}+4(G_{1}G_{2}\eta \cos \theta)^{2}}  \nonumber \\
=&\frac{4({\mathcal{C}}_{1}{\mathcal{C}}_{2}+{\mathcal{C}}_{3}-2\sqrt{{\mathcal{C}}_{1}{\mathcal{C}}_{2}{\mathcal{C}}_{3}}\sin \theta )}{({\mathcal{C}}_{1}+{\mathcal{C}}_{2}+{\mathcal{C}}_{3}+1)^{2}+4{\mathcal{C}}_{1}{\mathcal{C}}_{2}{\mathcal{C}}_{3}\cos ^{2}\theta},\label{b1b2} \\
T_{b_{2}b_{1}} \approx &\vert U_{32}\vert^{2}=\frac{4\gamma _{1}\gamma _{2}[(G_{1}G_{2})^{2}+(\kappa \eta)^{2}+2G_{1}G_{2}\kappa \eta \sin \theta ]}{(G_{2}^{2}\gamma
_{1}+G_{1}^{2}\gamma _{2}+\kappa \gamma _{1}\gamma_{2}+\kappa \eta^{2})^{2}+4(G_{1}G_{2}\eta \cos \theta )^{2}}  \nonumber \\
=&\frac{4({\mathcal{C}}_{1}{\mathcal{C}}_{2}+{\mathcal{C}}_{3}+2\sqrt{{\mathcal{C}}_{1}{\mathcal{C}}_{2}{\mathcal{C}}_{3}}\sin \theta )}{({\mathcal{C}}_{1}+{\mathcal{C}}_{2}+{\mathcal{C}}_{3}+1)^{2}+4{\mathcal{C}}_{1}{\mathcal{C}}_{2}{\mathcal{C}}_{3}\cos ^{2}\theta},\label{b2b1}
\end{align}
\end{subequations}
where we introduce the cooperativities between any two subsystems in this two-mechanical-mode optomechanical system as
\begin{subequations}
\begin{align}
{\mathcal{C}}_{1}=&\frac{G_{1}^{2}}{\gamma_{1}\kappa},\\
{\mathcal{C}}_{2}=&\frac{G_{2}^{2}}{\gamma_{2}\kappa},\\
{\mathcal{C}}_{3}=&\frac{\eta^{2}}{\gamma_{1}\gamma_{2}}.
\end{align}
\end{subequations}
According to Eqs.~(\ref{b1b2}) and~(\ref{b2b1}), the maximum transmittance for either $\theta=\pi/2$ or $\theta=3\pi/2$ can be obtained as
\begin{equation}
(T_{b_{2}b_{1}})_{\text{max}}=(T_{b_{1}b_{2}})_{\text{max}}=\frac{4(\sqrt{{\mathcal{C}}_{1}{\mathcal{C}}_{2}}
+\sqrt{{\mathcal{C}}_{3}})^{2}}{({\mathcal{C}}_{1}+{\mathcal{C}}_{2}+{\mathcal{C}}_{3}+1)^{2}}.
\end{equation}
By introducing a relative phonon-scattering rate from the mechanical modes $w$ to $v$ as
\begin{equation}
\Lambda_{vw}=\frac{T_{vw}-T_{wv}}{(T_{vw})_{\text{max}}},\label{defLambdavw}
\end{equation}
we can then obtain the rates between the two mechanical modes $b_{1}$ and $b_{2}$ as
\begin{subequations}
\begin{align}
\Lambda_{b_{2}b_{1}} =&\frac{T_{b_{2}b_{1}}-T_{b_{1}b_{2}}}{(T_{b_{2}b_{1}})_{\text{max}}}=\frac{4\sqrt{{\mathcal{C}}_{1}{\mathcal{C}}_{2}{\mathcal{C}}_{3}}\sin \theta }{(\sqrt{{\mathcal{C}}_{1}{\mathcal{C}}_{2}}+\sqrt{{\mathcal{C}}_{3}})^{2}\left(1+\frac{4{\mathcal{C}}_{1}{\mathcal{C}}_{2}{\mathcal{C}}_{3}\cos ^{2}\theta }{
({\mathcal{C}}_{1}+{\mathcal{C}}_{2}+{\mathcal{C}}_{3}+1)^{2}}\right)},\label{TTb2b1}\\
\Lambda_{b_{1}b_{2}} =&\frac{T_{b_{1}b_{2}}-T_{b_{2}b_{1}}}{(T_{b_{1}b_{2}})_{\text{max}}}=-\Lambda_{b_{2}b_{1}}.\label{TTb1b2}
\end{align}
\end{subequations}

In Figs.~\ref{scattering}(a) and~\ref{scattering}(b), the relative phonon-scattering rates $\Lambda_{b_{2}b_{1}}$ (blue curves) and $\Lambda_{b_{1}b_{2}}$ (red curves) are plotted as functions of the scaled frequency $\omega/\omega_{m}$ when the phase $\theta$ takes different values: (a) $\theta=\pi/2$ and (b) $\theta=3\pi/2$. It is obviously shown that the reciprocity of the phonon transfer between the two mechanical resonators is broken ($\Lambda_{b_{2}b_{1}}\neq0$) in a wide range of $\omega$ and the phonon transfer exhibits a perfect nonreciprocal response when $\theta=\pi/2$ and $\theta=3\pi/2$. When $\theta=\pi/2$ ($\theta=3\pi/2$), we have $T_{b_{2}b_{1}}>0$ and $T_{b_{1}b_{2}}<0$ ($T_{b_{2}b_{1}}<0$ and $T_{b_{1}b_{2}}>0$). In particular, when $\omega=\omega_{m}$ and $\theta=\pi/2$, we have $\Lambda_{b_{2}b_{1}}=1$, i.e., $T_{b_{1}b_{2}}=0$. This means that the unidirectional flow of the phonons from $b_{1}$ to $b_{2}$ is achieved. When $\omega=\omega_{m}$ and $\theta=3\pi/2$, we have $\Lambda_{b_{1}b_{2}}=1$, i.e., $T_{b_{2}b_{1}}=0$. This means the phonons can only be transferred from $b_{2}$ to $b_{1}$. Based on the above results, we can see that the phase-dependent phonon-exchange coupling plays an effective role on the relative phonon scattering between the two mechanical resonators. In Figs.~\ref{scattering}(c) and~\ref{scattering}(d), we show the dependence of the relative resonant-phonon-scattering rates on the phonon-exchange coupling parameters $\eta$ and $\theta$. The results indicate that a perfect nonreciprocal phonon transfer requires both $\eta\approx0.05\omega_{m}$ and $\theta=\pi/2$ or $3\pi/2$. Moreover, the exact calculations and the approximate analytical results match well with each other. Here, the solid ($\Lambda_{b_{2}b_{1}}$) and dashed lines ($\Lambda_{b_{1}b_{2}}$) are plotted using the exact solutions, while the symbols are based on the analytical calculations given in Eqs.~(\ref{TTb2b1}) and~(\ref{TTb1b2}). In Fig.~\ref{scattering}(d), when $0<\theta<\pi$, it shows $\Lambda_{b_{2}b_{1}}>0$, i.e., $T_{b_{2}b_{1}}>T_{b_{1}b_{2}}$. In the region $\pi<\theta<2\pi$, it exhibits $\Lambda_{b_{1}b_{2}}>0$, i.e., $T_{b_{1}b_{2}}>T_{b_{2}b_{1}}$. Meanwhile, the phonon transmission satisfies the reciprocity [$\Lambda_{b_{2}b_{1}}=\Lambda_{b_{1}b_{2}}=0$, i.e., $T_{b_{1}b_{2}}=T_{b_{2}b_{1}}$] at $\theta=n\pi$. Moreover, the transmittance is optimal for the process from $b_{1}$ ($b_{2}$) to $b_{2}$ ($b_{1}$) and is zero for the opposite process when $\theta=\pi/2$ ($\theta=3\pi/2$), namely, $T_{b_{1}b_{2}}=0$ and $T_{b_{2}b_{1}}=0$ at $\theta=\pi/2$ and $\theta=3\pi/2$, respectively.

In order to analyze the optomechanical cooperativities among the two subsystems in this three-mode optomechanical system, we introduce a new parameter defined by $\Pi={\mathcal{C}}_{3}/({\mathcal{C}}_{1}{\mathcal{C}}_{2})$, which is the ratio of the optomechanical cooperativities. Thus, the analytical solutions given in Eqs.~(\ref{TTb2b1}) and~(\ref{TTb1b2}) become
\begin{subequations}
\begin{align}
\Lambda_{b_{2}b_{1}}=&\frac{4\sqrt{\Pi }\sin \theta }{(1+\sqrt{\Pi })^{2}\left[1+\frac{4\Pi \cos ^{2}\theta}{\left( \frac{{\mathcal{C}}_{1}+{\mathcal{C}}_{2}+1}{{\mathcal{C}}_{1}{\mathcal{C}}_{2}}+\Pi \right)^{2}}\right]},\label{TTTb2b1}\\
\Lambda_{b_{1}b_{2}}=&-\Lambda_{b_{2}b_{1}}.\label{TTTb1b2}
\end{align}
\end{subequations}
It can be seen from Eqs.~(\ref{TTTb2b1}) and~(\ref{TTTb1b2}) that the relative nonreciprocal phonon transfer $\Lambda_{b_{2}b_{1}}=1$ ($\Lambda_{b_{1}b_{2}}=1$) is obtained at $\Pi=1$ and $\theta=\pi/2$ ($3\pi/2$). In Figs.~\ref{scattering2}(a) and~\ref{scattering2}(b), we plot the relative phonon-scattering rates $\Lambda_{b_{2}b_{1}}$ and $\Lambda_{b_{1}b_{2}}$ as functions of the phase $\theta$ when $\Pi$ takes various values: $\Pi=0.1, 0.5$, and $1$. The results show that the optimal nonreciprocity appears at $\Pi=1$ and either $\theta=\pi/2$ or $3\pi/2$. When $\Pi\neq 1$, the absolute value of the relative phonon-scattering rate will be decreased at a given phase $\theta$. We also plot the relative phonon-scattering rates $\Lambda_{b_{2}b_{1}}$ and $\Lambda_{b_{1}b_{2}}$ versus the ratio $\Pi$ when the phase takes $\theta=\pi/2$ (solid lines) and $\theta=3\pi/2$ (dashed lines), as shown in Figs.~\ref{scattering2}(c) and~\ref{scattering2}(d). In the region $0<\Pi<1$, the nonreciprocal phonon-transfer rate $\Lambda_{b_{2}b_{1}}$ increases with the increase of $\Pi$. In the region $\Pi>1$, the relative nonreciprocal phonon-transfer rate is suppressed. The optimal nonreciprocity emerges at $\Pi=1$, which indicates directional flow of phonons between the two mechanical resonators.

\section{The cooling limits of the two mechanical resonators   \label{appendixb}}

In this section, we present a detailed derivation of the cooling limits of the two mechanical resonators, which are obtained by adiabatically eliminating the cavity-field mode in the large cavity-field decay regime. In this case, the system is reduced to a two-coupled mechanical resonator system. The derivation of the cooling limits is based on the Langevin equations~(\ref{lineLangevineq}) for the quantum fluctuations of the system operators. To obtain the cooling limits, we consider the case where the linearized optomechanical coupling strengths $G_{1,2}$ are real and the system works in the parameter regime:
\begin{eqnarray}
\omega_{1,2}\gg \kappa\gg G_{1,2}\gg \gamma_{1,2}.
\end{eqnarray}
In this case, the cavity field can be eliminated adiabatically, and then the solution of the cavity-field fluctuation operator $\delta a(t)$ at the time scale $t\gg 1/\kappa$ can be obtained as
\begin{eqnarray}
\delta a(t)&\approx &-\frac{iG_{1}}{\kappa +i(\Delta +\omega_{1})}\delta b_{1}^{\dagger}(t)-\frac{iG_{1}}{\kappa +i(\Delta-\omega_{1})}\delta b_{1}(t)-\frac{iG_{2}}{\kappa +i(\Delta+\omega_{2})}\delta b_{2}^{\dagger}(t)-\frac{iG_{2}}{\kappa +i(\Delta -\omega_{2})}\delta b_{2}(t)+F_{a,\text{in}}(t),\label{dletaaadiaelim}\nonumber\\
\end{eqnarray}
where we introduce the new noise operator
\begin{equation}
F_{a,\text{in}}(t)=\sqrt{2\kappa }e^{-(\kappa +i\Delta)t}\int_{0}^{t}a_{\text{in}}(s)e^{(\kappa +i\Delta)s}ds.
\end{equation}
Substitution of Eq.~(\ref{dletaaadiaelim}) into Eqs.~(\ref{lineLangevineq}b) and~(\ref{lineLangevineq}c) leads to the equations of motion
\begin{subequations}
\begin{align}
\delta\dot{b}_{1}(t)=&\left(\frac{G_{1}^{2}}{\kappa-i(\Delta-\omega_{1})}-\frac{G_{1}^{2}}{\kappa +i(\Delta+\omega _{1})}\right)\delta b_{1}^{\dagger}(t)+\left(\frac{G_{1}^{2}}{\kappa-i(\Delta+\omega_{1})}-
\frac{G_{1}^{2}}{\kappa+i(\Delta-\omega_{1})}-(\gamma_{1}+i\omega_{1})\right)\delta b_{1}(t)\nonumber \\
&+\left(\frac{G_{1}G_{2}}{\kappa-i(\Delta-\omega_{2})}-\frac{G_{1}G_{2}}{\kappa +i(\Delta +\omega_{2})}\right)\delta b_{2}^{\dagger }(t)+\left(\frac{G_{1}G_{2}}{\kappa-i(\Delta+\omega_{2})}-\frac{G_{1}G_{2}}{\kappa+i(\Delta-\omega_{2})}-i\eta e^{i\theta}\right)\delta b_{2}(t)\nonumber \\
&-iG_{1}F_{a,\text{in}}(t)-iG_{1}F_{a,\text{in}}^{\dagger }(t)+\sqrt{2\gamma _{1}}b_{1,\text{in}}(t),\label{fulleqofb1}\\
\delta\dot{b}_{2}(t)=&\left(\frac{G_{1}G_{2}}{\kappa-i(\Delta-\omega_{1})}-\frac{G_{1}G_{2}}{\kappa+i(\Delta+\omega _{1})}\right)\delta b_{1}^{\dagger}(t)+\left(\frac{G_{1}G_{2}}{\kappa-i(\Delta+\omega_{1})}-\frac{G_{1}G_{2}}{\kappa+i(\Delta -\omega_{1})}-i\eta e^{-i\theta}\right)\delta b_{1}(t)\nonumber\\
&+\left(\frac{G_{2}^{2}}{\kappa-i(\Delta-\omega_{2})}-\frac{G_{2}^{2}}{\kappa +i(\Delta+\omega_{2})}\right)\delta b_{2}^{\dagger }(t)+\left(-(\gamma_{2}+i\omega_{2})+\frac{G_{2}^{2}}{\kappa-i(\Delta +\omega_{2})}-\frac{G_{2}^{2}}{\kappa+i(\Delta-\omega _{2})}\right)\delta b_{2}(t)\nonumber \\
&-iG_{2}F_{a,\text{in}}(t)-iG_{2}F_{a,\text{in}}^{\dagger}(t)+\sqrt{2\gamma_{2}}b_{2,\text{in}}(t).\label{fulleqofb2}
\end{align}
\end{subequations}
By making the RWA in Eqs.~(\ref{fulleqofb1}) and~(\ref{fulleqofb2}), we have
\begin{subequations}
\label{redeqofb1b2}
\begin{align}
\delta\dot{b}_{1}(t)=&-(\Gamma_{1}+i\Omega_{1})\delta b_{1}(t)+\xi_{1}\delta b_{2}(t)-iG_{1}F_{a,\text{in}}(t)-iG_{1}F_{a,\text{in}}^{\dagger}(t)+\sqrt{2\gamma_{1}}b_{1,\text{in}}(t),\\
\delta\dot{b}_{2}(t)=&\xi_{2}\delta b_{1}(t)-(\Gamma_{2}+i\Omega_{2}) \delta b_{2}(t)-iG_{2}F_{a,\text{in}}(t)-iG_{2}F_{a,\text{in}}^{\dagger}(t)+\sqrt{2\gamma_{2}}b_{2,\text{in}}(t),
\end{align}
\end{subequations}
where we introduce the effective resonance frequency $\Omega_{l}$ and decay rate $\Gamma_{l}$ for the $l$th mechanical resonator
\begin{subequations}
\begin{align}
\Omega_{l}=&\omega_{l}-\omega_{l,\text{opt}}, \\
\Gamma_{l}=&\gamma_{l}+\gamma_{l,\text{opt}},
\end{align}
\end{subequations}
with
\begin{subequations}
\begin{align}
\omega_{l,\text{opt}}=&\frac{G_{l}^{2}(\Delta+\omega_{l})}{\kappa^{2}+(\Delta +\omega_{l})^{2}}+\frac{G_{l}^{2}(\Delta -\omega_{l})}{\kappa^{2}+(\Delta-\omega_{l})^{2}},\\
\gamma_{l,\text{opt}}=&\frac{G_{l}^{2}\kappa}{\kappa^{2}+(\Delta-\omega_{l})^{2}}-\frac{G_{l}^{2}\kappa}{\kappa^{2}+(\Delta+\omega _{l})^{2}},\hspace{0.5 cm} l=1,2.
\end{align}
\end{subequations}
Here, $\omega_{l,\text{opt}}$ and $\gamma_{l, \text{opt}}$ denote the resonance frequency shift and the additional energy decay rate induced by the optomechanical couplings, respectively. We also introduce the effective coupling strengths between the two mechanical modes $b_{1}$ and $b_{2}$ after adiabatically eliminating the cavity mode as
\begin{subequations}
\begin{align}
\xi_{1}=&\frac{G_{1}G_{2}[\kappa +i(\Delta +\omega _{2})]}{\kappa^{2}+(\Delta +\omega _{2})^{2}}-\frac{G_{1}G_{2}[\kappa -i(\Delta -\omega_{2})]}{\kappa ^{2}+(\Delta -\omega _{2})^{2}}-i\eta e^{i\theta}, \\
\xi_{2}=&\frac{G_{1}G_{2}[\kappa +i(\Delta +\omega _{1})]}{\kappa^{2}+(\Delta +\omega _{1})^{2}}-\frac{G_{1}G_{2}[\kappa -i(\Delta -\omega_{1})]}{\kappa ^{2}+(\Delta -\omega _{1})^{2}}-i\eta e^{-i\theta}.
\end{align}
\end{subequations}
Under the parameter condition $\omega_{1,2}\gg\kappa\gg G_{1,2}$ and at resonance $\Delta=\omega_{1}=\omega_{2}$, we have
\begin{subequations}
\begin{align}
\xi_{1}\approx&-\left[\frac{G_{1}G_{2}}{\kappa} +i\left(\eta e^{i\theta}-\frac{G_{1}G_{2}}{2\omega_{2}}\right)\right],\\
\xi_{2}\approx&-\left[\frac{G_{1}G_{2}}{\kappa} +i\left(\eta e^{-i\theta}-\frac{G_{1}G_{2}}{2\omega_{1}}\right)\right],
\end{align}
\end{subequations}
and
\begin{subequations}
\begin{align}
\gamma_{l,\text{opt}}\approx& \frac{G_{l}^{2}}{\kappa},\\
\omega_{l,\text{opt}}\approx& \frac{G_{l}^{2}}{2\omega_{l}},\hspace{0.5 cm} l=1,2.
\end{align}
\end{subequations}
The final average phonon numbers (namely the steady-state values of the phonon numbers) can be obtained by solving Eq.~(\ref{redeqofb1b2}). To be concise, we reexpress Eq.~(\ref{redeqofb1b2}) as
\begin{equation}
\mathbf{\dot{v}}(t)=-\mathbf{Mv}(t)+\mathbf{N}(t),\label{eqcomptform}
\end{equation}
where $\mathbf{v}(t)=(\delta b_{1}(t), \delta b_{2}(t))^{T}$, $\mathbf{M}$ is defined by
\begin{equation}
\mathbf{M}=\left(\begin{array}{cc}\Gamma_{1}+i\Omega_{1} & -\xi_{1} \\-\xi_{2} & \Gamma_{2}+i\Omega_{2}
\end{array}
\right),
\end{equation}
and $\mathbf{N}(t) $ reads
\begin{equation}
\mathbf{N}(t) =\left(\begin{array}{c}
-iG_{1}F_{a,\text{in}}(t)-iG_{1}F_{a,\text{in}}^{\dagger}(t)+\sqrt{2\gamma_{1}}b_{\text{in,}1}(t)  \\
-iG_{2}F_{a,\text{in}}(t)-iG_{2}F_{a,\text{in}}^{\dagger}(t)+\sqrt{2\gamma_{2}}b_{\text{in,}2}(t)
\end{array}
\right).
\end{equation}
The formal solution of Eq.~(\ref{eqcomptform}) can be expressed as
\begin{equation}
\mathbf{v}(t)=e^{-\mathbf{M}t}\mathbf{v}(0) +e^{-\mathbf{M}t}\int_{0}^{t}e^{\mathbf{Ms}}\mathbf{N}(s)ds.
\end{equation}
The final average phonon numbers can be obtained by calculating the elements of the variance matrix. By a lengthy calculation, we obtain the approximate analytical expressions for the final average phonon numbers as
\begin{eqnarray}
n_{1}^{f} &=&\frac{\gamma_{1}\bar{n}_{1}}{2\vert u\vert ^{2}}\bigg[\frac{\vert[u-\Gamma_{1}+\Gamma_{2}-i(\Omega _{1}-\Omega_{2})]\vert^{2}}{\lambda _{1}^{\ast}+\lambda _{1}}+2\mathrm{Re}\Big[\frac{[u^{\ast}-\Gamma _{1}+\Gamma _{2}+i(\Omega _{1}-\Omega _{2})] [u+\Gamma_{1}-\Gamma _{2}+i(\Omega _{1}-\Omega_{2})]}{\lambda _{1}^{\ast}+\lambda_{2}}\Big]\nonumber\\
&& +\frac{\vert [u+\Gamma_{1}-\Gamma_{2}+i(\Omega_{1}-\Omega_{2})] \vert ^{2}}{\lambda_{2}^{\ast}+\lambda _{2}}\bigg]+\frac{G_{1}^{2}}{4\vert u\vert^{2}}\bigg[\frac{\vert [u-\Gamma _{1}+\Gamma _{2}-i(\Omega _{1}-\Omega _{2})] \vert^{2}}{\lambda_{1}^{\ast }+\lambda _{1}}\Big(\frac{1}{\kappa+\lambda _{1}+i\Delta }+\frac{1}{\kappa +\lambda_{1}^{\ast}-i\Delta}\Big) \nonumber\\
&&+2\mathrm{Re}\Big[\frac{[u^{\ast }-\Gamma _{1}+\Gamma_{2}+i(\Omega _{1}-\Omega_{2})][u+\Gamma _{1}-\Gamma _{2}+i(\Omega_{1}-\Omega_{2})]}{\lambda_{1}^{\ast}+\lambda_{2}}\Big(\frac{1}{\kappa +\lambda _{2}+i\Delta }+\frac{1}{\kappa +\lambda _{1}^{\ast}-i\Delta}\Big)\Big] \nonumber\\
&&+\frac{\vert [u+\Gamma _{1}-\Gamma _{2}+i(\Omega _{1}-\Omega_{2})] \vert ^{2}}{\lambda _{2}^{\ast }+\lambda _{2}}\Big(\frac{1}{\kappa +\lambda _{2}+i\Delta}+\frac{1}{\kappa +\lambda_{2}^{\ast}-i\Delta}\Big)\bigg]\nonumber\\
&&+\frac{\vert \xi_{1}\vert^{2}}{\vert u\vert^{2}}\Bigg[G_{2}^{2}\bigg[\frac{1}{\lambda_{1}^{\ast}+\lambda _{1}}\Big(\frac{1}{\kappa+\lambda _{1}+i\Delta }+\frac{1}{\kappa +\lambda _{1}^{\ast }-i\Delta}\Big)-2\mathrm{Re}\Big[\frac{1}{\lambda _{1}^{\ast }+\lambda_{2}}\Big(\frac{1}{\kappa+\lambda _{2}+i\Delta }+\frac{1}{\kappa+\lambda _{1}^{\ast }-i\Delta}\Big)\Big]\nonumber\\
&&+\frac{1}{\lambda_{2}^{\ast }+\lambda _{2}}\Big(\frac{1}{\kappa+\lambda _{2}+i\Delta }+\frac{1}{\kappa +\lambda_{2}^{\ast}-i\Delta}\Big)\bigg]+2\gamma_{2}\bar{n}_{2}\bigg(\frac{1}{\lambda _{1}^{\ast }+\lambda _{1}}-\frac{1}{\lambda _{1}^{\ast }+\lambda _{2}}-\frac{1}{\lambda _{2}^{\ast }+\lambda _{1}}+\frac{1}{\lambda _{2}^{\ast }+\lambda _{2}}\bigg)\Bigg],
\end{eqnarray}
and
\begin{eqnarray}
n_{2}^{f} &=&\frac{\gamma _{2}\bar{n}_{2}}{2\vert u\vert ^{2}}\bigg[\frac{\vert u+\Gamma _{1}-\Gamma _{2}+i(\Omega _{1}-\Omega_{2})\vert ^{2}}{\lambda _{1}^{\ast }+\lambda _{1}}+2\mathrm{Re}\Big[\frac{[u^{\ast }+\Gamma _{1}-\Gamma_{2}-i(\Omega _{1}-\Omega _{2})][u-\Gamma_{1}+\Gamma _{2}-i(\Omega _{1}-\Omega _{2})]}{\lambda _{1}^{\ast }+\lambda _{2}}\Big]  \nonumber\\
&&+\frac{\vert u-\Gamma _{1}+\Gamma _{2}-i(\Omega_{1}-\Omega _{2})\vert ^{2}}{\lambda _{2}^{\ast }+\lambda _{2}}\bigg]+\frac{G_{2}^{2}}{4\vert u\vert ^{2}}\bigg[\frac{\vert u+\Gamma _{1}-\Gamma _{2}+i(\Omega_{1}-\Omega_{2})\vert ^{2}}{\lambda _{1}^{\ast }+\lambda_{1}}\Big(\frac{1}{\kappa +\lambda _{1}+i\Delta}+\frac{1}{\kappa +\lambda _{1}^{\ast}-i\Delta }\Big)  \nonumber\\
&&+2\mathrm{Re}\Big[\frac{[u-\Gamma _{1}+\Gamma _{2}-i(\Omega _{1}-\Omega _{2})][u^{\ast}+\Gamma _{1}-\Gamma _{2}-i(\Omega _{1}-\Omega _{2})]}{\lambda_{1}^{\ast }+\lambda_{2}}\Big(\frac{1}{\kappa +\lambda _{2}+i\Delta}+\frac{1}{\kappa +\lambda _{1}^{\ast }-i\Delta }\Big)\Big]  \nonumber\\
&&+\frac{\vert u-\Gamma _{1}+\Gamma _{2}-i(\Omega _{1}-\Omega_{2})\vert ^{2}}{\lambda^{\ast}_{2}+\lambda_{2}}\Big(\frac{1}{\kappa +\lambda_{2}+i\Delta }+\frac{1}{\kappa +\lambda^{\ast}_{2}-i\Delta }\Big)\bigg]  \nonumber\\
&&+\frac{\vert \xi_{2}\vert^{2}}{\vert u\vert ^{2}}\Bigg[G^{2}_{1}\bigg[\frac{1}{\lambda _{1}^{\ast}+\lambda _{1}}\Big(\frac{1}{\kappa +\lambda _{1}+i\Delta }+\frac{1}{\kappa
+\lambda _{1}^{\ast }-i\Delta }\Big)-2\mathrm{Re}\Big[\frac{1}{\lambda _{2}+\lambda_{1}^{\ast }}\Big(\frac{1}{\kappa +\lambda _{2}+i\Delta }+\frac{1}{\kappa+\lambda _{1}^{\ast }-i\Delta }\Big)\Big] \nonumber\\
&&+\frac{1}{\lambda _{2}^{\ast }+\lambda _{2}}\Big(\frac{1}{\kappa +\lambda_{2}+i\Delta}+\frac{1}{\kappa +\lambda _{2}^{\ast }-i\Delta }\Big)\bigg]+2\gamma _{1}\bar{n}_{1}\Big(\frac{1}{\lambda_{1}^{\ast }+\lambda _{1}}-\frac{1}{\lambda _{1}^{\ast }+\lambda _{2}}-\frac{1}{\lambda _{2}^{\ast}+\lambda _{1}}+\frac{1}{\lambda _{2}^{\ast }+\lambda _{2}}\Big)\Bigg],
\end{eqnarray}
where $\lambda_{1}$ and $\lambda_{2}$ ($\lambda^{\ast}_{1}$ and $\lambda^{\ast}_{2}$ being complex conjugate) are the eigenvalues of the coefficient matrix $\mathbf{M}$,
\begin{subequations}
\label{eigenvalues}
\begin{align}
\lambda_{1}=&\frac{1}{2}[\Gamma_{1}+\Gamma_{2}+i(\Omega_{1}+\Omega_{2})- u],\\
\lambda_{2}=&\frac{1}{2}[\Gamma_{1}+\Gamma_{2}+i(\Omega_{1}+\Omega_{2})+ u]
\end{align}
\end{subequations}
where
\begin{equation}
u=\sqrt{4\xi_{1}\xi_{2}+[\Gamma_{1}-\Gamma_{2}+i(\Omega_{1}-\Omega_{2})]^{2}}.
\end{equation}
For the case $\omega_{1}\approx\omega_{2}$ and $\Gamma_{1}\approx\Gamma_{2}$, the approximate analytical expressions of the final average phonon numbers can be reduced as
\begin{eqnarray}
n_{1}^{f} &\approx &\frac{\gamma_{1}\bar{n}_{1}}{2}\bigg(\frac{1}{\lambda_{1}^{\ast }+\lambda_{1}}+2\mathrm{Re}\Big[\frac{1}{\lambda_{1}^{\ast}+\lambda_{2}}\Big]+\frac{1}{\lambda _{2}^{\ast }+\lambda
_{2}}\bigg)+\frac{G_{1}^{2}}{4}\bigg[\frac{1}{\lambda _{1}^{\ast }+\lambda_{1}}\Big(\frac{1}{\kappa +\lambda _{1}+i\Delta }+\frac{1}{\kappa +\lambda_{1}^{\ast }-i\Delta }\Big)   \nonumber\\
&&+2\mathrm{Re}\Big[\frac{1}{\lambda _{1}^{\ast }+\lambda _{2}}\Big(\frac{1}{\kappa +\lambda_{2}+i\Delta }+\frac{1}{\kappa +\lambda _{1}^{\ast }-i\Delta }\Big)\Big]+\frac{1}{\lambda _{2}^{\ast }+\lambda _{2}}\Big(\frac{1}{\kappa +\lambda_{2}+i\Delta }+\frac{1}{\kappa +\lambda _{2}^{\ast }-i\Delta }\Big)\bigg]\nonumber\\
&&+\frac{\vert \xi_{1}\vert }{4|\xi _{2}|}\Bigg[G_{2}^{2}\bigg[\frac{1}{\lambda _{1}^{\ast }+\lambda _{1}}\Big(\frac{1}{\kappa +\lambda _{1}+i\Delta }+\frac{1}{\kappa +\lambda_{1}^{\ast }-i\Delta}\Big)-2\mathrm{Re}\Big[\frac{1}{\lambda _{1}^{\ast }+\lambda _{2}}\Big(\frac{1}{\kappa+\lambda_{2}+i\Delta }+\frac{1}{\kappa +\lambda _{1}^{\ast }-i\Delta }\Big)\Big]\nonumber\\
&&+\frac{1}{\lambda _{2}^{\ast}+\lambda _{2}}\Big(\frac{1}{\kappa +\lambda _{2}+i\Delta}+\frac{1}{\kappa +\lambda_{2}^{\ast}-i\Delta }\Big)\bigg]+2\gamma _{2}\bar{n}_{2}\bigg(\frac{1}{\lambda _{1}^{\ast }+\lambda _{1}}-2\mathrm{Re}\Big[\frac{1}{\lambda_{1}^{\ast }+\lambda_{2}}\Big]+\frac{1}{\lambda _{2}^{\ast }+\lambda _{2}}\bigg)\Bigg],\label{finala}
\end{eqnarray}
and
\begin{eqnarray}
n_{2}^{f} &\approx&\frac{\gamma _{2}\bar{n}_{2}}{2}\bigg(\frac{1}{\lambda _{1}^{\ast}+\lambda _{1}}+2\mathrm{Re}\Big[\frac{1}{\lambda _{1}^{\ast }+\lambda _{2}}\Big]+\frac{1}{\lambda _{2}^{\ast }+\lambda _{2}}\bigg)+
\frac{G_{2}^{2}}{4}\bigg[\frac{1}{\lambda_{1}^{\ast }+\lambda _{1}}\Big(\frac{1}{\kappa +\lambda _{1}+i\Delta}+\frac{1}{\kappa +\lambda _{1}^{\ast }-i\Delta }\Big)\nonumber\\
&&+2\mathrm{Re}\Big[\frac{1}{\lambda _{1}^{\ast }+\lambda _{2}}\Big(\frac{1}{\kappa +\lambda _{2}+i\Delta }+\frac{1}{\kappa +\lambda _{1}^{\ast }-i\Delta}\Big)\Big] +\frac{1}{\lambda^{\ast}_{2}+\lambda _{2}}\Big(\frac{1}{\kappa+\lambda _{2}+i\Delta }+\frac{1}{\kappa +\lambda^{\ast}_{2}-i\Delta }\Big)\bigg]\nonumber\\
&&+\frac{\vert \xi _{2}\vert }{4\vert \xi_{1}\vert }\Bigg[G_{1}^{2}\bigg[\frac{1}{\lambda _{1}^{\ast}+\lambda _{1}}\Big(\frac{1}{\kappa +\lambda _{1}+i\Delta }+\frac{1}{\kappa
+\lambda _{1}^{\ast }-i\Delta }\Big)-2\mathrm{Re}\Big[\frac{1}{\lambda_{1}^{\ast }+\lambda_{2}}\Big(\frac{1}{\kappa +\lambda _{2}+i\Delta }+\frac{1}{\kappa+\lambda _{1}^{\ast }-i\Delta }\Big)\Big]  \nonumber\\
&&+\frac{1}{\lambda _{2}^{\ast }+\lambda _{2}}\Big(\frac{1}{\kappa +\lambda_{2}+i\Delta}+\frac{1}{\kappa +\lambda _{2}^{\ast }-i\Delta }\Big)\bigg]+2\gamma _{1}\bar{n}_{1}\Big(\frac{1}{\lambda _{1}^{\ast }+\lambda _{1}}-2\mathrm{Re}\Big[\frac{1}{\lambda _{1}^{\ast }+\lambda _{2}}\Big]+\frac{1}{\lambda _{2}^{\ast }+\lambda _{2}}\Big)\Bigg].\label{finalb}
\end{eqnarray}
By substituting Eq.~(\ref{eigenvalues}) into Eqs.~(\ref{finala}) and~(\ref{finalb}) and considering the parameters relations $\omega_{1,2}\gg\kappa\gg G_{1,2}\gg\{\gamma_{1,\text{opt}}\approx\gamma_{2,\text{opt}}\}\gg\gamma_{1,2}$, the final average phonon numbers can be simplified as
\begin{eqnarray}
n_{l=1,2}^{f}&\approx&\frac{\gamma_{l}\bar{n}_{l}+\gamma_{l,\text{opt}}n_{\text{opt}}}{\Gamma_{l}+\chi_{+}}+\frac{(-1)^{l-1}\sqrt{\chi_{l}}(\sqrt{\chi_{1}}n_{\chi_{1}}
-\sqrt{\chi_{2}}n_{\chi_{2}})}{\Gamma_{l}+\chi_{-}},\label{coolimit}
\end{eqnarray}
where we introduce the following variables
\begin{subequations}
\begin{align}
n_{\text{opt}}=&\frac{4\kappa ^{2}}{(\omega_{1}+\omega_{2}+2\Delta)^{2}},\\
n_{\chi_{1(2)}}=&\frac{2(\gamma_{2(1)}\bar{n}_{2(1)}+\gamma_{2(1)\text{opt}}n_{\text{opt}})}{\Gamma_{1}+\Gamma_{2}+2\chi_{+}}, \\
\chi_{\pm}=&\mp \sqrt{\chi_{1}\chi_{2}}-\text{Re}\left[\frac{\xi_{1}\xi_{2}}{\Gamma_{1}+\Gamma_{2}}\right],\\
\chi_{l=1,2}=&\frac{\vert \xi_{l}\vert ^{2}}{\Gamma_{1}+\Gamma_{2}}.
\end{align}
\end{subequations}
Here, $n_{\text{opt}}$ stands for the effective phonon number in the optomechanical cooling bath, and $\chi_{1}$ and $\chi_{2}$ are the effective phonon-transfer rates from $b_{2}$ to $b_{1}$ and from $b_{1}$ to $b_{2}$, respectively. The corresponding cooling limits ($n_{1}^{\lim }$, $n_{2}^{\lim}$) are obtained by taking the optimal driving detuning $\Delta=\omega_{l}$ in Eq.~(\ref{coolimit}). In particular, the first term in Eq.~(\ref{coolimit}) is contributed by the thermal bath and the effective optical bath connected by the $l$th mechanical resonator, while the phonon extraction contribution induced by the phonon-exchange channel is presented by the last term. Physically, the nonreciprocity of the phonon transfer is decided by the phonon-exchange rate $\chi_{l}$ which depends on the phase $\theta$. In the case $\bar{n}_{1}\approx\bar{n}_{2}$ and $\gamma_{1}\approx \gamma_{2}$, we have $n_{\chi_{1}}\approx n_{\chi_{2}}=n_{\chi}$ and thus $(\sqrt{\chi_{1}}n_{\chi_{1}}-\sqrt{\chi_{2}}n_{\chi_{2}})\approx (\sqrt{\chi_{1}}-\sqrt{\chi_{2}})n_{\chi}$. In the region $0<\theta<\pi$ ($\pi<\theta<2\pi$), we obtain $\sqrt{\chi_{1}}<\sqrt{\chi_{2}}$ ($\sqrt{\chi_{1}}>\sqrt{\chi_{2}}$). This means that the phonon-transfer rate from $b_{1}$ ($b_{2}$) to $b_{2}$ ($b_{1}$) is larger than that for the opposite case. According to Eq.~(\ref{coolimit}), we then have the relation $n^{f}_{1}<n^{f}_{2}$ ($n^{f}_{1}>n^{f}_{2}$) in the region $0<\theta<\pi$ ($\pi<\theta<2\pi$). When $\theta=\pi/2$ ($3\pi/2$) and $\sqrt{{\mathcal{C}}_{1}{\mathcal{C}}_{2}}=\sqrt{{\mathcal{C}}_{3}}$, the unidirectional flow of the phonons between the two mechanical resonators is achieved [$\chi_{1}\approx0$ ($\chi_{2}\approx0$)]. For $\theta=n\pi$, the phonon transfer between the two mechanical resonators is reciprocal ($\sqrt{\chi_{1}}=\sqrt{\chi_{2}}$), due to the emergence of the dark mode. Once the phonon-transfer channel is turned off ($\eta=0$), the ground-state cooling is unfeasible owing to the invalid effective cooling channel ($\Gamma_{l}+\chi_{+}\rightarrow\gamma_{l}$). In the absence of the optomechanical cooling channels ($G_{1,2}=0$), Eq.~(\ref{coolimit}) becomes $n_{l=1,2}^{f}\approx \bar{n}_{l}+(-1)^{l-1}(n_{\chi_{1}}-n_{\chi_{2}})/2$, which indicates quantum thermalization in this coupled mechanical system.

%%%%%%%%%%%%%%%%%%%%%%%%%%%%%%
\begin{figure}[tbp]
\centering
\includegraphics[bb=0 0 533 235, width=0.7 \textwidth]{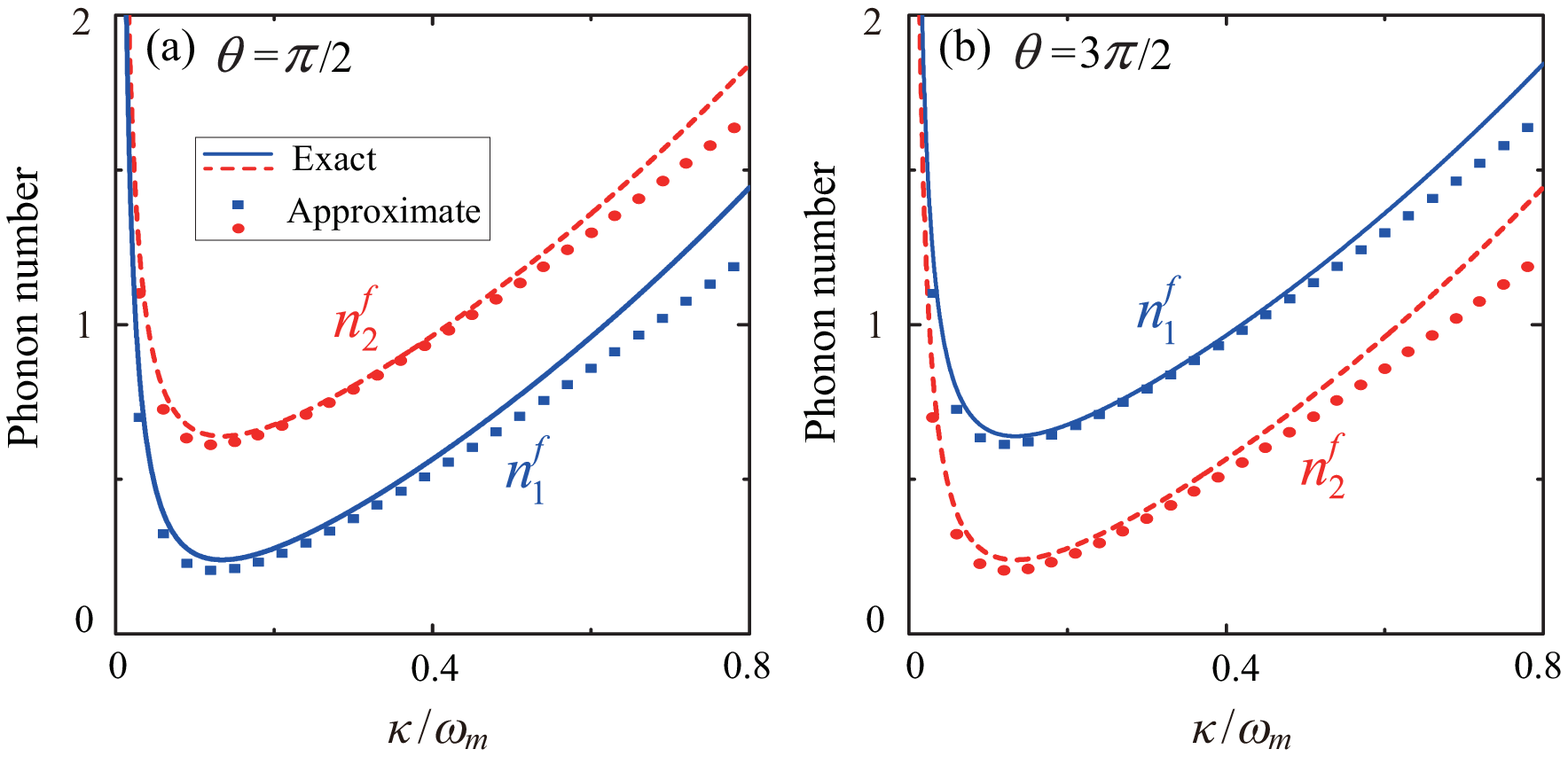}
\caption{(Color online) The final average phonon numbers $n^{f}_{1}$ and $n^{f}_{2}$ are plotted as functions of $\kappa/\omega_{m}$ when the phase $\theta$ takes the values: (a) $\theta=\pi/2$ and (b) $\theta=3\pi/2$. The exact results are given by Eq.~(\ref{finalexact}) (solid curves) and the approximate results obtained by the adiabatic elimination method are given by Eq.~(\ref{coolimit}) (symbols). Here, the used parameters are $\Delta=\omega_{1}=\omega_{2}=\omega_{m}$, $G_{1}/\omega_{m}=G_{2}/\omega_{m}=0.08$, $\eta/\omega_{m}=0.05$, $\gamma_{1}/\omega_{m}=\gamma_{2}/\omega_{m}=10^{-5}$, and $\bar{n}_{1}=\bar{n}_{2}=1000$.}\label{kexapp}
\end{figure}
%%%%%%%%%%%%%%%%%%%%%%%%%%%%%%

Moreover, both the exact and approximate final average phonon numbers $n^{f}_{1}$ and $n^{f}_{2}$ are plotted in Fig.~\ref{kexapp} as functions of the cavity-field decay rate $\kappa$ at the optimal driving detuning $\Delta=\omega_{m}$ when the modulation phase $\theta$ takes various values: (a) $\theta=\pi/2$ and (b) $\theta=3\pi/2$. Here, the blue solid curves ($n^{f}_{1}$) and the red dashed curves ($n^{f}_{2}$) are plotted using the exact solutions given in Eq.~(\ref{finalexact}), while the symbols are based on the analytical calculations given in Eq.~(\ref{coolimit}). We can see from Fig.~\ref{kexapp} that the analytical cooling limits and the exact results match well with each other when $\kappa/\omega_{m}<0.4$, and the difference between the numerical simulation and approximate results increases when $\kappa/\omega_{m}>0.4$. This means that the cooling performances of the two mechanical resonators are excellent in the resolved-sideband regime ($\kappa\ll\omega_{m}$). This result is consistent with the sideband cooling results in the typical optomechanical systems. We also see from Fig.~\ref{kexapp}(a) that the cooling performance of the first resonator is better than that of the second resonator ($n^{f}_{1}<n^{f}_{2}$) when $\theta=\pi/2$. However, when $\theta=3\pi/2$, the opposite cooling performance ($n^{f}_{1}>n^{f}_{2}$) has been displayed in comparison with the case of $\theta=\pi/2$, as shown in Fig.~\ref{kexapp}(b). Physically, the nonreciprocal phonon-transfer mechanism is more helpful to cool the first (second) resonator when $0<\theta<\pi$ ($\pi<\theta<2\pi$). In particular, the optimal cooling performances of the two mechanical resonators require that the working value of cavity-field decay rate is around $\kappa/\omega_{m}=0.1\sim0.2$, as shown in Fig.~\ref{kexapp}. This is a result of the competition between the efficiency of extraction of the thermal excitations and the phonon-sideband resolution condition.  When $\kappa/\omega_{m}<0.1$, the cooling performances of the two mechanical resonators become worse. Physically, the vacuum bath of the cavity field extracts the thermal excitations in the two mechanical resonators through a manner of nonequilibrium dynamics, and then the total system reaches a steady state. When the cavity-field decay rate $\kappa$ is equal to $0$, the vacuum bath cannot extract the thermal phonons in these two mechanical resonators, and then this system will be thermalized to a thermal equilibrium state.

\section{The dark-mode effect and its breaking in a multiple-mechanical-resonator optomechanical system}

\begin{figure}[tbp]
\centering
\includegraphics[bb=0 0 391 143, width=0.7\textwidth]{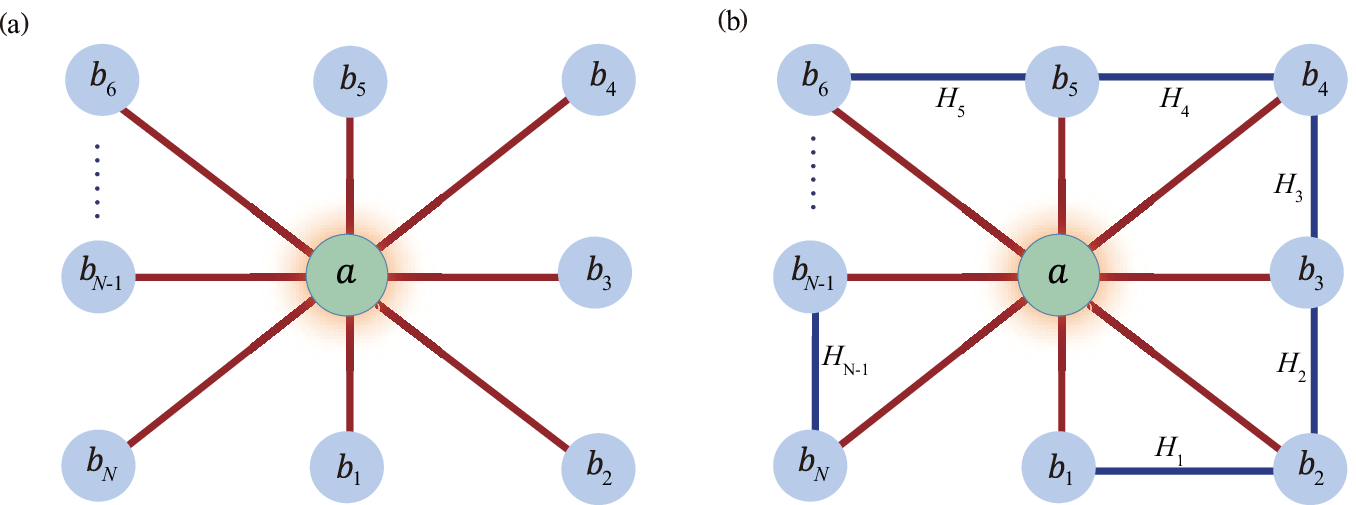}
\caption{(Color online) (a) The $N$-mechanical-resonator optomechanical system: a cavity-field mode simultaneously couples to $N$ mechanical resonators through the optomechanical interactions. (b) The phonon-exchange interactions between two neighboring mechanical resonators are introduced into the $N$-mechanical-resonator optomechanical system described by panel (a). Note that there is no direct coupling between the first resonator and the $N$th resonator.}\label{net}
\end{figure}

In this section, we study the dark-mode effect in a multiple-mechanical-resonator optomechanical system, which consists of one cavity mode and $N$ $(N\geq 3)$ mechanical resonators [see Figs.~\ref{net}(a) and~\ref{net}(b)]. The Hamiltonian of this system can be written in a frame rotating at the driving frequency $\omega_{L}$ as
\begin{eqnarray}
H_{I}&=&\Delta_{c}a^{\dagger}a+\sum_{j=1}^{N}\omega_{j}b_{j}^{\dagger}b_{j}+\sum_{j=1}^{N}g_{j}a^{\dagger }a(b_{j}+b_{j}^{\dagger})+(\Omega a+\Omega^{\ast}a^{\dagger})+\sum_{j=1}^{N-1}\eta_{j}(e^{i\theta_{j}}b_{j}^{\dagger}b_{j+1}+\mathrm{H.c.}),\label{eq1iniHIN}
\end{eqnarray}
where $\Delta_{c}=\omega_{c}-\omega_{L}$ is the detuning of the cavity-field resonance frequency $\omega_{c}$ with respect to the driving frequency $\omega_{L}$. The operators $a$ ($a^{\dagger}$) and $b_{j}$ ($b^{\dagger}_{j}$) are, respectively, the annihilation (creation) operators of the cavity-field mode and the $j$th mechanical resonator (with the resonance frequency $\omega_{j}$). The optomechanical interactions between the cavity mode and the $j$th mechanical resonator are described by the $g_{j}$ terms (with $g_{j}$ being the single-photon optomechanical-coupling strength). The cavity-field driving is denoted by the $\Omega$ term (with $\Omega$ being the driving amplitude). To manipulate the energy exchange between the neighboring mechanical resonators, we introduce a phase-dependent phonon-exchange interaction between the neighboring mechanical resonators, with the coupling strength $\eta_{j}$ and the phase $\theta_{j}$. By phenomenologically adding the damping and noise terms into the Heisenberg equations obtained based on the Hamiltonian in Eq.~(\ref{eq1iniHIN}), the quantum Langevin equations for the operators of the optical and mechanical modes can be obtained as
\begin{eqnarray}
\dot{a} &=&-ia[\Delta_{c}+g_{1}(b_{1}+b_{1}^{\dagger})+g_{2}(b_{2}+b_{2}^{\dagger})+\cdots+g_{N}(b_{N}+b_{N}^{\dagger})]-i\Omega -\kappa a+\sqrt{2\kappa }a_{\text{in}},  \nonumber \\
\dot{b}_{1} &=&-(\gamma_{1}+i\omega_{1})b_{1}-ig_{1}a^{\dagger}a-i\eta_{1}e^{i\theta_{1}}b_{2}+\sqrt{2\gamma _{1}}b_{1,\text{in}},  \nonumber \\
\dot{b}_{2} &=&-(\gamma_{2}+i\omega_{2})b_{2}-ig_{2}a^{\dagger}a-i\eta_{1}e^{-i\theta_{1}}b_{1}-i\eta_{2}e^{i\theta_{2}}b_{3}+\sqrt{2\gamma_{2}}b_{2,\text{in}},  \nonumber \\
\dot{b}_{3} &=&-(\gamma_{3}+i\omega_{3})b_{3}-ig_{3}a^{\dagger}a-i\eta_{2}e^{-i\theta_{2}}b_{2}-i\eta_{3}e^{i\theta_{3}}b_{4}+\sqrt{2\gamma_{3}}b_{3,\text{in}},  \nonumber \\
\dot{b}_{4} &=&-(\gamma_{4}+i\omega_{4})b_{4}-ig_{4}a^{\dagger}a-i\eta_{3}e^{-i\theta_{3}}b_{3}-i\eta_{4}e^{i\theta_{4}}b_{5}+\sqrt{2\gamma _{4}}b_{4,\text{in}},  \nonumber \\
&&\vdots  \nonumber \\
\dot{b}_{N-1} &=&-(\gamma _{N-1}+i\omega _{N-1})b_{N-1}-ig_{N-1}a^{\dagger}a-i\eta _{N-2}e^{-i\theta _{N-2}}b_{N-2}-i\eta _{N-1}e^{i\theta_{N-1}}b_{N}+\sqrt{2\gamma _{N-1}}b_{N-1,\text{in}},  \nonumber \\
\dot{b}_{N} &=&-(\gamma _{N}+i\omega _{N})b_{N}-ig_{N}a^{\dagger }a-i\eta_{N-1}e^{-i\theta_{N-1}}b_{N-1}+\sqrt{2\gamma_{N}}b_{N,\text{in}}.
\label{LangevineqorigNmr}
\end{eqnarray}

To cool the mechanical resonators, we consider the strong-driving regime of the cavity field such that the average photon number in the cavity is sufficiently large and then the linearization procedure can be used to simplify the physical model. To this end, we express the operators in Eq.~(\ref{LangevineqorigNmr}) as the sum of their steady-state mean values and quantum fluctuations, namely $o=\langle o \rangle_{\text{ss}} +\delta o$
for operators $a$, $a^{\dagger}$, $b_{j=1-N}$, and $b^{\dagger}_{j}$. By separating the classical motion and the quantum fluctuation, the linearized equations of motion for the quantum fluctuations can be written as
\begin{eqnarray}
\label{NlineLangevineq}
\frac{d}{dt}\delta a &=&-(\kappa +i\Delta )\delta a-i\alpha \lbrack g_{1}(\delta b_{1}+\delta b_{1}^{\dagger })+g_{2}(\delta b_{2}+\delta b_{2}^{\dagger})+\cdots+g_{N-1}(\delta b_{N-1}+\delta b_{N-1}^{\dagger }) \nonumber \\
&&+g_{N}(\delta b_{N}+\delta b_{N}^{\dagger })]+\sqrt{2\kappa }a_{\text{in}},  \nonumber \\
\frac{d}{dt}\delta b_{1} &=&-(\gamma _{1}+i\omega _{1})\delta b_{1}-ig_{1}\alpha ^{\ast }\delta a-ig_{1}\alpha \delta a^{\dagger }-i\eta_{1}e^{i\theta _{1}}\delta b_{2}+\sqrt{2\gamma _{1}}b_{1,\text{in}},  \nonumber \\
\frac{d}{dt}\delta b_{2} &=&-(\gamma _{2}+i\omega _{2})\delta b_{2}-ig_{2}\alpha ^{\ast }\delta a-ig_{2}\alpha \delta a^{\dagger }-i\eta_{1}e^{-i\theta _{1}}\delta b_{1}-i\eta _{2}e^{i\theta _{2}}\delta b_{3}+\sqrt{2\gamma _{2}}b_{2,\text{in}},  \nonumber \\
\frac{d}{dt}\delta b_{3} &=&-(\gamma _{3}+i\omega _{3})\delta b_{3}-ig_{3}\alpha ^{\ast }\delta a-ig_{3}\alpha \delta a^{\dagger }-i\eta_{2}e^{-i\theta _{2}}\delta b_{2}-i\eta _{3}e^{i\theta _{3}}\delta b_{4}+\sqrt{2\gamma _{3}}b_{3,\text{in}},  \nonumber \\
\frac{d}{dt}\delta b_{4} &=&-(\gamma _{4}+i\omega _{4})\delta b_{4}-ig_{4}\alpha ^{\ast }\delta a-ig_{4}\alpha \delta a^{\dagger }-i\eta_{3}e^{-i\theta _{3}}\delta b_{3}-i\eta _{4}e^{i\theta _{4}}\delta b_{5}+\sqrt{2\gamma _{4}}b_{4,\text{in}}  \nonumber \\
&&\vdots  \nonumber \\
\frac{d}{dt}\delta b_{N-1} &=&-(\gamma _{N-1}+i\omega _{N-1})\delta b_{N-1}-ig_{N-1}\alpha ^{\ast }\delta a-ig_{N-1}\alpha \delta a^{\dagger}-i\eta _{N-2}e^{-i\theta _{N-2}}\delta b_{N-2}\nonumber\\
&&-i\eta _{N-1}e^{i\theta_{N-1}}\delta b_{N}+\sqrt{2\gamma _{N-1}}b_{N-1,\text{in}},  \nonumber \\
\frac{d}{dt}\delta b_{N} &=&-(\gamma _{N}+i\omega _{N})\delta b_{N}-ig_{N}\alpha ^{\ast }\delta a-ig_{N}\alpha \delta a^{\dagger }-i\eta_{N-1}e^{-i\theta _{N-1}}\delta b_{N-1}+\sqrt{2\gamma _{N}}b_{N,\text{in}}.
\end{eqnarray}
Based on Eqs.~(\ref{NlineLangevineq}), we adopt the same procedure as that used in the two-mechanical-resonator case to infer a linearized optomechanical Hamiltonian governing the evolution of quantum fluctuations. For studying quantum cooling of these mechanical resonators, we focus on the beam-splitting-type interactions (i.e., the rotating-wave interaction term) between these bosonic modes because these terms dominate the linearized couplings in this system, and hence we can simplify the Hamiltonian of the system by making the RWA. The linearized optomechanical Hamiltonian under the RWA is given by
\begin{eqnarray}
H_{I}=&\Delta \delta a^{\dagger}\delta a+\omega_{j}\sum_{j=1}^{N}\delta b_{j}^{\dagger}\delta b_{j}+\sum_{j=1}^{N}G_{j}(\delta a^{\dagger}\delta b_{j}+\delta b_{j}^{\dagger}\delta a)+H_{\text{mrc}}, \label{eq1iniH1Nm}
\end{eqnarray}
where $\Delta =\Delta_{c}+\sum_{j=1}^{N}g_{j}(\beta_{j}+\beta_{j}^{\ast})$ is the normalized driving detuning after the linearization, and $G_{j}=g_{j}|\alpha|$ is the linearized optomechanical coupling strength between the $j$th mechanical resonator and the cavity-field mode. The interaction Hamiltonians between the neighboring mechanical resonators are given by
\begin{eqnarray}
\label{mrca}
H_{\text{mrc}}=&\sum_{j=1}^{N-1}H_{j},
\end{eqnarray}
with
\begin{eqnarray}
\label{mrcb}
H_{j}=&\eta_{j}(e^{-i\theta_{j}}\delta b_{j}\delta b_{j+1}^{\dagger}+e^{i\theta_{j}}\delta b_{j+1}\delta b_{j}^{\dagger}),
\end{eqnarray}
which describes the phonon-exchange interaction between the $j$th resonator and the $(j+1)$th resonator.

In order to investigate the dark-mode effect in the $N$-mechanical-resonator optomechanical system, we firstly consider the case where the phonon-exchange interaction between the neighbouring mechanical resonators is absent, i.e., $H_{\text{mrc}}=0$, as shown in Fig.~\ref{net}(a). For convenience, we assume that all the mechanical resonators have the same resonance frequencies ($\omega_{j}=\omega_{m}$) and optomechanical coupling strengths ($G_{j}=G$). In this system, there exists a bright mode $B_{+}=\sum_{j=1}^{N}\delta b_{j}/\sqrt{N}$ and ($N-1$) dark modes which decouple from the cavity-field mode. As a result, the phonons stored in these dark modes cannot be extracted though the optomechanical cooling channel, and then these mechanical resonators cannot be cooled to their quantum ground states. Here, we can obtain the cooling limits of the $N$ mechanical resonators, which are given by $\bar{n}(N-1)/N$. The result shows that in the presence of the dark-mode effect, the final average phonon numbers in these mechanical resonators depend on the number of the mechanical resonators. In this case, the ground-state cooling cannot be realized in these mechanical resonators. In particular, the final average phonon numbers in these mechanical resonators are approximately equal to the thermal excitations in their heat baths when $N\gg1$ and hence $\bar{n}(N-1)/N\approx \bar{n}$.

%%%%%%%%%%%%%%%%%%%%%%%%%%%%%%
%%%%%%%%%%%%%%%%%%%%%%%%%%%%%%
To break the dark-mode effect and realize the simultaneous ground-state cooling in the $N$-mechanical-resonator optomechanical system, the phase-dependent phonon-exchange interaction $H_{\text{mrc}}$ should be introduced, as shown in Fig.~\ref{net}(b). Without loss of generality, we assume that all the coupling strengths of the phonon-exchange interactions are same $\eta_{j}=\eta$. Thus, we can diagonalize the Hamiltonian of these coupled mechanical resonators as
\begin{eqnarray}
H_{\text{mrt}} &=&\omega_{m}\sum_{j=1}^{N}\delta b_{j}^{\dagger}\delta b_{j}+\eta\sum_{j=1}^{N-1}(e^{-i\theta_{j}}\delta b_{j}\delta b_{j+1}^{\dagger}+e^{i\theta_{j}}\delta b_{j+1}\delta b_{j}^{\dagger})=\sum_{k=1}^{N}\Omega_{k}B_{k}^{\dagger}B_{k},
\end{eqnarray}
where $B_{k}$ is the $k$th mechanical normal mode with the resonance frequency $\Omega_{k}$ given by
\begin{eqnarray}
\Omega_{k}=\omega_{m}+2\eta\cos \left(\frac{k\pi}{N+1}\right),\hspace{1 cm}k=1,2,3,...,N.
\end{eqnarray}
The relationship between the mechanical modes $\delta b_{j}$ and the normal modes $B_{k}$ is given by
\begin{equation}
\delta b_{j}=\bigg\{\begin{array}{c}
\frac{1}{A}\sum_{k=1}^{N}\sin \left(\frac{k\pi}{N+1}\right)B_{k}, \hspace{2.6 cm} j=1, \\
\frac{1}{A}e^{-i\sum_{\nu =1}^{j-1}\theta_{\nu }}\sum_{k=1}^{N}\sin\left(\frac{jk\pi}{N+1}\right)B_{k}, \hspace{1.3 cm} j\geq 2,
\end{array}
\end{equation}
where we introduce the variable
\begin{eqnarray}
A=\sqrt{\frac{N+1}{2}}.
\end{eqnarray}
The Hamiltonian in Eq.~(\ref{eq1iniH1Nm}) can be rewritten with these mechanical normal modes as
\begin{eqnarray}
H_{I}=&\Delta\delta a^{\dagger}\delta a+\sum_{k=1}^{N}\Omega_{k}B_{k}^{\dagger}B_{k}+H_{\text{om}},
\end{eqnarray}
where the optomechanical Hamiltonian $H_{\text{om}}$ reads
\begin{equation}
H_{\text{om}}=\frac{G}{A}\sum_{k=1}^{N}\left[\sin\left(\frac{k\pi}{N+1}\right)+\sum_{j=2}^{N}e^{i\sum_{\nu =1}^{j-1}\theta_{\nu }}\sin \left(\frac{jk\pi}{N+1}\right)\right] aB_{k}^{\dagger}+\mathrm{H.c.}.\label{Hom}
\end{equation}
It can be seen from Eq.~(\ref{Hom}) that the function of these phases in the optomechanical interactions is determined by the term $\sum_{\nu =1}^{j-1}\theta_{\nu}$. Hence, we can apply a single phase to realize the dark-mode-breaking task. For simplicity, we assume $\theta_{j}=0$ for $j=2$-$(N-1)$ in the following discussions.

As a special case, we first analyze the case of $N=2$. In this case, the multiple-mechanical-resonator optomechanical system is reduced to the two-mechanical-resonator optomechanical system, which has been analyzed before. When $N=2$, the optomechanical interaction reads
\begin{eqnarray}
H_{\text{om}}&=&\frac{\sqrt{2}G}{2}(1+e^{i\theta_{1}})aB_{1}^{\dagger}+\frac{\sqrt{2}G}{2}(1-e^{i\theta_{1}})aB_{2}^{\dagger}
+\mathrm{H.c.}.
\end{eqnarray}
It is obvious that when $\theta=n\pi$ for an integer $n$, the cavity field is decoupled from one of the two hybrid mechanical modes: either $B_{1}$ or $B_{2}$. This hybrid mechanical mode decoupled from the cavity mode is the dark mode. However, in a general case $\theta\neq n\pi$, the dark-mode effect is broken, and then the ground-state cooling becomes accessible under proper parameter conditions.

For the case of $N\geq3$, the effective coupling coefficient between the cavity-field mode $a$ and the $k$th normal mode $B_{k}$ in Eq.~(\ref{Hom}) can be expressed as
\begin{eqnarray}
&&\frac{G}{A}\left[\sin \left(\frac{k\pi}{N+1}\right)+\sum_{j=2}^{N}e^{i\sum_{\nu =1}^{j-1}\theta _{\nu }}\sin \left(\frac{jk\pi }{N+1}\right)\right] \nonumber\\
&=&\frac{G}{A}\bigg[\sin \left(\frac{k\pi }{N+1}\right)+e^{i\theta _{1}}\sin\left(\frac{2k\pi}{N+1}\right)+e^{i\left( \theta _{1}+\theta _{2}\right) }\sin \left(\frac{3k\pi}{N+1}\right)+\cdots \nonumber\\
&&+e^{i\sum_{\nu =1}^{N-3}\theta _{\nu }}\sin \left(\frac{N-2}{N+1}k\pi\right)+e^{i\sum_{\nu =1}^{N-2}\theta _{\nu }}\sin \left(\frac{N-1}{N+1}k\pi\right)+e^{i\sum_{\nu =1}^{N-1}\theta _{\nu }}\sin \left(\frac{Nk\pi}{N+1}\right)\bigg] \nonumber\\
&=&\frac{G}{A}\bigg\{\Big[\sin \left(\frac{1}{N+1}k\pi \right)+e^{i\sum_{\nu=1}^{N-1}\theta _{\nu }}\sin \left(\frac{N}{N+1}k\pi\right)\Big]+\Big[e^{i\theta_{1}}\sin \left(\frac{2}{N+1}k\pi\right) +e^{i\sum_{\nu =1}^{N-2}\theta _{\nu }}\sin \left(\frac{N-1}{N+1}k\pi\right)\Big]\nonumber\\
&&+\Big[e^{i(\theta_{1}+\theta _{2})}\sin \left(\frac{3}{N+1}k\pi\right)+e^{i\sum_{\nu =1}^{N-3}\theta_{\nu }}\sin \left(\frac{N-2}{N+1}k\pi\right)\Big]+\cdots\bigg\}.
\end{eqnarray}
Below, we consider two cases corresponding to odd and even numbers $N$, respectively.

(i) For an odd number $N$ and $\theta_{j}=0$ (for $j=2$-$(N-1)$), the coefficient becomes
\begin{eqnarray}
&&\frac{G}{A}\left[\sin \left(\frac{k\pi }{N+1}\right)+\sum_{j=2}^{N}e^{i\sum_{\nu =1}^{j-1}\theta _{\nu }}\sin \left(\frac{jk\pi}{N+1}\right)\right] \nonumber\\
&=&\frac{G}{A}\bigg\{\left[\sin \left(\frac{k\pi}{N+1}\right)+e^{i\theta _{1}}\sin\left(\frac{Nk\pi}{N+1}\right)\right]+e^{i\theta _{1}}\left[\sin \left(\frac{2k\pi}{N+1}\right)+\sin \left(\frac{N-1}{N+1}k\pi\right) \right] \nonumber\\
&&+e^{i\theta_{1}}\left[\sin \left(\frac{3k\pi}{N+1}\right)+\sin \left(\frac{N-2}{N+1}k\pi\right)\right]+\cdots+e^{i\theta _{1}}\sin \left(\frac{k\pi }{2}\right)\bigg\}.\label{odd}
\end{eqnarray}
On one hand, if $k$ is an odd number, we have
\begin{eqnarray}
&&\frac{G}{A}\left[\sin \left(\frac{k\pi }{N+1}\right)+\sum_{j=2}^{N}e^{i\sum_{\nu =1}^{j-1}\theta_{\nu }}\sin \left(\frac{jk\pi }{N+1}\right)\right] \nonumber\\
&=&\frac{G}{A}\left[(1+e^{i\theta _{1}}) \sin \left(\frac{k\pi }{N+1}\right)+2e^{i\theta _{1}}\sin \left(\frac{2k\pi }{N+1}\right)+2e^{i\theta _{1}}\sin \left(\frac{3k\pi}{N+1}\right)+\cdots+e^{i\theta _{1}}\sin\left (\frac{k\pi}{2}\right)\right];\label{odd1}
\end{eqnarray}
On the other hand, if $k$ is an even number, we have
\begin{eqnarray}
\frac{G}{A}\left[\sin \left(\frac{k\pi}{N+1}\right)+\sum_{j=2}^{N}e^{i\sum_{\nu =1}^{j-1}\theta_{\nu }}\sin \left(\frac{jk\pi }{N+1}\right)\right]=\frac{G}{A}\left( 1-e^{i\theta _{1}}\right)\sin\left(\frac{k\pi}{N+1}\right).\label{odd2}
\end{eqnarray}

(ii) For an even number $N$ and $\theta_{j}=0$ (for $j=2$-$(N-1)$), the coefficient can be simplified as
\begin{eqnarray}
&&\frac{G}{A}\left[\sin \left(\frac{k\pi}{N+1}\right)+\sum_{j=2}^{N}e^{i\sum_{\nu =1}^{j-1}\theta _{\nu }}\sin \left(\frac{jk\pi }{N+1}\right)\right] \nonumber\\
&=&\frac{G}{A}\bigg\{\left[\sin \left(\frac{k\pi }{N+1}\right)+e^{i\theta_{1}}\sin\left(\frac{N}{N+1}k\pi\right)\right]+e^{i\theta _{1}}\left[\sin \left(\frac{2k\pi }{N+1}\right)+\sin \left(\frac{N-1}{N+1}k\pi\right)\right]\nonumber\\
&&+e^{i\theta_{1}}\left[\sin \left(\frac{3k\pi }{N+1}\right)+\sin \left(\frac{N-2}{N+1}k\pi\right)\right]+\cdots\bigg\}.\label{even}
\end{eqnarray}
In this case, when $k$ is an odd number, we have
\begin{eqnarray}
&&\frac{G}{A}\left[\sin \left(\frac{k\pi }{N+1}\right)+\sum_{j=2}^{N}e^{i\sum_{\nu =1}^{j-1}\theta _{\nu }}\sin \left(\frac{jk\pi }{N+1}\right)\right] \nonumber\\
&=&\frac{G}{A}\left[(1+e^{i\theta _{1}})\sin \left(\frac{k\pi }{N+1}\right)+2e^{i\theta _{1}}\sin \left(\frac{2k\pi }{N+1}\right)+2e^{i\theta_{1}}\sin \left(\frac{3k\pi}{N+1}\right)+\cdots\right];\label{even1}
\end{eqnarray}
In addition, when $k$ is an even number, we have
\begin{eqnarray}
\frac{G}{A}\left[\sin\left(\frac{k\pi}{N+1}\right)+\sum_{j=2}^{N}e^{i\sum_{\nu =1}^{j-1}\theta _{\nu }}\sin \left(\frac{jk\pi}{N+1}\right)\right]
=\frac{G}{A}\left(1-e^{i\theta_{1}}\right)\sin \left(\frac{k\pi}{N+1}\right).\label{even2}
\end{eqnarray}

According to Eqs.~(\ref{odd}-\ref{even2}), we can see that for odd numbers $k$, the coupling strength between the cavity-field mode and the $k$th normal mode $B_{k}$ is nonzero.
However, for even numbers $k$, the coupling strength between the cavity-field mode and the $k$th normal mode $B_{k}$
can be expressed as
\begin{equation}
H_{\text{ck}}=\frac{G}{A}\left[\left(1-e^{i\theta_{1}}\right) \sin \left(\frac{k\pi}{N+1}\right)\right]aB_{k}^{\dagger }+\mathrm{H.c.},\hspace{1 cm}k=\text{even number}.\label{darkmodes}
\end{equation}
Obviously, when $\theta_{1}=2n\pi$, the coupling strength between the $k$th mechanical normal mode ($B_{k=\text{even}}$) and the cavity mode ($a$) is equal to zero. In this case, all the even normal modes are decoupled from the cavity field. Then ground-state cooling cannot be realized in this system due to the dark-mode effect. Nevertheless, we can cool these mechanical resonators by choosing proper parameters to break the dark-mode effect ($\theta_{1}\neq2n\pi$).

\section{Ground-state cooling of the multiple mechanical resonators}

In this section, we study the simultaneous cooling of multiple mechanical resonators in the $N$-mechanical-resonator optomechanical system. To evaluate the cooling performance of the multiple mechanical resonators, we calculate the final average phonon numbers in these mechanical resonators. To this end, we re-express the linearized quantum Langevin equations~(\ref{NlineLangevineq}) as
\begin{eqnarray}
\mathbf{\dot{u}}(t)=\mathbf{Au}(t)+\mathbf{N}(t),\label{MatrixLangevin}
\end{eqnarray}
where we introduce the vectors of the system operators
\begin{eqnarray}
\mathbf{u}(t)=[\delta a(t),\delta b_{1}(t), \delta b_{2}(t),\cdots,\delta b_{N}(t),\delta a^{\dagger}(t),\delta b^{\dagger}_{1}(t), \delta b^{\dagger}_{2}(t),\cdots,\delta b^{\dagger}_{N}(t)]^{T},\label{MatrixuNmode}
\end{eqnarray}
the vector of the noise operators
\begin{eqnarray}
\mathbf{N}(t)=&\sqrt{2}[\sqrt{\kappa}a_{\text{in}}(t),\sqrt{\gamma_{1}}b_{1,\text{in}}(t), \sqrt{\gamma_{2}}b_{2,\text{in}}(t) ,\cdots, \sqrt{\gamma_{N}}b_{N,\text{in}}(t) ,\sqrt{\kappa}a^{\dagger}_{\text{in}}(t),\sqrt{\gamma_{1}}b^{\dagger}_{1,\text{in}}(t), \sqrt{\gamma_{2}}b^{\dagger}_{2,\text{in}}(t),\cdots, \sqrt{\gamma_{N}}b^{\dagger}_{N,\text{in}}(t)]^{T},\label{MatrixNmode}\nonumber\\
\end{eqnarray}
and the coefficient matrix
\small
\begin{equation}
\mathbf{A}=\left(
\begin{array}{cccccccccc}
-(\kappa +i\Delta ) & -iG_{1} & -iG_{2} & \cdots  & -iG_{N} & 0 & -iG_{1} &
-iG_{2} & \cdots  & -iG_{N} \\
-iG_{1}^{\ast } & -(\gamma _{1}+i\omega _{1}) & -i\eta _{1}e^{i\theta _{1}}
& \cdots  & -i\eta _{N-1}e^{-i\theta _{N-1}} & -iG_{1} & 0 & 0 & \cdots  & 0 \\
-iG_{2}^{\ast } & -i\eta _{1}e^{-i\theta _{1}} & -(\gamma _{2}+i\omega _{2})
& \cdots  & 0 & -iG_{2} & 0 & 0 & \cdots  & 0 \\
\vdots  & \vdots  & \vdots  & \ddots  & \vdots  & \vdots  & \vdots  & \vdots
& \ddots  & \vdots  \\
-iG_{N}^{\ast } & -i\eta _{N-1}e^{i\theta _{N-1}} & 0 & \cdots  & -(\gamma
_{N}+i\omega _{N}) & -iG_{4} & 0 & 0 & \cdots  & 0 \\
0 & iG_{1}^{\ast } & iG_{2}^{\ast } & \cdots  & iG_{N}^{\ast } & -(\kappa
-i\Delta ) & iG_{1}^{\ast } & iG_{2}^{\ast } & \cdots  & iG_{N}^{\ast } \\
iG_{1}^{\ast } & 0 & 0 & \cdots  & 0 & iG_{1} & -(\gamma _{1}-i\omega _{1})
& i\eta _{1}e^{-i\theta _{1}} & \cdots  & i\eta _{N-1}e^{i\theta _{N-1}} \\
iG_{2}^{\ast } & 0 & 0 & \cdots  & 0 & iG_{2} & i\eta _{1}e^{i\theta _{1}} &
-(\gamma _{2}-i\omega _{2}) & \cdots  & 0 \\
\vdots  & \vdots  & \vdots  & \vdots  & \vdots  & \vdots  & \vdots  & \vdots
& \ddots  & \vdots  \\
iG_{N}^{\ast } & 0 & 0 & \cdots  & 0 & iG_{N} & i\eta _{N-1}e^{-i\theta _{N-1}}
& 0 & \cdots  & -(\gamma _{N}-i\omega _{N})
\end{array}
\right).
\end{equation}
\normalsize
The formal solution of the linearized quantum Langevin equations Eq.~(\ref{MatrixLangevin}) can be obtained as
\begin{equation}
\mathbf{u}(t) =\mathbf{M}(t) \mathbf{u}(0)+\int_{0}^{t}\mathbf{M}(t-s) \mathbf{N}(s)ds,
\end{equation}
where the matrix $\mathbf{M}(t)$ is given by $\mathbf{M}(t)=\text{exp}(\mathbf{A}t)$, and hence the stability conditions derived from the Routh-Hurwitz criterion have satisfied. Note that in our simulations the real part of the eigenvalues of the coefficient matrix $\mathbf{A}$ is negative.
%%%%%%%%%%%%%%%%%%%%%%%%%%%%%%
\begin{figure}[tbp]
\centering
\includegraphics[bb=0 0 568 233, width=0.7 \textwidth]{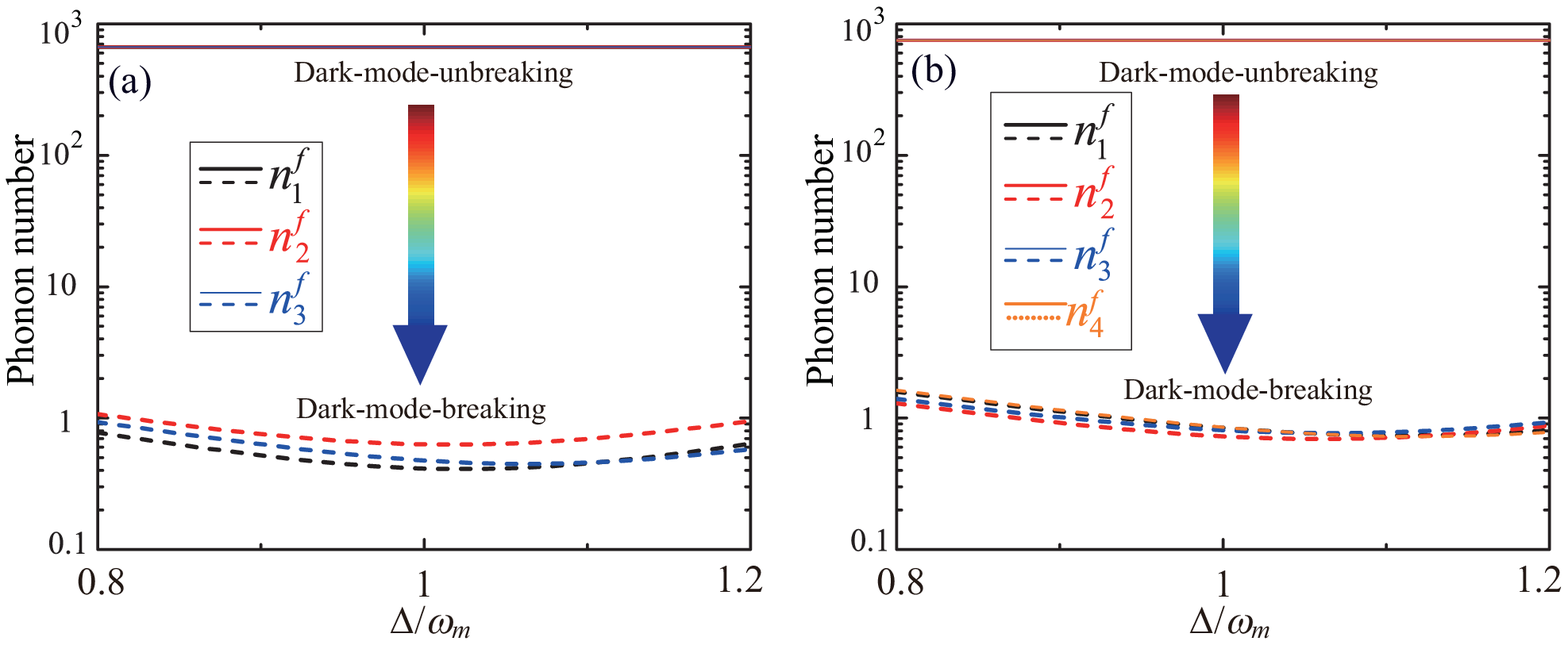}
\caption{(Color online) The final average phonon numbers $n^{f}_{j}$ in these mechanical resonators as functions of the effective driving detuning $\Delta$ in the dark-mode-unbreaking case ($\eta_{j}=\eta=0$) and the dark-mode-breaking case ($\eta_{j}/\omega_{m}=\eta/\omega_{m}=0.1$, $\theta_{1}=\pi/2$, and $\theta_{j\neq1}=0$) for $N=3$ and $N=4$. Here we take $G_{j}/\omega_{m}=G/\omega_{m}=0.1$, $\kappa/\omega_{m}=0.2$, $\gamma_{j}/\omega_{m}=10^{-5}$, and $\bar{n}_{j}=10^{3}$.}\label{Deta}
\end{figure}
%%%%%%%%%%%%%%%%%%%%%%%%%%%%%%

For studying the quantum cooling of these mechanical resonators, we calculate the steady-state average phonon numbers in these mechanical resonators. This can be realized by calculating the steady-state values of the covariance matrix $\mathbf{V}$, which is defined by the matrix elements
\begin{equation}
\mathbf{V}_{ij}=\frac{1}{2}[\langle\mathbf{u}_{i}(\infty)\mathbf{u}_{j}(\infty)\rangle+\langle\mathbf{u}_{j}(\infty)\mathbf{u}_{i}(\infty)\rangle].
\end{equation}
In the linearized optomechanical system, the covariance matrix $\mathbf{V}$ satisfies the Lyapunov equation
\begin{equation}
\mathbf{A}\mathbf{V}+\mathbf{V}\mathbf{A}^{T}=-\mathbf{Q},
\end{equation}
where
\begin{equation}
\mathbf{Q}=\frac{1}{2}(\mathbf{C}+\mathbf{C}^{T}).
\end{equation}
Here $\mathbf{C}$ is the noise correlation matrix which is defined by the elements
\begin{eqnarray}
\langle \mathbf{N}_{k}(s) \mathbf{N}_{l}(s^{\prime})\rangle =\mathbf{C}_{k,l}\delta (s-s^{\prime }).
\end{eqnarray}
For the Markovian baths as considered in this work, we have $\mathbf{C}(s,s') =\mathbf{C}\delta(s-s')$, where the constant matrix $\mathbf{C}$ is given by
\begin{equation}
\mathbf{C=}\left(
\begin{array}{cccccccccc}
0 & 0 & 0 & \cdots  & 0 & 2\kappa  & 0 & 0 & \cdots  & 0 \\
0 & 0 & 0 & \cdots  & 0 & 0 & 2\gamma _{1}\left( \bar{n}_{1}+1\right)  & 0 &
\cdots  & 0 \\
0 & 0 & 0 & \cdots  & 0 & 0 & 0 & 2\gamma _{2}\left( \bar{n}_{2}+1\right)  &
\cdots  & 0 \\
\vdots  & \vdots  & \vdots  & \vdots  & \vdots  & \vdots  & \vdots  & \vdots
& \ddots  & 0 \\
0 & 0 & 0 & \cdots  & 0 & 0 & 0 & 0 & \cdots  & 2\gamma _{N}\left( \bar{n}%
_{N}+1\right)  \\
0 & 0 & 0 & \cdots  & 0 & 0 & 0 & 0 & \cdots  & 0 \\
0 & 2\gamma _{1}\bar{n}_{1} & 0 & \cdots  & 0 & 0 & 0 & 0 & \cdots  & 0 \\
0 & 0 & 2\gamma _{2}\bar{n}_{2} & \cdots  & 0 & 0 & 0 & 0 & \cdots  & 0 \\
\vdots  & \vdots  & \vdots  & \ddots  & \vdots  & \vdots  & \vdots  & \vdots
& \ddots  & \vdots  \\
0 & 0 & 0 & \cdots  & 2\gamma _{N}\bar{n}_{N} & 0 & 0 & 0 & \cdots  & 0
\end{array}
\right).
\end{equation}
%%%%%%%%%%%%%%%%%%%%%%%%%%%%%%
\begin{figure}[tbp]
\centering
\includegraphics[bb=0 0 359 145, width=0.7\textwidth]{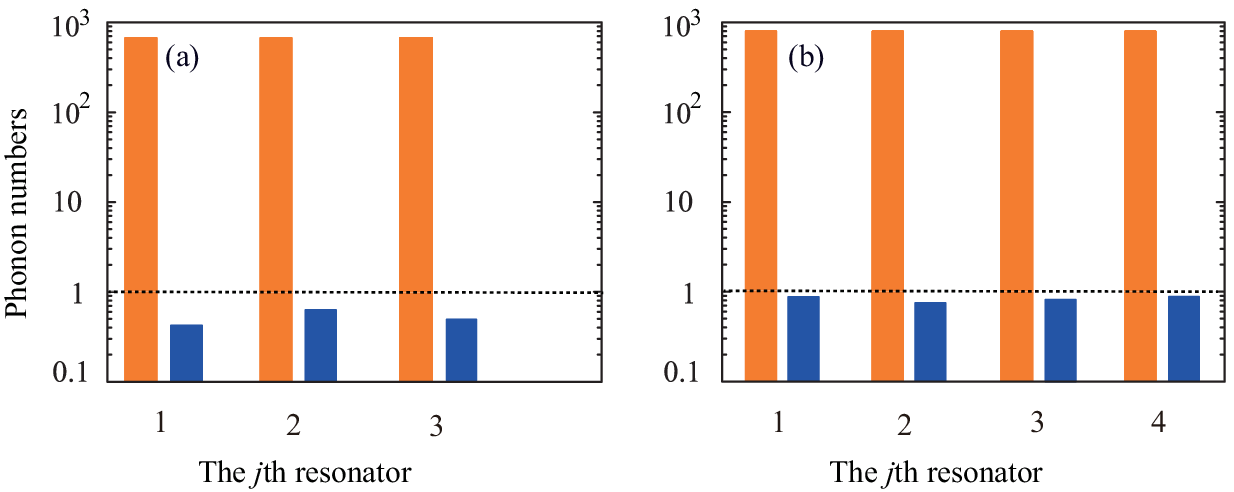}
\caption{(Color online) The final average phonon numbers $n^{f}_{j}$ in these mechanical resonators are plotted in the dark-mode-unbreaking [$\eta_{j}=0$ (orange bars)] and -breaking [$\eta_{j}=0.1\omega_{m}$ and $\theta_{1}=\pi/2$ (blue bars)] cases for (a) $N=3$ and (b) $N=4$. Here $\Delta=\omega_{m}$, and other used parameters are the same as those given in Fig.~\ref{Deta}.}\label{FigS11}
\end{figure}
%%%%%%%%%%%%%%%%%%%%%%%%%%%%%%
Based on the covariance matrix $\mathbf{V}$, the final average phonon number in the $j$th mechanical resonator can be obtained as
\begin{eqnarray}
\label{finalexactNmr}
\langle \delta b_{j}^{\dagger}\delta b_{j}\rangle=\mathbf{V}_{N+j+2,j+1}-\frac{1}{2},
\end{eqnarray}
where $\mathbf{V}_{N+j+2,j+1}$ can be obtained by solving the Lyapunov equation.
%%%%%%%%%%%%%%%%%%%%%%%%%%%%%%
\begin{figure}[bp]
\centering
\includegraphics[bb=0 0 569 245, width=0.7\textwidth]{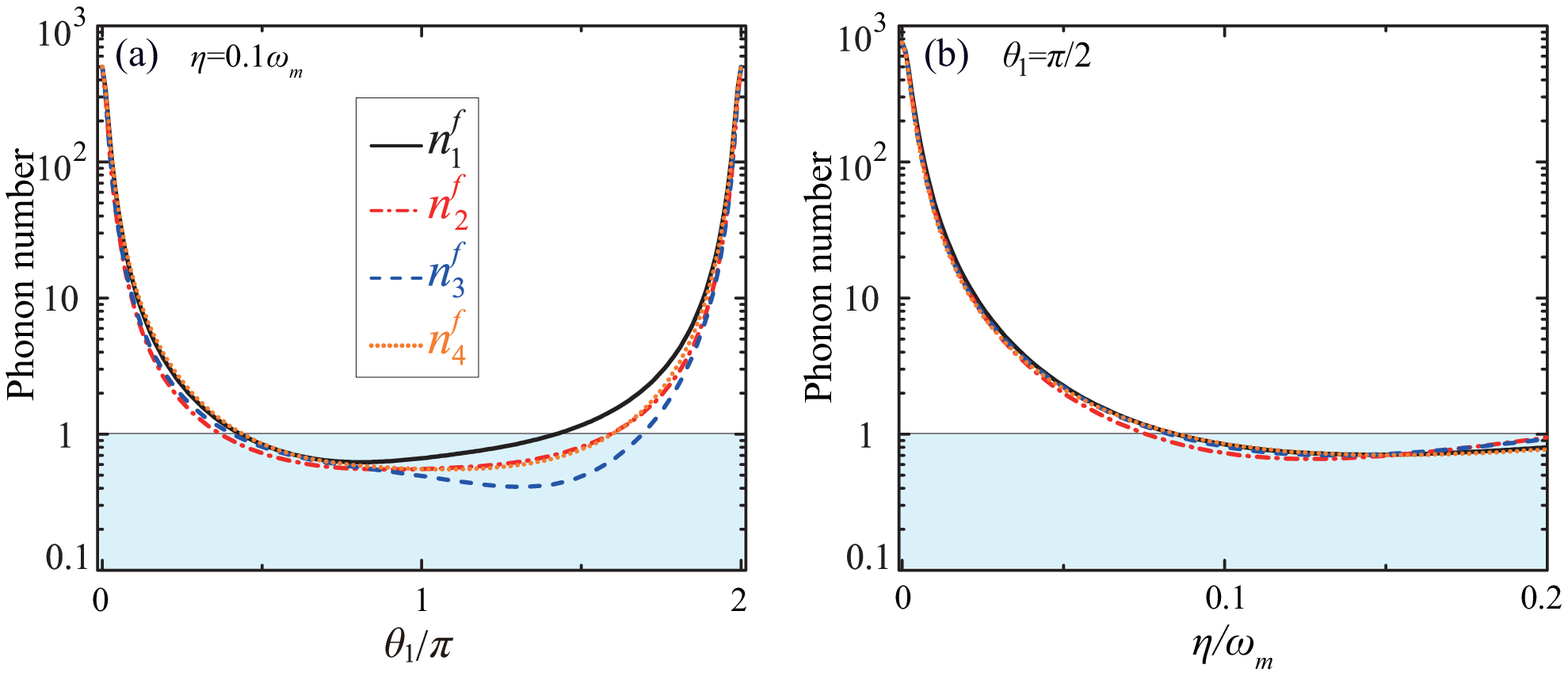}
\caption{(Color online) The final average phonon numbers $n^{f}_{j}$ in the mechanical resonators as functions of (a) the phase $\theta_{1}$ when $\eta/\omega_{m}=0.1$ and (b) the phonon-exchange coupling $\eta$ when $\theta_{1}=\pi/2$ for $N=4$. Here $G_{j}/\omega_{m}=G/\omega_{m}=0.1$. Other used parameters are the same as those given in Fig.~\ref{Deta}.}\label{thetaeta}
\end{figure}
%%%%%%%%%%%%%%%%%%%%%%%%%%%%%%

Below we simulate the cooling performance of the mechanical resonators for the cases of $N=3$ and $4$. For convenience, we assume that all the mechanical resonators have the same resonance frequencies ($\omega_{j}=\omega_{m}$ for $j=1$-$N$), optomechanical coupling strengths ($G_{j}=G$ for $j=1$-$N$), and phonon-exchange coupling strengths [$\eta_{j}=\eta$ for $j=1$-$(N-1)$]. Moreover, we consider the case of $\theta_{1}=\pi/2$ and $\theta_{j>2}=0$. In Fig.~\ref{Deta}, we plot the final average phonon numbers $n^{f}_{j}$ in these mechanical resonators as functions of the scaled driving detuning $\Delta/\omega_{m}$ in both the dark-mode-breaking ($\eta_{j}=0.1\omega_{m}$ and $\theta_{1}=\pi/2$) and -unbreaking ($\eta_{j}=\eta=0$) cases. The results show that the ground-state cooling is unfeasible for these mechanical resonators when the phonon-exchange interactions are absent ($\eta_{j}=\eta=0$) [the upper curves in Figs.~\ref{Deta}(a) and~\ref{Deta}(b)]. This is because the phonon excitation energy stored in the dark modes cannot be extracted through the optomechanical cooling channel. When the couplings among these mechanical resonators are introduced, the dark modes are broken and then the ground-state cooling can be realized, as shown in Figs.~\ref{Deta}(a) and~\ref{Deta}(b). In particular, the optimal driving detuning is located at $\Delta\approx\omega_{m}$, in consistent with the resolved-sideband cooling case.

To see the cooling performance more clearly, we compare the cooling results of these mechanical resonators in the presence of mechanical couplings with the results corresponding to the absence of the mechanical couplings. In Fig.~\ref{FigS11}, we plot the final average phonon numbers of these mechanical resonators in the two cases. Here we can see that final average phonon numbers could be smaller than $1$ when the mechanical couplings are introduced into the system, which means that the simultaneous ground-state cooling of these mechanical resonators can be achieved by breaking the dark-mode effect.

We also investigate the dependence of the cooling performance on the mechanical coupling parameters $\eta$ and $\theta$. In Fig.~\ref{thetaeta}, we plot the final average phonon numbers $n^{f}_{j}$ in these mechanical resonators as functions of the phase $\theta$ and the scaled phonon-exchange coupling strength $\eta/\omega_{m}$. The results show that ground-state cooling can be realized for proper values of the phase $\theta_{1}\neq2n\pi$ when $\eta=0.1\omega_{m}$ [Fig.~\ref{thetaeta}(a)]. In addition, the cooling efficiency of the multiple mechanical resonators can be controlled by tuning the phonon-exchange interaction strength $\eta$ when the phase is fixed at $\theta_{1}=\pi/2$ [Fig.~\ref{thetaeta}(b)]. Under these parameters, the dark-mode effect is broken and then the thermal occupations can be extracted through the optomechanical-cooling channels.

\section{Discussions on the justification of performing the RWA \label{RWA}}

%%%%%%%%%%%%%%%%%%%%%%%%%%%%%%
\begin{figure}[tbp]
\centering
\includegraphics[bb=0 0 538 209, width=0.7 \textwidth]{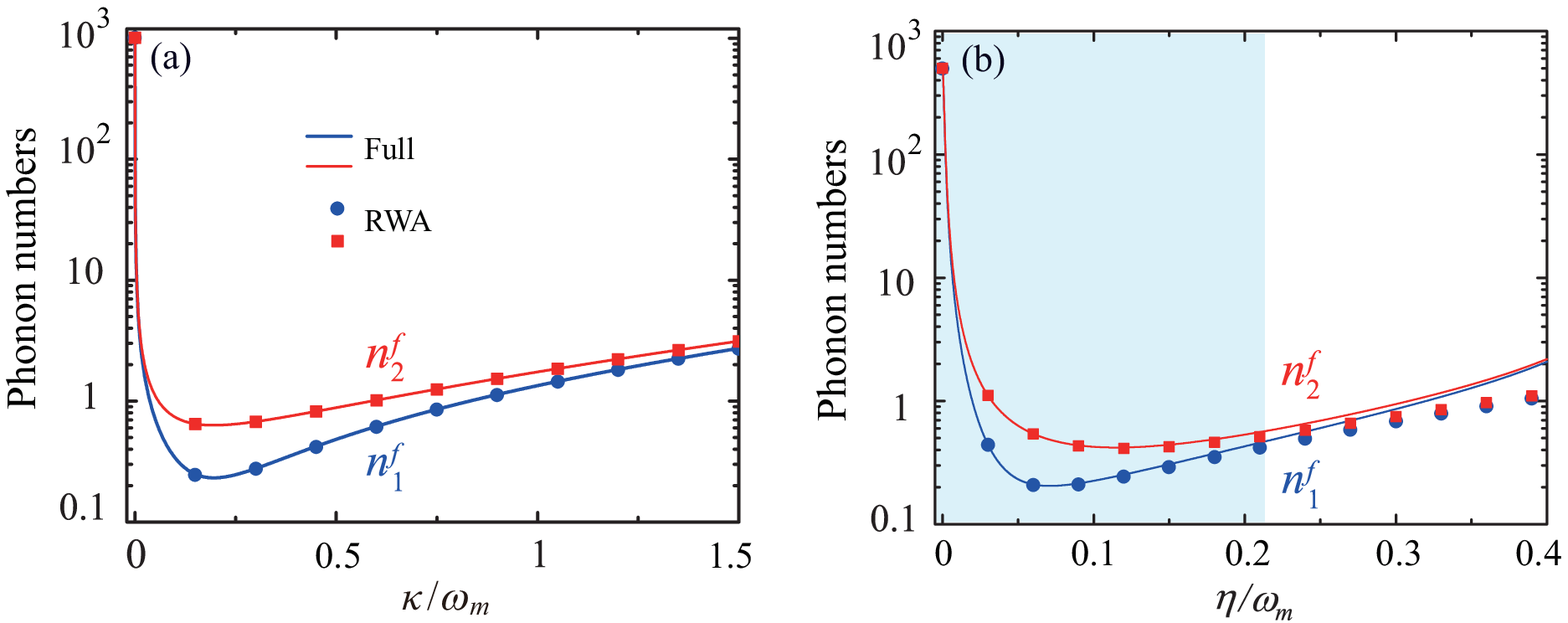}
\caption{(Color online) The final average phonon numbers $n^{f}_{1}$ and $n^{f}_{2}$ versus (a) the cavity-field decay rate $\kappa$ when $\eta=0.05\omega_{m}$ and (b) the phonon-phonon coupling strength $\eta$ when $\kappa=0.2\omega_{m}$. Here the symbols and the solid curves correspond to the Hamiltonian under the RWA and the full Hamiltonian, respectively. Other parameters used are given by $\omega_{2}=\omega_{1}=\omega_{m}$, $\Delta=\omega_{m}$, $\theta=\pi/2$, $G_{1}=G_{2}=0.1\omega_{m}$, $\gamma_{1}=\gamma_{2}=10^{-5}\omega_{m}$, and $\bar{n}_{1}=\bar{n}_{2}=10^{3}$.}\label{FigS13}
\end{figure}
%%%%%%%%%%%%%%%%%%%%%%%%%%%%%%
In our model, we consider an excitation-number-conservation-type phonon-phonon interaction $\eta(e^{i\theta}b_{1}^{\dagger}b_{2}+e^{-i\theta}b_{2}^{\dagger}b_{1})$, which is obtained by making the rotating-wave approximation (RWA) in the full phonon-exchange interaction Hamiltonian $\eta(e^{i\theta}b_{1}^{\dagger}+e^{-i\theta}b_{1})(b_{2}^{\dagger}+b_{2})$. To evaluate the validity of the RWA, we compare the results obtained based on the approximate Hamiltonian and the full Hamiltonian including the counterrotating term. In Figs.~\ref{FigS13}(a) and~\ref{FigS13}(b), we show the final average phonon numbers $n_{1}^{f}$ and $n_{2}^{f}$ as functions of the cavity-field decay rate $\kappa$ and the mechanical coupling strength $\eta$. Here, the symbols and the solid curves correspond to the Hamiltonian under the RWA and the full Hamiltonian, respectively. Figure~\ref{FigS13}(a) shows an excellent agreement between the results obtained with the approximate Hamiltonian and the full Hamiltonian in both the resolved- and unresolved-sideband regimes. We can also see from Fig.~\ref{FigS13}(b) that the approximate results match well with the exact results when $\eta<0.2\omega_{m}$. Physically, the optomechanical cooling and heating are governed by the rotating-wave and the CR terms, respectively. In the weak-coupling regime ($\eta\ll\omega_{m}$) and under the near-resonance condition ($\omega_{2}$ around $\omega_{1}$), the CR term in the phonon-phonon interaction can be safely omitted by applying the RWA. The difference between these two treatments becomes non-negligible when $\eta>0.2\omega_{m}$. The reason is that the CR term, which simultaneously creates phonon excitations in the two mechanical resonators, becomes important for a large phonon-phonon coupling strength $\eta$. These features indicate that the RWA performed in the phonon-phonon interaction is justified in our simulations, and that the CR interaction can be omitted safely under the condition $\eta\ll\omega_{m}$.

\section{Simultaneous cooling of the mechanical supermodes}

In this section, we discuss the simultaneous cooling of the mechanical supermodes in cavity optomechanical systems. Note that the notations used in this section are independent of the those used in other sections. We consider the case where the two mechanical resonators are coupled to each other by a phonon-hopping coupling. Then, two mechanical supermodes are formed and the cavity field is coupled to the two supermodes. In the presence of the phonon-hopping coupling between the two mechanical resonators, the Hamiltonian of this coupled mechanical system reads ($\hbar=1$)
\begin{equation}
H_{\text{c}}=\omega_{m}c_{1}^{\dagger }c_{1}+\omega_{m}c_{2}^{\dagger}c_{2}+\lambda (c_{1}^{\dagger}c_{2}+c_{2}^{\dagger}c_{1}),\label{coupledM}
\end{equation}
where the operators $c_{l=1,2}$ ($c^{\dagger}_{l}$) are the annihilation (creation) operators of the $l$th mechanical resonator, with the corresponding resonance frequencies $\omega_{m}$, and the parameter $\lambda$ is a coupling constant of the mechanical interaction between the two mechanical resonators. In the weak-coupling regime ($\lambda\ll\omega_{m}$), the counter-rotating term in the phonon-phonon interaction can be safely omitted by making the rotating-wave approximation. Below, we diagonalize this coupled mechanical system by introducing two mechanical supermodes $C_{\pm}$, given by
\begin{subequations}
\begin{align}
C_{+}=&\frac{1}{\sqrt{2}}(c_{1}+c_{2}),  \\
C_{-}=&\frac{1}{\sqrt{2}}(-c_{1}+c_{2}),
\end{align}
\end{subequations}
where these new operators satisfy the bosonic commutation relations $[C_{+},C^{\dagger}_{+}]=1$ and $[C_{-},C^{\dagger}_{-}]=1$. Thus, Hamiltonian ~(\ref{coupledM}) becomes
\begin{eqnarray}
H_{\text{c}}&=&\omega_{C,+}C_{+}^{\dagger }C_{+}+\omega_{C,-}C_{-}^{\dagger }C_{-},
\end{eqnarray}
where we introduce the resonance frequencies of these supermodes as
\begin{equation}
\omega_{C,\pm }=\omega_{m}\pm\lambda.
\end{equation}

To cool the two mechanical supermodes, we couple the two mechanical supermodes to a common optical cavity-field mode by the optomechanical interactions ~\cite{Ramos2014APL1}. In the strong-driving regime, the linearized optomechanical Hamiltonian in the RWA takes the form as
\begin{eqnarray}
H_{\text{RWA}} &=&\Delta \delta a^{\dagger}\delta a+\omega_{C,+}\delta C_{+}^{\dagger}\delta C_{+}+\omega_{C,-}\delta C_{-}^{\dagger }\delta C_{-}+G_{+}(\delta a\delta C_{+}^{\dagger}+\delta a^{\dagger}\delta C_{+}) +G_{-}(\delta a\delta C_{-}^{\dagger}+\delta a^{\dagger}\delta C_{-}),\label{HRWA1}
\end{eqnarray}
where $\Delta$ is the normalized driving detuning of the cavity field, and the parameters $G_{\pm}$ are the optomechanical couplings between the cavity-field mode and the two mechanical supermodes. It can be seen from Eq.~(\ref{HRWA1}) that the couplings between the cavity field and the two supermodes are the same as the three-mode optomechanical model considered in the main text. Therefore, all the analyses in the three-mode optomechanical system are suitable to the coupled cavity-supermode case. Based on the fact that the mechanical coupling between the two mechanical resonators is much smaller than the resonant frequencies of the two resonators ($\lambda\ll\omega_{m}$), the frequencies $\omega_{C,\pm}$ of the two mechanical supermodes are close to each other. We proceed to analyze the cooling performance of the two mechanical supermodes. Concretely, we consider two special cases.

(i) When the frequency difference between the two mechanical supermodes is larger than the effective mechanical linewidth ($\Delta\omega=|\omega_{C,+}-\omega_{C,-}|>\Gamma_{l=+,-}$), the simultaneous ground-state cooling of the two mechanical supermodes is accessible under proper parameter conditions. Physically, when the two mechanical supermodes are well separated in frequency, there is no dark mode, then the ground-state cooling can be realized when this system works in the resolved-sideband regime and under a proper driving (red-sideband resonance). This cooling situation is similar to the case shown in Fig.~2(b) and Fig.~S1(e) [see blank area].

(ii) When the frequency difference between the two mechanical supermodes is smaller than the effective mechanical linewidth ($\Delta\omega=|\omega_{C,+}-\omega_{C,-}|\leq\Gamma_{l=+,-}$), the cooling of the two mechanical supermodes is suppressed. This is because, though the dark mode exists theoretically only in the degenerate-resonator case, the dark-mode effect actually works for a wider detuning range in the near-degenerate-resonator case. The suppression region of the ground-state for the mechanical supermodes is characterized by the effective mechanical linewidth. The cooling of the individual mechanical supermodes is suppressed in this region, i.e., the individual mechanical supermodes have significant spectral overlap and become effectively degenerate. This cooling situation is similar to the case shown in Fig.~2(b) and Fig.~S1(e) [see shadow area].

In the case (ii), for achieving quantum ground-state cooling of the two mechanical supermodes, we need to introduce a phase-dependent phonon-hopping coupling between the two mechanical supermodes. Thus, the linearized optomechanical Hamiltonian including a phase-dependent coupling between the two supermodes takes the following form
\begin{eqnarray}
H_{\text{RWA}} &=&\Delta \delta a^{\dagger}\delta a+\omega_{C,+}\delta C_{+}^{\dagger}\delta C_{+}+\omega _{C,-}\delta C_{-}^{\dagger}\delta C_{-}+G_{+}(\delta a\delta C_{+}^{\dagger }+\delta a^{\dagger}\delta C_{+})+G_{-}(\delta a\delta C_{-}^{\dagger}+\delta a^{\dagger}\delta C_{-}) \nonumber\\
&&+\tilde{\lambda} (e^{i\phi}\delta C_{+}^{\dagger}\delta C_{-}+e^{-i\phi }\delta C_{-}^{\dagger}\delta C_{+}).\label{HRWA}
\end{eqnarray}
By introducing two new bosonic modes $\tilde{C}_{+}$ and $\tilde{C}_{-}$ defined by
\begin{subequations}
\begin{align}
\tilde{C}_{+}=&\tilde{f^{\prime}}C_{+}-e^{i\phi}\tilde{h^{\prime}}C_{-}, \\
\tilde{C}_{-}=&e^{-i\phi }\tilde{h^{\prime}}C_{+}+\tilde{f^{\prime}}C_{-},
\end{align}
\end{subequations}
Hamiltonian ~(\ref{HRWA}) becomes
\begin{equation}
H_{\text{RWA}}=\Delta \delta a^{\dagger}\delta a+\tilde{\omega}_{C,+}\tilde{C}_{+}^{\dagger}\tilde{C}_{+}+\tilde{\omega}_{C,-}\tilde{C}_{-}^{\dagger }\tilde{C}_{-}+\tilde{G}_{+}(\delta a\tilde{C}_{+}^{\dagger }+\delta a^{\dagger}\tilde{C}_{+}) +\tilde{G}_{-}(\delta a\tilde{C}_{-}^{\dagger }+\delta a^{\dagger }\tilde{C}_{-}),
\end{equation}
where we introduce the resonance frequencies $\tilde{\omega}_{C,\pm}$, the coupling strengths $\tilde{G}_{\pm}$, and  the coefficients $\tilde{f^{\prime}}$ and $\tilde{h^{\prime}}$ as
\begin{subequations}
\begin{align}
\tilde{\omega}_{C,+}=&\frac{1}{2}\left( \omega _{C,+}+\omega _{C,-}+\sqrt{\left(\omega_{C,+}-\omega _{C,-}\right)^{2}+4\tilde{\lambda} ^{2}}\right), \\
\tilde{\omega}_{C,-}=&\frac{1}{2}\left( \omega _{C,+}+\omega _{C,-}-\sqrt{\left(\omega_{C,+}-\omega _{C,-}\right)^{2}+4\tilde{\lambda} ^{2}}\right),\\
\tilde{G}_{+}=&(\tilde{f^{\prime}}G_{+}-e^{-i\phi}\tilde{h^{\prime}}G_{-}), \\
\tilde{G}_{-}=&(e^{i\phi}\tilde{h^{\prime}}G_{+}+\tilde{f^{\prime}}G_{-}),
\end{align}
\end{subequations}
with
\begin{subequations}
\begin{align}
\tilde{f^{\prime}}=&\frac{\left\vert \tilde{\omega}_{C,-}-\omega _{C,+}\right\vert}{\sqrt{\left( \tilde{\omega}_{C,-}-\omega_{C,+}\right) ^{2}+\tilde{\lambda}^{2}}}, \\
\tilde{h^{\prime}}=&\frac{\tilde{\lambda} }{\left( \tilde{\omega}_{C,-}-\omega_{C,+}\right)}\tilde{f^{\prime}}.
\end{align}
\end{subequations}
We note that the cooling of the two mechanical supermodes can also be explained by the physical mechanism proposed in this manuscript. By combining this phase-dependent phonon-exchange interaction with the optomechanical couplings, the interference effect works and the dark-mode effect is broken, which can lead to the ground-state cooling of the two mechanical supermodes.

\section{Physical mechanism for breaking the dark-state effect in a Lambda-type three-level system}
%%%%%%%%%%%%%%%%%%%%%
\begin{figure}[tbp]
\center
\includegraphics[width=6 cm]{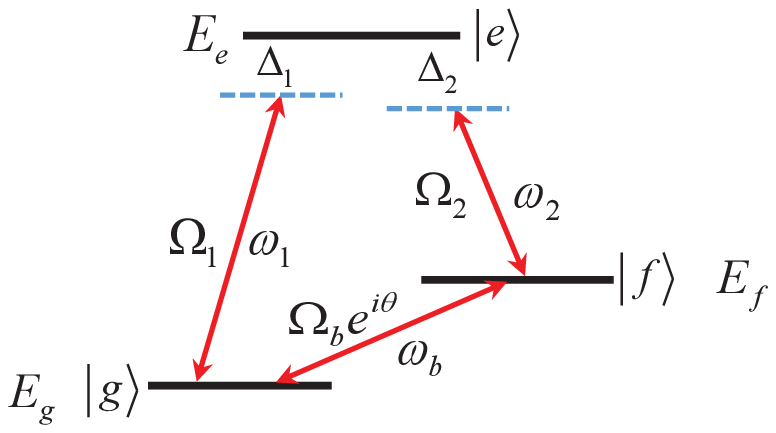}
\caption{ Schematic diagram of the three-level system with these states $\vert g\rangle$, $\vert f\rangle$, and $\vert e\rangle$ (with the corresponding energies $E_{g}$, $E_{f}$, and $E_{e}$). A Lambda-type coupling configuration is formed by the transition processes $\vert g\rangle\rightarrow\vert e\rangle$ and $\vert f\rangle \rightarrow\vert e\rangle$ with the coupling strengthes $\Omega_{1}$ and $\Omega_{2}$, and the detunings $\Delta_{1}$ and $\Delta_{2}$. A phase-dependent resonant coupling (with the coupling strength $\Omega_{b}e^{i\theta}$) between the two lower states $\vert g\rangle$ and $\vert f\rangle$ is introduced to break the dark-state effect existing in the Lambda-type three-level system working in the two-photon resonance regime $\Delta_{1}=\Delta_{2}=\Delta$.}
\label{3levelsystem}
\end{figure}
%%%%%%%%%%%%%%%%%%%%%%%

In this section, we show the physical mechanism for breaking the dark-state effect in a Lambda-type three-level system by introducing a phase-dependent transition between the two lower levels (as shown in Fig.~\ref{3levelsystem}). It is well known that there exists a dark state in the Lambda-type three-level system in the two-photon resonance regime. For the dark state, the superposition coefficient of the excited state is zero.
Below, we show that this dark state will be broken by introducing a phase-dependent transition coupling between the two lower states. Note that in a typical natural atom, the direct transition between the two lower states of a Lambda three-level atom is forbidden due to the transition selection rule. However, this transition is accessible either in artificial cycle three-level systems~\cite{Liu2005PRL1} or induced by indirect transition.
The Hamiltonian of the system reads
\begin{eqnarray}
H &=&E_{e}\vert e\rangle \langle e\vert+E_{f}\vert f\rangle \langle f\vert +E_{g}\vert g\rangle \langle g\vert
+\Omega _{1}(\vert e\rangle \langle g\vert e^{-i\omega _{1}t}+\vert g\rangle \langle e\vert e^{i\omega_{1}t})
+\Omega_{2}(\vert e\rangle \langle f\vert e^{-i\omega_{2}t}+\vert f\rangle \langle e\vert e^{i\omega_{2}t})  \notag \\
&&+\Omega _{b}(\vert f\rangle \langle g\vert e^{i\theta }e^{-i\omega _{b}t}+\vert g\rangle \langle f\vert e^{-i\theta }e^{i\omega _{b}t}),
\end{eqnarray}
where $E_{e}$, $E_{f}$, and $E_{g}$ are, respectively, the energies of these three energy levels $\vert e\rangle$, $\vert f\rangle$, and $\vert g\rangle$. Two monochromatic fields with frequencies $\omega_{1}$ and $\omega_{2}$ are coupled to the atomic transitions $\vert g\rangle\rightarrow\vert e\rangle$ and $\vert f\rangle \rightarrow\vert e\rangle$ (forming a Lambda configuration of couplings), respectively, with $\Omega_{1}$ and $\Omega_{2}$ being the corresponding real transition amplitudes. In this system, corresponding to these two transitions $\vert g\rangle\rightarrow\vert e\rangle$ and $\vert f\rangle \rightarrow\vert e\rangle$, we introduce the transition detunings as $\Delta_{1}=E_{e}-E_{g}-\omega_{1}$ and $\Delta_{2}=E_{e}-E_{f}-\omega_{2}$. We know that the Labmda-type couplings support a dark state in this system when the transitions satisfy the two-photon resonance condition [$\Delta_{1}=\Delta_{2}$ in Fig.~\ref{3levelsystem}]. Below, we will focus on the two-photon resonant transition case, $\Delta_{1}=\Delta_{2}=\Delta$. To exhibit our dark-state-breaking idea, we introduce a field to resonantly couple the two lower states $\vert f\rangle$ and $\vert g\rangle$. In particular, this coupling has a phase-dependent coupling strength, which is the critical factor for this dark-state-breaking approach.
In a rotating frame with respect to
\begin{equation}
H_{0}=(E_{g}+\omega_{1})\vert e\rangle \langle e\vert+E_{f}\vert f\rangle \langle f\vert +E_{g}\vert g\rangle \langle g\vert,
\end{equation}
the Hamiltonian of the system becomes
\begin{eqnarray}
V_{I} &=&\Delta \vert e\rangle \langle e\vert +\Omega_{1}( \vert e\rangle \langle g\vert +\vert g\rangle \langle e\vert )
+\Omega _{2}(\vert e\rangle \langle f\vert +\vert f\rangle \langle e\vert ) +\Omega _{b}(\vert f\rangle \langle g\vert e^{i\theta }+\vert g\rangle \langle f\vert e^{-i\theta}).
\end{eqnarray}
By defining these three basis states with the following vectors
\begin{equation}
\vert e\rangle=(1,0,0)^{T},\hspace{0.5 cm}\vert f\rangle=(0,1,0)^{T},\hspace{0.5 cm}\vert g\rangle=(0,0,1)^{T},
\end{equation}
where ``$T$" denotes the matrix transpose, the interaction Hamiltonian $V_{I}$ can be expressed as
\begin{eqnarray}
V_{I}&=&\left(\begin{array}{ccc}
\Delta & \Omega _{2} & \Omega _{1} \\
\Omega _{2} & 0 & \Omega _{b}e^{i\theta } \\
\Omega _{1} & \Omega _{b}e^{-i\theta } & 0
\end{array}
\right).\label{HamVI}
\end{eqnarray}
For the sake of simplicity and without loss of generality, we consider the symmetric coupling case $\Omega_{1}=\Omega_{2}=\Omega$ and the single- and two-photon resonance case $\Delta_{1}=\Delta_{2}=\Delta=0$, then the Hamiltonian~(\ref{HamVI}) becomes
\begin{equation}
V_{I}=\Omega\left(
\begin{array}{ccc}
0 & 1 & 1 \\
1 & 0 & \eta e^{i\theta } \\
1 & \eta e^{-i\theta } & 0
\end{array}
\right),\label{VIDelta0mat}
\end{equation}
where we introduce the ratio $\eta=\Omega_{b}/\Omega$.

The dark-state effect can be analyzed by investigating the eigensystem of the matrix $V_{I}$ in Eq.~(\ref{VIDelta0mat}). The eigen-equation can be expressed as
\begin{equation}
\frac{1}{\Omega}V_{I}\vert\lambda_{s}\rangle=\lambda_{s}\vert\lambda_{s}\rangle,\hspace{0.5 cm}s=1,2,3,\label{eigeneqVI}
\end{equation}
where $\lambda_{s}$ are the eigenvalues, which are determined by the secular (cubic) equation
\begin{equation}
\lambda^{3}-(2+\eta^{2})\lambda-2\eta\cos\theta=0.\label{seculareqcubic}
\end{equation}
Using the Cardano formula, the solutions of the cubic equation~(\ref{seculareqcubic}) can be obtained as
\begin{eqnarray}
\lambda_{1}=s_{1}+s_{2}, \hspace{0.5 cm}
\lambda_{2}=-\frac{1}{2}\left( s_{1}+s_{2}\right) +\frac{i\sqrt{3}}{2}\left(s_{1}-s_{2}\right), \hspace{0.5 cm}
\lambda_{3}=-\frac{1}{2}\left( s_{1}+s_{2}\right) -\frac{i\sqrt{3}}{2}\left(s_{1}-s_{2}\right),\label{cubiceigenvalues}
\end{eqnarray}
where
\begin{eqnarray}
s_{1}=\left(r+\sqrt{q^{3}+r^{2}}\right)^{\frac{1}{3}},\hspace{0.5 cm}
s_{2}=\left(r-\sqrt{q^{3}+r^{2}}\right)^{\frac{1}{3}},
\end{eqnarray}
with $q=-(2+\eta^{2})/3$ and $r=\eta\cos\theta$.
%%%%%%%%%%%%%%%%%%%%%
\begin{figure}[tbp]
\center
\includegraphics[width=0.95\textwidth]{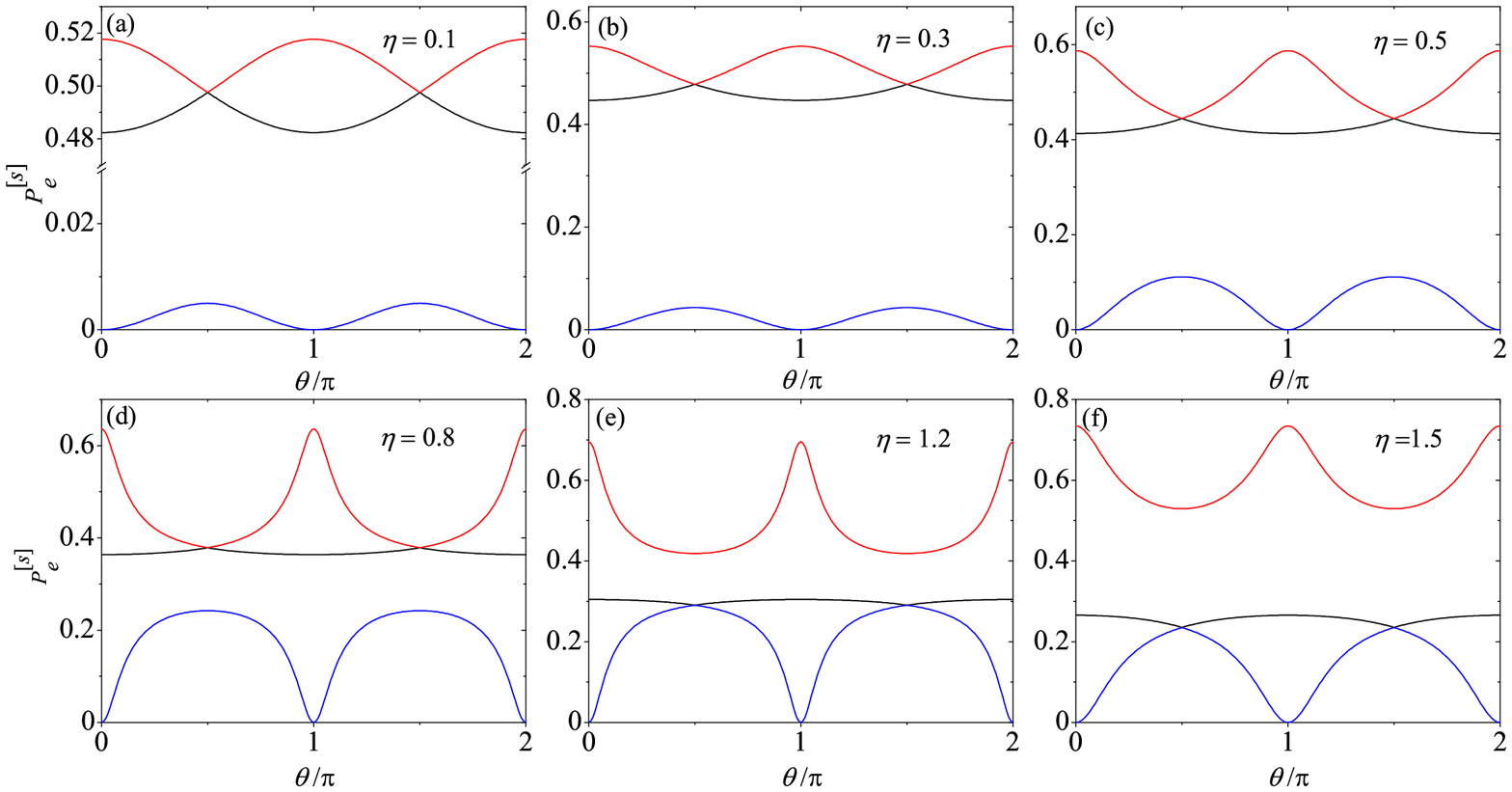}
\caption{The probability $P_{e}^{[s]}$ of the excited state $|e\rangle$ in these eigenstates $\vert\lambda_{s}\rangle$ as a function of $\theta$ when (a) $\eta=\Omega_{b}/\Omega=0.1$, (b) $0.3$, (c) $0.5$, (d) $0.8$, (e) $1.2$, and (f) $1.5$. Here, we can see that one of these three eigenstates has no excited state probability at $\theta=n\pi$, which means that there is a dark state at $\theta=n\pi$ and hence the dark-state effect is broken when $\theta\neq n\pi$.}
\label{darkstate3level}
\end{figure}
%%%%%%%%%%%%%%%%%%%%%%%

In general, the form of these eigenstates defined in Eq.~(\ref{eigeneqVI}) can be expressed as
\begin{equation}
\vert\lambda_{s}\rangle=c_{g}^{[s]}\vert g\rangle+c_{f}^{[s]}\vert f\rangle+c_{e}^{[s]}\vert e\rangle,\hspace{0.5 cm} s=1,2,3.
\end{equation}
The dark state can be checked by calculating the probability of the excited state $|e\rangle$ in these eigenstates as follows
\begin{equation}
P_{e}^{[s]}=\vert \langle e\vert\lambda_{s}\rangle \vert^{2}=\vert c_{e}^{[s]}\vert^{2},\hspace{0.5 cm} s=1,2,3.
\end{equation}
The case $P_{e}^{[s]}=0$ implies a dark state of this system. In Fig.~\ref{darkstate3level},
we plot the probability $P_{e}^{[s]}$ of the excited state $|e\rangle$ in these three eigenstates $\vert\lambda_{s}\rangle$ as a function of $\theta$ when the ratio $\eta=\Omega_{b}/\Omega$ takes various values.
Here we can see that when $\theta=n\pi$ for an integer $n$, one of the eigenstates becomes a dark state. In other cases, there are no dark states. Therefore, the phase-dependent resonant transition $|g\rangle\leftrightarrow|f\rangle$ can be used to break the dark-state effect in this Lambda-type three-level system.

The analytical expressions of these eigenstates can be obtained as
\begin{equation}
\left\vert\lambda_{s}\right\rangle=\Lambda_{s}\left[\vert g\rangle+\frac{\lambda_{s}\eta e^{i\theta}+1}{\lambda_{s}^{2}-1}\vert f\rangle+\frac{\eta e^{i\theta}+\lambda_{s}}{\lambda_{s}^{2}-1}\vert e\rangle\right],\hspace{0.5 cm}s=1,2,3,\label{eigstanalyr}
\end{equation}
where the corresponding eigenvalue $\lambda_{s}$ is given by Eq.~(\ref{cubiceigenvalues}), and the normalization constant is
\begin{equation}
\Lambda_{s}=\left\vert\frac{\left(1-\lambda_{s}^{2}\right)}{\sqrt{\lambda_{s}^{4}+\eta^{2}-\lambda_{s}^{2}+4\lambda_{s}\eta\cos\theta+\lambda_{s}^{2}\eta^{2}+2}}\right\vert,\hspace{0.5 cm}s=1,2,3.
\end{equation}
When one of these eigenstates is a dark state, then the probability amplitude of the excited state $\vert e\rangle$ in this eigenstate~(\ref{eigstanalyr}) is zero, and we have the relation
\begin{equation}
\lambda=-\eta e^{i\theta}.
\end{equation}
By substituting the above relation into the secular equation Eq.~(\ref{seculareqcubic}),
we have
\begin{equation}
\eta^{3}\left(-e^{i3\theta}+e^{i\theta}\right) +i2\eta\sin\theta=0,
\end{equation}
which leads to these two equations
\begin{eqnarray}
\left[\cos\theta-\cos\left(3\theta\right)\right]\eta^{3}=0,\hspace{0.5 cm}\eta^{3}\left[\sin\theta-\sin\left(3\theta\right)\right]+2\eta\sin\theta=0.
\end{eqnarray}%
For a nonzero $\eta$, the solutions of these two equations are
\begin{equation}
\theta =n\pi, \hspace{0.5 cm}n=0,\pm1,\pm2, \cdots.
\end{equation}

When $\theta=n\pi$, we have $e^{i\theta}=e^{-i\theta}=(-1)^{n}$, then the eigenvalues of the matrix~(\ref{VIDelta0mat}) are given by
\begin{equation}
\lambda_{1}=(-1)^{n+1}\eta,\hspace{0.5 cm} \lambda_{2}=\frac{1}{2}\left[(-1)^{n}\eta-\sqrt{8+\eta^{2}}\right],\hspace{0.5 cm}\lambda_{3}=\frac{1}{2}\left[(-1)^{n}\eta+\sqrt{8+\eta^{2}}\right].
\end{equation}
The corresponding eigenstates are given by
\begin{eqnarray}
\vert\lambda_{1}\rangle&=&\frac{1}{\sqrt{2}}(-|f\rangle+|g\rangle),\notag \\
\vert\lambda_{2}\rangle&=&\Lambda_{2}\left\{\frac{1}{2}\left[(-1)^{n+1}\eta-\sqrt{8+\eta^{2}}\right]|e\rangle+|f\rangle+|g\rangle\right\},\notag \\
\vert\lambda_{3}\rangle&=&\Lambda_{3}\left\{\frac{1}{2}\left[(-1)^{n+1}\eta+\sqrt{8+\eta^{2}}\right]|e\rangle+|f\rangle+|g\rangle\right\},
\end{eqnarray}
where $\Lambda_{2,3}=[\frac{1}{4}(\sqrt{8+\eta^{2}}\pm(-1)^{n}\eta)^2+2]^{-1/2}$ are normalization constants. In this case, the eigenstate $\vert\lambda_{1}\rangle$ is a dark state.

\section{A possible experimental realization and derivation of a phase-dependent phonon-hopping interaction between two mechanical resonators}

\subsection{A possible experimental realization}
%%%%%%%%%%%%%%%%%%%%%%%
\begin{figure}[tbp]
\center
\includegraphics[bb=11 430 527 724, width=12 cm]{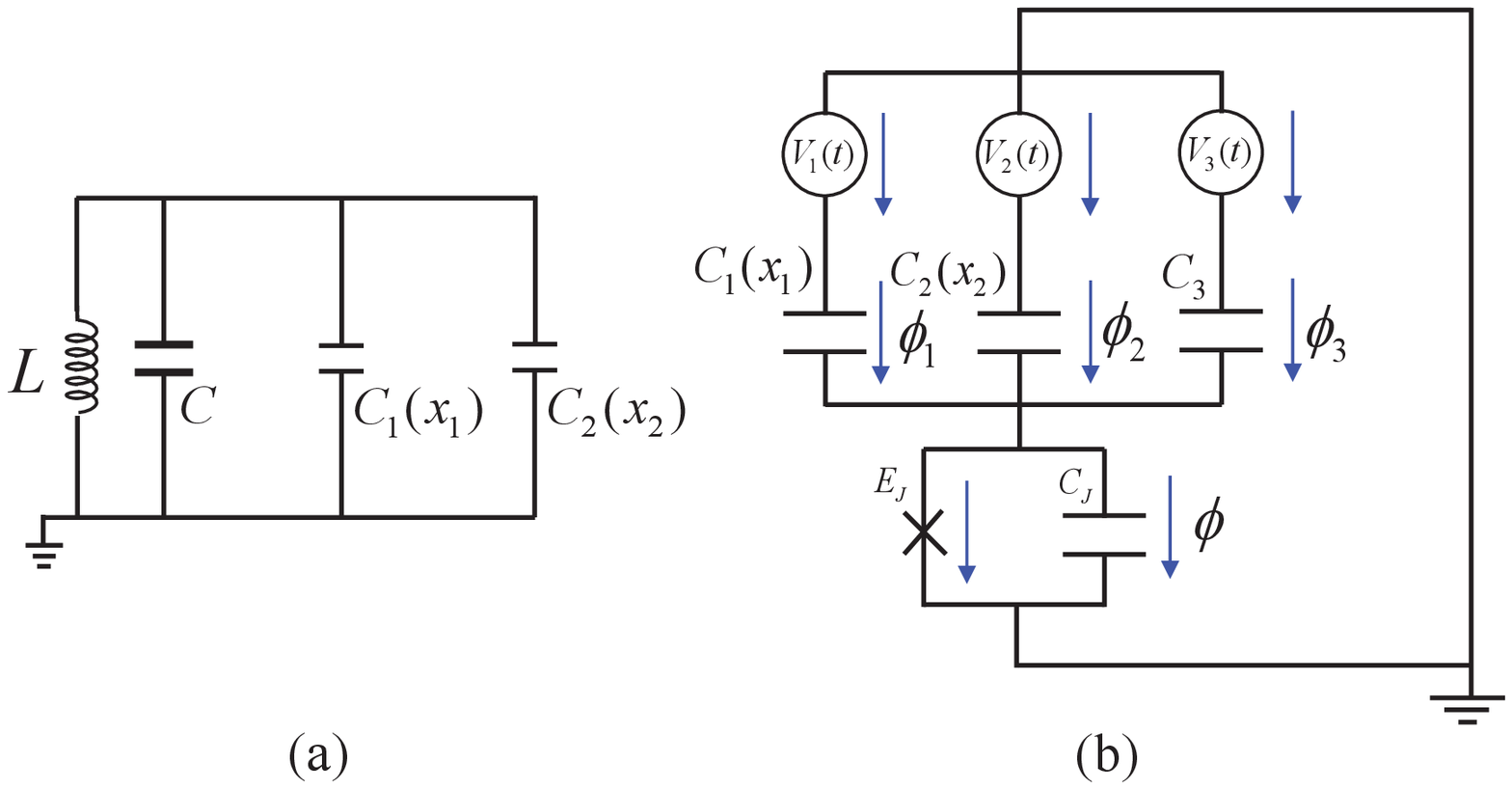}
\caption{(a) The circuit electromechanical system consists of a microwave cavity represented by an inductance $L$ and three capacitances: $C$ and $C_{j=1,2}(x_{j})$. Here, the two capacitances $C_{j=1,2}(x_{j})$ depend on the two micromechanical resonators $b_{j=1,2}$. The displacement $x_{j=1,2}$ of each mechanical resonator modulates the total capacitance and hence the cavity frequency $\omega_{c}$. A phase-dependent phonon-hopping interaction $\eta(e^{i\theta}b_{1}^{\dagger}b_{2}+e^{-i\theta}b_{2}^{\dagger}b_{1})$ between the two micromechanical resonators is generated via a superconducting quantum circuit given in panel (b). (b) Schematic diagram of the superconducting quantum circuit: A Josephson junction with the Josephson energy $E_{J}$ and the capacitance $C_{J}$ is connected to three gate voltages $V_{j=1,2,3}(t)$ through the corresponding gate capacitances $C_{j=1,2}(x_{j})$ and $C_{3}$. Two mechanical resonators are coupled to the superconducting charge qubit through the gate capacitances $C_{j=1,2}(x_{j})$. The gate voltages are properly designed such that a phase-dependent phonon-hopping interaction between the two mechanical resonators can be induced. The phase drops across these capacitor $C_{j=1,2,3}$ and the Josephson junction are marked as $\phi_{j}$ and $\phi$, respectively.}
\label{circuits}
\end{figure}
%%%%%%%%%%%%%%%%%%%%%%%

In this section, we propose a possible experimental implementation of our scheme based on the circuit electromechanical system, as shown in Fig.~\ref{circuits}(a). The circuit electromechanical system~\cite{Massel2011Nature1,Massel2012Nc2} consists of a microwave cavity described by the equivalent inductance $L$ and capacitance $C$ and two micromechanical resonators $b_{j=1,2}$. The electromechanical coupling arises when the displacement $x_{j=1,2}$ of each mechanical resonator independently modulates the total capacitance through $C_{j=1,2}(x_{j})$, and therefore the resonance frequency of the cavity $\omega_{c}$. This electromechanical coupling can be described by $g_{j}=(\omega_{c}/2C)\partial C_{j}/\partial x_{j}$. Meanwhile, an effective phase-dependent phonon-hopping interaction between the two mechanical resonators is introduced by coupling them to a superconducting charge qubit, as shown in Fig.~\ref{circuits}(b). The detailed derivation of the phase-dependent phonon-exchange interaction is presented in the next subsection.

\subsection{Derivation of a phase-dependent phonon-hopping interaction between two mechanical resonators}

In this section, we present a detailed derivation of an effective phase-dependent phonon-hopping interaction between two mechanical resonators. Here the two mechanical resonators are coupled to a superconducting charge qubit, which is described by the circuit given in Fig.~\ref{circuits}(b). In this circuit, a Josephson junction with the Josephson energy $E_{J}$ and the capacitance $C_{J}$ is connected to three gate voltages $V_{j=1,2,3}(t)$ through the corresponding gate capacitances $C_{j=1,2}(x_{j})$ and $C_{3}$. Here the two gate capacitors with capacitances $C_{j=1,2}(x_{j})$ are formed by one fixed plate and one mechanical resonator. The third capacitor has a constant capacitance. We denote the phase drops across these capacitor $C_{j=1,2,3}$ and the Josephson junction as $\phi_{j}$ and $\phi$, respectively. In this circuit, the energy stored in these capacitors is the total kinetic energy~\cite{Nakaharabook1}, which can be written as
\begin{equation}
T=\frac{1}{2}C_{1}(x_{1}) \dot{\Phi}_{1}^{2}+\frac{1}{2}C_{2}(x_{2}) \dot{\Phi}_{2}^{2}+\frac{1}{2}C_{3}\dot{\Phi}_{3}^{2}+\frac{1}{2}C_{J}\dot{\Phi}^{2},
\end{equation}
where $\Phi_{j=1,2,3}$ and $\Phi$ are the generalized magnetic fluxes associated with the phase drops $\phi_{j}$ and $\phi$ across the capacitances $C_{j}$ and the Josephson junction. The relation between the generalized magnetic flux and the phase drop is defined by $\phi_{j=1,2,3}=2\pi\Phi_{j}/\Phi_{0}$, where $\Phi_{0}$ is the magnetic flux quanta. The Josephson energy is identified as the potential energy, which takes the form as~\cite{Nakaharabook1}
\begin{equation}
U=-E_{J}\cos\left(\frac{2\pi}{\Phi_{0}}\Phi\right),
\end{equation}
where $E_{J}$ is the Josephson energy of this junction.

Based on these voltages relations in these loops, we have the relations
\begin{eqnarray}
V_{j}(t) +\dot{\Phi}_{j}+\dot{\Phi} &=&0,\hspace{0.5 cm} j=1,2,3,
\end{eqnarray}
then the Lagrangian of this system can be expressed as
\begin{eqnarray}
L&=&T-U\nonumber\\
&=&\frac{1}{2}C_{1}\left( x_{1}\right) V_{1}^{2}\left( t\right) +\frac{1}{2}C_{2}\left( x_{2}\right) V_{2}^{2}\left( t\right) +\frac{1}{2}
C_{3}V_{3}^{2}\left( t\right) +\frac{1}{2}\left( C_{1}\left( x_{1}\right)+C_{2}\left( x_{2}\right) +C_{3}+C_{J}\right) \dot{\Phi}^{2}\notag \\
&&+\left[ C_{1}\left( x_{1}\right) V_{1}\left( t\right) +C_{2}\left(x_{2}\right) V_{2}\left( t\right) +C_{3}V_{3}\left( t\right) \right]\dot{\Phi}+E_{J}\cos \left( \frac{2\pi }{\Phi _{0}}\Phi \right).
\end{eqnarray}
We introduce the momentum canonically conjugate to $\Phi$ as
\begin{eqnarray}
P &=&\frac{\partial L}{\partial \dot{\Phi}}
=\left[ C_{1}\left( x_{1}\right) V_{1}\left( t\right) +C_{2}\left(
x_{2}\right) V_{2}\left( t\right) +C_{3}V_{3}\left( t\right) \right] +\left[
C_{1}\left( x_{1}\right) +C_{2}\left( x_{2}\right) +C_{3}+C_{J}\right] \dot{
\Phi}.
\end{eqnarray}
Then the Hamiltonian of this circuit can be derived as~\cite{Nakaharabook1}
\begin{eqnarray}
H &=&\frac{1}{2}\frac{4e^{2}}{C_{\Sigma }\left( x_{1},x_{2}\right) }\left[
\hat{n}-n_{g}\left( x_{1},x_{2},t\right) \right] ^{2}-E_{J}\cos \left( \frac{%
2\pi }{\Phi _{0}}\Phi \right)  \notag \\
&&-\frac{1}{2}\left[ C_{1}\left( x_{1}\right) V_{1}^{2}\left( t\right)
+C_{2}\left( x_{2}\right) V_{2}^{2}\left( t\right) +C_{3}V_{3}^{2}\left(
t\right) \right],
\end{eqnarray}%
where we introduce the Cooper-pair number $n$, the gate capacitance $C_{\Sigma}\left( x_{1},x_{1}\right)$, and the gate Cooper-pair number $n_{g}$, which are defined by
\begin{equation}
P=2en,\hspace{1 cm}
C_{\Sigma }\left(x_{1},x_{2}\right)=C_{1}\left(x_{1}\right)+C_{2}\left(x_{2}\right)+C_{3}+C_{J},
\end{equation}
and
\begin{equation}
n_{g}\left( x_{1},x_{2},t\right) =\frac{1}{2e}\left[ C_{1}\left(
x_{1}\right) V_{1}\left( t\right) +C_{2}\left( x_{2}\right) V_{2}\left(
t\right) +C_{3}V_{3}\left( t\right) \right].
\end{equation}

The quantization of this circuit can be performed by introducing the commutative relation between the number operator $\hat{n}$ and the phase operator $\hat{\phi}$ as $[\hat{\phi},\hat{n}]=i$.
Then we can express the Hamiltonian in the eigen-representation of the number operator as
\begin{eqnarray}
H &=&\frac{1}{2}\frac{4e^{2}}{C_{\Sigma }\left( x_{1},x_{2}\right) }\sum_{n\in Z}
\left[ n-n_{g}\left( x_{1},x_{2},t\right) \right] ^{2}\left\vert
n\right\rangle \left\langle n\right\vert -\frac{E_{J}}{2}\sum_{n\in Z}\left(
\left\vert n\right\rangle \left\langle n+1\right\vert +\left\vert
n+1\right\rangle \left\langle n\right\vert \right)  \notag \\
&&-\frac{1}{2}\left[ C_{1}\left( x_{1}\right) V_{1}^{2}\left( t\right)
+C_{2}\left( x_{2}\right) V_{2}^{2}\left( t\right) +C_{3}V_{3}^{2}\left(
t\right) \right].
\end{eqnarray}

In this work, we consider the case where this circuit works in the charge qubit regime $E_{C}\gg E_{J}$, with $E_{C}=4e^{2}/C_{\Sigma}$ being the Coulomb energy.
In particular, we choose the gate charge in the vicinity of $1/2$, so that the states $\vert 0\rangle$ and $\vert 1\rangle$ have almost degenerate energies. In this case, other states have higher energies and can be
ignored in the our discussions. Then the Hamiltonian becomes
\begin{eqnarray}
H &\approx&\frac{1}{2}\frac{4e^{2}}{C_{\Sigma }\left( x_{1},x_{2}\right) }\left[
n_{g}\left( x_{1},x_{2},t\right) ^{2}\left\vert 0\right\rangle \left\langle 0\right\vert +\left[ 1-n_{g}\left( x_{1},x_{2},t\right) \right]^{2}\left\vert 1\right\rangle \left\langle 1\right\vert \right]-\frac{E_{J}}{2}\left(\left\vert 0\right\rangle \left\langle 1\right\vert+\left\vert 1\right\rangle \left\langle 0\right\vert \right) \notag \\
&&-\frac{1}{2}\left[ C_{1}\left( x_{1}\right) V_{1}^{2}\left( t\right) +C_{2}\left(
x_{2}\right) V_{2}^{2}\left( t\right) +C_{3}V_{3}^{2}\left( t\right) \right].
\end{eqnarray}
By introducing the Pauli operators $\left\vert 0\right\rangle \left\langle 0\right\vert -\left\vert
1\right\rangle \left\langle 1\right\vert=\sigma _{z}$ and $
\left\vert 0\right\rangle \left\langle 0\right\vert +\left\vert
1\right\rangle \left\langle 1\right\vert=I$, we can express the Hamiltonian as
\begin{eqnarray}
H &=&\frac{1}{2}\frac{4e^{2}}{C_{\Sigma }\left( x_{1},x_{2}\right) }\left[
n_{g}\left( x_{1},x_{2},t\right) -\frac{1}{2}\right] \sigma _{z}-\frac{E_{J}
}{2}\sigma _{x}+M,
\end{eqnarray}
where the term $M$ stands for the ac voltage driving term on these two mechanical resonators
\begin{eqnarray}
M=\frac{1}{4}\frac{4e^{2}}{C_{\Sigma }\left( x_{1},x_{2}\right) }\left[
1-2n_{g}\left(x_{1},x_{2},t\right) +2n_{g}^{2}\left( x_{1},x_{2},t\right)
\right]-\frac{1}{2}\left[ C_{1}\left( x_{1}\right) V_{1}^{2}\left( t\right)
+C_{2}\left( x_{2}\right)V_{2}^{2}\left( t\right)+C_{3}V_{3}^{2}\left(
t\right) \right].
\end{eqnarray}
We consider the case in which the voltage drivings are far-off-resonance to these two mechanical resonators (namely the driving frequencies of the two voltages are much smaller than the resonance frequencies of the two mechanical resonators) and then the term $M$ will be discarded in our following discussions.
When the vibration amplitudes of the mechanical resonators are much smaller than the distances between the fixed plate and the rest mechanical resonator of the capacitors, we can approximate the capacitances as
\begin{eqnarray}
C_{1}\left( x_{1}\right)\approx C_{10}\left( 1-\frac{x_{1}}{l_{1}}\right),\hspace{0.5 cm} C_{2}\left( x_{2}\right)\approx C_{20}\left( 1-\frac{x_{2}}{l_{2}}\right),
\end{eqnarray}
where $C_{j0}$ (for $j=1,2$) are the capacitances of the gate capacitors when the mechanical resonators are rest, and $l_{j=1,2}$ are the rest distances between the fixed plate and the mechanical resonators in these gate capacitors. In addition, we choose the following gate voltages for our purpose,
\begin{eqnarray}
V_{1}\left(t\right)&=&V_{10}\cos \left( \omega _{1}t+\varphi _{1}\right),\hspace{0.5 cm}V_{2}\left( t\right)=V_{20}\cos \left( \omega _{2}t+\varphi _{2}\right),\hspace{0.5 cm}
V_{3}\left(t\right)=\frac{e-C_{10}V_{1}(t) -C_{20}V_{2}(t)}{C_{3}}.
\end{eqnarray}
In this case, we can obtain the relation
\begin{equation}
n_{g}\left( x_{1},x_{2},t\right)-\frac{1}{2}=-\left[ \frac{C_{10}V_{10}}{2e}\frac{x_{1}}{l_{1}}\cos\left(\omega_{1}t+\varphi_{1}\right)
+\frac{C_{20}V_{20}}{2e}\frac{x_{2}}{l_{2}}\cos\left(\omega_{2}t+\varphi_{2}\right) \right].
\end{equation}%
By making the rotation for the qubit $-\sigma _{x} \rightarrow \tau _{z}$ and $\sigma _{z}\rightarrow\tau _{x}$, we can express the Hamiltonian upto the first order of the mechanical displacements $x_{1}$ and $x_{2}$ as
\begin{equation}
H_{I}\approx \frac{E_{J}}{2}\tau_{z}-\frac{E_{C}}{2}\left[\frac{C_{10}V_{10}}{2e}\frac{x_{1}}{l_{1}}\cos\left(\omega_{1}t+\varphi_{1}\right) +\frac{C_{20}V_{20}}{2e}\frac{x_{2}}{l_{2}}\cos\left(\omega_{2}t+\varphi_{2}\right)\right]\tau_{x},
\end{equation}
where $E_{C}=4e^{2}/C_{\Sigma0}$ under the approximation $C_{\Sigma}\left(x_{1},x_{2}\right)\approx\left( C_{10}+C_{20}+C_{J}\right)\equiv C_{\Sigma0}$. We should point out that the mechanical displacement terms in $C_{\Sigma}\left(x_{1},x_{2}\right)$ only introduce the second-order terms of $x_{j=1,2}/l_{j}$, which have been neglected in our considerations.

By including the free Hamiltonian of the two mechanical resonators and using the relations $x_{j=1,2}=\sqrt{\hbar/(2m\omega_{m})}(b_{j}+b_{j}^{\dagger})$ and $p_{j=1,2}=-i\sqrt{\hbar m\omega_{m}/2}(b_{j}-b_{j}^{\dagger})$, the total Hamiltonian of this circuit system becomes
\begin{eqnarray}
H_{I} &\approx &\omega _{m}b_{1}^{\dagger }b_{1}+\omega _{m}b_{2}^{\dagger}b_{2}+\frac{\omega_{0}}{2}\tau _{z}  \notag \\
&&-\left[ g_{1}\left( b_{1}+b_{1}^{\dagger }\right) \left( e^{i\left( \omega
_{d}t+\varphi _{1}\right) }+e^{-i\left( \omega _{d}t+\varphi _{1}\right)
}\right) +g_{2}\left( b_{2}+b_{2}^{\dagger }\right) \left( e^{i\left( \omega
_{d}t+\varphi _{2}\right) }+e^{-i\left( \omega _{d}t+\varphi _{2}\right)
}\right) \right] \left( \tau _{+}+\tau _{-}\right),
\end{eqnarray}%
where we consider the case $\omega_{1}=\omega_{2}=\omega_{d}$ and introduce these parameters
\begin{eqnarray}
g_{1}=\frac{E_{C}}{4}\frac{C_{10}V_{10}}{2e}\frac{x_{10}}{l_{1}},\hspace{0.5 cm}
g_{2}=\frac{E_{C}}{4}\frac{C_{20}V_{20}}{2e}\frac{x_{20}}{l_{2}}, \hspace{0.5 cm}
\omega_{0}=E_{J},
\end{eqnarray}
with $x_{j0}=\sqrt{\hbar/(2m\omega_{m})}$ being the zero-point fluctuation of these mechanical resonators.
%%%%%%%%%%%%%%%%%%%%%%%
\begin{figure}[tbp]
\center
\includegraphics[bb=35 515 260 680, width=5 cm]{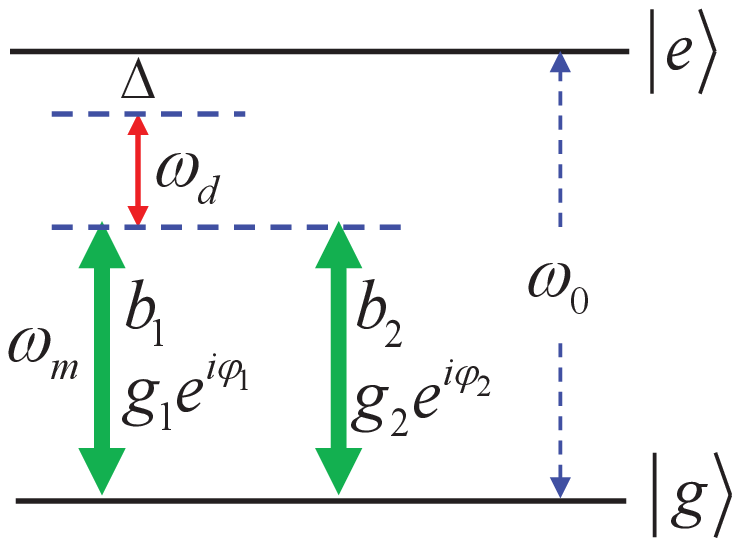}
\caption{Schematic diagram of the energy levels and these involved resonance frequencies of this coupled qubit-resonator system. Two mechanical resonators with resonance frequency $\omega_{m}$ are phase-dependently coupled to the superconducting charge qubit with the energy separation $\omega_{0}$. The ac gate voltages with frequency $\omega_{d}$ are applied to the Josephson junction through the gate capacitors.}
\label{levels}
\end{figure}
%%%%%%%%%%%%%%%%%%%%%%%

To analyze the physical processes in this system, we now work in the rotating frame with respect to
\begin{equation}
H_{0}=\omega_{m}b_{1}^{\dagger}b_{1}+\omega_{m}b_{2}^{\dagger}b_{2}+\frac{\omega_{0}}{2}\tau_{z},
\end{equation}
then the Hamiltonian becomes
\begin{eqnarray}
V_{I}(t)&=&-g_{1}(\tau_{+}b_{1}^{\dagger}e^{i\left(\omega_{0}+\omega_{m}+\omega_{d}\right)t}e^{i\varphi_{1}}+b_{1}\tau_{-}e^{-i\left(\omega_{0}+\omega_{m}+\omega_{d}\right)t}e^{-i\varphi_{1}}) \nonumber\\
&&-g_{2}(\tau_{+}b_{2}^{\dagger}e^{i\left(\omega_{0}+\omega_{m}+\omega_{d}\right)t}e^{i\varphi_{2}}+b_{2}\tau_{-}e^{-i\left(\omega_{0}+\omega_{m}+\omega_{d}\right)t}e^{-i\varphi_{2}})\nonumber \\
&&-g_{1}(\tau_{+}b_{1}^{\dagger}e^{i\left(\omega_{0}+\omega_{m}-\omega_{d}\right)t}e^{-i\varphi_{1}}+b_{1}\tau_{-}e^{-i\left(\omega_{0}+\omega_{m}-\omega_{d}\right)t}e^{i\varphi_{1}})\nonumber\\
&&-g_{2}(\tau_{+}b_{2}^{\dagger}e^{i\left(\omega_{0}+\omega_{m}-\omega_{d}\right)t}e^{-i\varphi_{2}}+b_{2}\tau_{-}e^{-i\left(\omega_{0}+\omega_{m}-\omega_{d}\right)t}e^{i\varphi_{2}}) \nonumber\\
&&-g_{1}(\tau_{+}b_{1}e^{i\left(\omega_{0}-\omega_{m}+\omega_{d}\right)t}e^{i\varphi_{1}}+b_{1}^{\dagger}\tau_{-}e^{-i\left(\omega_{0}-\omega_{m}+\omega_{d}\right)t}e^{-i\varphi_{1}})\nonumber\\
&&-g_{2}(\tau_{+}b_{2}e^{i\left(\omega_{0}-\omega_{m}+\omega_{d}\right)t}e^{i\varphi_{2}}+b_{2}^{\dagger}\tau_{-}e^{-i\left(\omega_{0}-\omega_{m}+\omega_{d}\right)t}e^{-i\varphi_{2}}) \nonumber\\
&&-g_{1}(\tau_{+}b_{1}e^{i\left(\omega_{0}-\omega_{m}-\omega_{d}\right)t}e^{-i\varphi_{1}}+b_{1}^{\dagger}\tau_{-}e^{-i\left(\omega_{0}-\omega_{m}-\omega_{d}\right)t}e^{i\varphi_{1}}) \nonumber\\
&&-g_{2}(\tau_{+}b_{2}e^{i\left(\omega_{0}-\omega_{m}-\omega_{d}\right)t}e^{-i\varphi_{2}}+b_{2}^{\dagger}\tau_{-}e^{-i\left(\omega_{0}-\omega_{m}-\omega_{d}\right)t}e^{i\varphi_{2}}).
\end{eqnarray}
Here we can see that in this system there are eight physical processes, which are determined by the four detunings $\omega_{0}+\omega_{m}\pm\omega_{d}$ and $\omega_{0}-\omega_{m}\pm\omega_{d}$. From the viewpoint of the qubit and the resonators, the terms including $\omega_{0}+\omega_{m}\pm\omega_{d}$ and $\omega_{0}-\omega_{m}\pm\omega_{d}$ are the counterrotating terms and the corotating terms, respectively.
In this work, the motivation for introducing the ac voltages $V_{1}(t)$ and $V_{2}(t)$ is to pick up the phase-sensitive interactions between the mechanical resonators and the charge qubit. For this purpose, we choose the ac voltages with the frequency $\omega_{d}$ to pick up the terms with $\omega_{0}-\omega_{m}-\omega_{d}$. Namely, we choose the
parameters to satisfy the following parameter conditions
\begin{equation}
\omega_{0}+\omega_{m}\pm\omega_{d}\gg\omega_{0}-\omega_{m}+\omega_{d}\gg\omega_{0}-\omega_{m}-\omega_{d}.
\end{equation}
The terms with $\omega_{0}+\omega_{m}\pm\omega_{d}$ and $\omega_{0}-\omega_{m}+\omega_{d}$ are the far-off-resonance terms and the terms with $\omega_{0}-\omega_{m}-\omega_{d}$ are the target terms which work in the large-detuning regime. The energy levels and these involved resonance frequencies of this coupled qubit-resonator system are shown in Fig.~\ref{levels}. In this case, the qubit-resonator interactions work in the large-detuning regime: $\Delta\gg g_{j=1,2}\sqrt{n_{j}}$, where $n_{j}$ is the maximal excitation number involved in the $j$th mechanical resonator, and then we can obtain a phase-dependent photon-hopping interaction between the two mechanical resonators. Here the phase is the difference between the two phases $\varphi_{1}$ and $\varphi_{2}$ associated with the qubit-resonator couplings.

Based on the above analyses, we can obtain the approximate Hamiltonian as
\begin{eqnarray}
V_{I}\left(t\right)&\approx&-\left[\tau_{+}\left(g_{1}b_{1}e^{-i\varphi_{1}}+g_{2}b_{2}e^{-i\varphi_{2}}\right)e^{i\Delta t}+\left(
g_{1}b_{1}^{\dagger}e^{i\varphi_{1}}+g_{2}b_{2}^{\dagger}e^{i\varphi_{2}}\right)\tau_{-}e^{-i\Delta t}\right],
\end{eqnarray}%
where we introduce the detuning $\Delta=\omega_{0}-\omega_{m}-\omega_{d}$.
The time factor can be eliminated by going back to the Schr\"{o}dinger representation, in which the Hamiltonian of the system can be written as
\begin{equation}
H_{\text{eff}}=\omega_{m}b_{1}^{\dagger}b_{1}+\omega_{m}b_{2}^{\dagger}b_{2}+\frac{\omega_{0}-\omega_{d}}{2}\tau_{z}-\tau_{+}\left(
g_{1}b_{1}e^{-i\varphi_{1}}+g_{2}b_{2}e^{-i\varphi_{2}}\right)-\left(g_{1}b_{1}^{\dagger}e^{i\varphi_{1}}+g_{2}b_{2}^{\dagger}e^{i\varphi_{2}}\right)\tau_{-}.
\end{equation}
In this work, we consider the physical process associated with the detuning $\Delta$ working in the large detuning case. Then we can adiabatically eliminate the qubit coherence in the physical processes and an effective phonon-phonon interaction between the two mechanical modes can be induced by the second-order perturbation. In this case, we can derive an effective Hamiltonian to describe the interactions using the method of the Frohlich-Nakajima transformation~\cite{Frohlich19501,Nakajima19531}. To this end, we express the effective Hamiltonian $H_{\text{eff}}$ as two parts
\begin{eqnarray}
H_{0} &=&\omega_{m}b_{1}^{\dagger }b_{1}+\omega_{m}b_{2}^{\dagger}b_{2}+\frac{\omega_{0}-\omega_{d}}{2}\tau_{z},\nonumber\\
H_{I} &=&-\tau_{+}\left(g_{1}b_{1}e^{-i\varphi_{1}}+g_{2}b_{2}e^{-i\varphi_{2}}\right)-\tau_{-}\left(g_{1}b_{1}^{\dagger}e^{i\varphi_{1}}+g_{2}b_{2}^{\dagger}e^{i\varphi_{2}}\right).
\end{eqnarray}
We also introduce the operator
\begin{eqnarray}
S&=&\frac{1}{\Delta}\tau_{+}\left(g_{1}b_{1}e^{-i\varphi_{1}}+g_{2}b_{2}e^{-i\varphi_{2}}\right)
-\frac{1}{\Delta}\left(g_{1}b_{1}^{\dagger}e^{i\varphi_{1}}+g_{2}b_{2}^{\dagger}e^{i\varphi_{2}}\right)\tau_{-},
\end{eqnarray}
which is determined by the equation
\begin{equation}
H_{I}+[H_{0},S]=0.
\end{equation}
This equation means that the first-order physical process is eliminated.
An effective Hamiltonian describing the second-order physical interaction can then be obtained as
\begin{eqnarray}
H'_{\textrm{eff}}&=&H_{0}+\frac{1}{2}\left[H_{I},S\right]\nonumber\\
&=&\omega_{m}b_{1}^{\dagger}b_{1}+\omega_{m}b_{2}^{\dagger}b_{2}+\frac{\omega_{0}-\omega _{d}}{2}\tau _{z}
+\frac{g_{1}^{2}}{\Delta}\tau_{z}b_{1}^{\dagger}b_{1}+\frac{g_{2}^{2}}{
\Delta}\tau_{z}b_{2}^{\dagger}b_{2}+\frac{\left(g_{1}^{2}+g_{2}^{2}\right)}{\Delta}\tau_{+}\tau_{-}\nonumber\\
&&+\frac{g_{1}g_{2}}{\Delta}\tau_{z}\left(b_{1}^{\dagger}b_{2}e^{i\left(\varphi_{1}-\varphi
_{2}\right)}+b_{2}^{\dagger}b_{1}e^{-i\left(\varphi_{1}-\varphi_{2}\right)}\right).
\end{eqnarray}
The above Hamiltonian shows that there is no transition in the qubit states, and that a conditional phase-dependent interaction between the two mechanical resonators is introduced.
We assume that the qubit is initial in its ground state $\vert g\rangle$ ($\tau_{z}\vert g\rangle=-\vert g\rangle$), then a phase-dependent phonon-hopping interaction is obtained.

\end{document}